\documentclass[%
               paper=a4,
               pagesize=auto,
               fontsize=12pt,
               twoside=off,
               titlepage=off,%
               abstract=off,%
               headings=small,
               captions=tableheading,
               captions=oneline,%
               headsepline=on,
               headinclude=off,%
               footinclude=off,%
               parskip=half+,%
              ]{scrartcl} 
\pdfoutput=1
 \usepackage[%
             pdfpagelabels=true,
             pdftex,
            ]{hyperref}
 \hypersetup{%
             colorlinks=false,
             pdftoolbar=true,
             pdfpagemode=UseOutlines,
             pdfpagelayout=TwoColumnRight,
             pdfstartview=Fit,
             bookmarksopen=true,
             bookmarksopenlevel=4,
             pdftitle = {Efficient prediction of broadband trailing edge noise and application to porous edge treatment},%
             pdfsubject = {Contribution to Special Issue International Journal of Aeroacoustics},%
             pdfauthor = {},%
             pdfkeywords = {Trailing Edge Noise, Porous Material, CAA, PIANO},%
             pdfcreator = {Adobe-Acrobat-Distiller},%
             pdfproducer = {LaTeX with hyperref}%
            }

\usepackage[utf8]{inputenc} 
\usepackage{subfigure}
\ifpdfoutput{\usepackage[pdftex]{graphicx}}{\usepackage{graphicx}} 
\graphicspath{{floats/}} 
\usepackage{booktabs}
\usepackage{multirow}
\usepackage{rotating}
\usepackage[intlimits,sumlimits,namelimits]{amsmath}
\usepackage{amsfonts,amssymb}
\usepackage{nicefrac}
\usepackage{xspace} 
\selectfont 
\areaset[7mm]{175mm}{240mm}  
\usepackage[autooneside,automark]{scrpage2} 
\usepackage{pdflscape}
\setcapindent{1em} 
\addtokomafont{caption}{\small\sffamily} 
\setkomafont{captionlabel}{\bfseries} 
\newcommand{\bfm}[1]{\mbox{\boldmath $#1$}}
\newcommand{\pp}[2]{\frac{\partial{#1}}{\partial{#2}}}
\newcommand{\ppsd}[2]{\frac{\partial^2 #1}{\partial #2^2}}
\newcommand{\ppsq}[3]{\frac{\partial^2{#1}}{\partial{#2} \partial{#3}}}

\newcommand{\nop}{\bfm \nabla}
\newcommand*{\comment}[2][note!]{}
\newcommand*{\blindcomment}[2][]{}
\newcommand*{\ind}[1]{\text{{#1}}}
\newcommand*{\Ma}{\text{Ma}\xspace}
\newcommand*{\Rey}{\text{Re}\xspace}
\newcommand*{\fig}[1]{Fig.~\ref{#1}\xspace}
\newcommand*{\Fig}[1]{Fig.~\ref{#1}\xspace}
\newcommand*{\figs}[3][and]{Figs.~\ref{#2}~#1~\ref{#3}\xspace}
\newcommand*{\eqn}[1]{Eq.~(\ref{#1})\xspace}
\newcommand*{\eqns}[3][and]{Eqs.~(\ref{#2})~#1~(\ref{#3})\xspace}
\newcommand*{\vM}[1]
  {\boldsymbol{{#1}}}
\DeclareMathOperator{\const}{\mathrm{const}\,}
\newcommand*{\Punkt}
  {\quad \text.}
\let\punkt\Punkt
\newcommand*{\Komma}
  {\quad \text,}
\let\komma\Komma
\let\vareps\varepsilon
\usepackage{units}
\usepackage{xspace}
\usepackage{gensymb}
\usepackage[numbers,sort]{natbib}%
\begin{document}
%
\titlehead{Preprint}
\title{Efficient prediction of broadband trailing edge noise and application to porous edge treatment}
\author{Benjamin W. Fa{\ss}mann \and Christof Rautmann \and Roland Ewert \and Jan W. Delfs}
\publishers{\normalsize\itshape Institute of Aerodynamics and Flow Technology, Technical Acoustics \\
            \normalsize\itshape German Aerospace Center (DLR) \\
            \normalsize\itshape Lilienthalplatz 7, 381208 Braunschweig, Germany}
\date{ }
%
\maketitle[1]
%
%
%
\begin{abstract}
  \bfseries \sffamily \footnotesize
\noindent Trailing edge noise generated by turbulent flow traveling past an edge of an airfoil represents one of the most essential paradigms of aeroacoustic sound generation at solid surfaces. It is of great interest for noise problems in various areas of industrial application. First principle based Computational Aeroacoustics (CAA) methods with short response time are needed in the industrial design process for reliable prediction of spectral differences in turbulent-boundary-layer trailing-edge noise (TBL-TEN) due to design modifications. In this work an aeroacoustic method is studied that rests on a hybrid two-step CFD/CAA procedure. In a first step Reynolds Averaged Navier-Stokes simulation provides the time-averaged solution to the turbulent flow, including the mean-flow and turbulence statistics such as length-scale, time-scale and turbulence kinetic energy. Fluctuating sound sources are stochastically generated from RANS statistics with the Fast Random Particle-Mesh Method (FRPM) to simulate in a second CAA step broadband aeroacoustic sound. From experimental findings it is well known that porous trailing edges significantly lower trailing edge noise level over a large bandwidth of frequencies reaching 6 to 8dB reduction. Furthermore, sound reduction depends on the porous material parameters, e.g. geometry, porosity, permeability and pore size. This paper presents first results for the extended hybrid CFD/CAA method to include the effect on noise of porous materials with specifically prescribed parameters. To incorporate the effect of porosity, an extended formulation of the Acoustic Perturbation Equations (APE) with source terms is derived based on a reformulation of the volume averaged Navier-Stokes equations into perturbation form. Proper implementation of the Darcy and Forchheimer terms is verified for sound propagation in homogeneous and anisotropic porous medium. Sound generation is studied for a generic symmetric NACA0012 airfoil without lift to separate secondary effects of lift and camber on sound from those of the basic edge noise treatments. The reference solid airfoil configurations are compared with published experimental data. Convincing agreement in the prediction of one-third-octave band spectra is found. Simulation with porous edge treatment reveals a broadband noise reduction capability of approximately 6dB with similar trends as seen in experiment.
\end{abstract}
%
%
\section{INTRODUCTION}

Airfoil self-noise radiated from an airfoil exposed to a flow is caused by different source mechanisms (see \citet{Brooks.1989}), e.g., vortical structures in a turbulent flow interacting with surfaces, see \citet[Ch. 6, 7 ]{Howe.2003} or \citet{Wagner.1996}. Turbulent-boundary-layer trailing-edge noise (TBL-TEN) represents one of the most prominent source mechanisms that is important for a wide range of technical applications like noise generated at the airframe of an aircraft with the high-lift system deployed, broadband fan noise generated in turbo machines, or noise generated by the turbulent flow at wind turbine rotor blades.

Over the last two decades the renewable energy share in the global energy mix has grown steadily. Among different sources the wind power sector has made significant progress. The maturation of the technology together with constant installation of new turbines have pushed the cumulative installed capacity to \unit[282.5]{GW} worldwide at the end of 2012. That represents an average annual growth rate in worldwide wind energy capacity of \unit[22]{\%} over the last 10 years, see \citet{GWEC.2012}. In its market forecast the Global Wind Energy Council (GWEC) predicts a cumulative installed capacity of about \unit[500]{GW} at the end of 2017. This process goes hand in hand with the construction of new wind farms and the replacement of older turbines with newer more efficient ones (repowering). Basically, two points are important in the development of new wind farms: annual energy production (AEP) of the turbines and the size and location of the wind farm. To achieve a high AEP, rotor diameters are increasing up to the structural limits resulting in rotor blade lengths of up to \unit[70]{m}. Turbine size becomes an even more important parameter at locations with low average wind speeds where the total amount of kinetic energy in the wind is low and therefore a bigger rotor cross sectional area is needed to harvest as much energy as possible, refer to \citet[Ch. 14, 19, 20]{Hau.2014}. Due to the shortage of new on-shore building cites wind farms are moving closer to residential areas. Besides the anesthetic circumstances noise immissions are one major issue for nearby living people.

The efficient and reliable prediction of wind turbine noise in the development process of a new turbine is crucial for quiet rotor blade design (see \citet{Schepers.2007}). As a first step towards establishing an aeroacoustic design capability one could think of evaluating the aeroacoustic properties of different rotor blade shapes already in the aerodynamic design phase. After the basic shape is determined, supplementary trailing-edge modification can be applied to achieve an even greater reduction of noise emission. Hence, the prediction-method should also be capable of calculating the noise reduction effect of trailing edge treatments based on porous materials.

For trailing-edge noise different prediction approaches are known ranging from fast and simple semi-empirical methods like the Brooks, Pope and Marcolini (BPM) model (see \citet{Brooks.1989}) to high fidelity scale resolving numerical simulation methods like Large Eddy Simulation (LES). As the semi empirical method is based on reference measurements, it is prone to inaccuracy when applied to airfoils which significantly deviate from the reference geometry. Furthermore, it can only be used for 2-D calculations, thus no 3-D modification of the trailing edge (e.g. serrations) can aeroacoustically be evaluated. Alternatively, one could use more sophisticated 3-D time and space resolved LES simulation. Significant for this approach is still its very high computational effort.  

In the field of aircraft application, future prospective of short range aircraft with short take off and landing (STOL) properties involve non-conventional high lift systems as overblown flaps taking advantage of the coanda effect, see \citet{Radespiel.2013}. In this context, additional sound generation must be avoided, at first place. In 1979, \citet{Hayden.1976} investigated several edge concepts for overblown flaps. \citet{Howe.1979} presented a basic theory on this. Besides simple trailing-edge modifications, the application of porous material is another promising means for reducing trailing edge noise, see \citet{Hayden.1974}. The positive effect of lengthwise slits applied to the trailing edge was shown by \citet{Herr.2007}. The resulting reduction compared to the solid reference is about \unit[6]{dB}. As the manufacturing of narrow slits is expensive it is of crucial interest to investigate the acoustic benefit of rigid porous material as sintered metal fiber felts or metal foams. A variety of porous materials applied to the trailing edge of a high lift airfoil is tested by \citet{Herr.2014}. Maximal sound reduction of about \unit[6]{dB} to \unit[8]{dB} is reported. \par

Prospective noise optimization in all fields should be integrated into the airfoil design process to enable short-term development cycles aspired by the industry. Today's computational power allows a detailed and high resolution investigation of a variety of approaches to conquer the edge noise problem. But full resolution of porous material with all its geometrical details at technical relevant Reynolds numbers still exceeds datum computer capability. Thus, a frequently used means of modeling porous material and multi-phase flow is the method of volume averaging, refer to \citet{Slattery.1969,Whitaker.1973,Gray.1975,Gray.1977,Hassanizadeh.1979a,Hassanizadeh.1979b,Drew.1983,Bear.1984,Ni.1991}. \citet{Breugem.2005b} uses the Volume Averaged Navier-Stokes (VANS) equations for hydrodynamic applications. 

In this article an efficient method with prospective aeroacoustics design quality is applied that utilizes a hybrid two-step CFD/CAA procedure with stochastically realized 4-D spatio-temporal turbulence. In a first step Reynolds Averaged Navier-Stokes simulation provides the time-averaged solution to the turbulent flow, including the mean-flow and turbulence statistics such as length- and time-scales and turbulence kinetic energy. With the Fast Random Particle-Mesh Method (FRPM) fluctuating sound sources are stochastically generated from RANS statistics to simulate in a second CAA step broadband aeroacoustic sound with Acoustic Perturbation Equations (APE) (see \citet{Ewert.2003}). The method has the capability to simulate airfoils of arbitrary geometries and variable flow regime. 

The present article aims at extending the application of the hybrid CFD/CAA approach to include the effect of porous trailing-edge modification on broadband sound. Section~\ref{sec:Num_Methods} gives an overview about the applied method and presents the main details of an extended APE reformulation from volume averaging of the Navier-Stokes equations to incorporate the effect of porous medium into CAA simulation. In Section~\ref{sec:Comp_Setup}, the computational set-up for CFD and CAA simulation is introduced, focusing on resolution constraints. Also a close look on the special issues regarding porous inlays is taken. Section~\ref{sec:Results} discusses the latest results for the prediction of broadband sound from a generic NACA0012 airfoil with solid trailing edge and its reliability is studied by comparison with experimental results. Further, CAA results for simulation with porous medium are presented. First, generic test cases for the propagation of acoustic waves in homogeneous porous material with spatially isotropic porosity properties on the one hand and spatially anisotropic porosity properties on the other hand are used to verify the implementation. Next, results for the NACA0012 airfoil with porous trailing edge treatment and stochastic sound sources are compared to the reference case without edge treatment. Finally, main conclusions are drawn in Section~\ref{sec:Conclusion}.

In the paper, only 2-D results are shown. But, for achieving the goal of predicting local 3-D modification of the rotor blade, the presented hybrid approach is performing in 3-D as well. The actual expenses directly depend on the level of detail what should be achieved. If the computations include an elaborate geometry or turbulent structures should be resolved, the computation takes longer than the investigation of some roughly assembled 2-D sections of a well known rotor blade. Beside this adjustment capability of the spatial resolution of the set-up, the implementation of the presented approach is fully parallelized to gain a speed up of the simulations. In addition, the sound field on a collecting surface in the near farfield of the rotor blade can be propagated to an arbitrary observer's position. Using Fast Multipole Boundary Element Method, the sourface pressure can serve as input to calculate the acoustic at a far field position. 
\section{NUMERICAL METHODS} \label{sec:Num_Methods}
In this chapter, the numerical methods will be presented which were used for prediction of trailing edge noise at standard and low noise trailing edges. First, the hybrid approach of sequential aerodynamic and aeroacoustic simulations is generally depicted. Second, volume averaged perturbation equations are derived. They will be applied for predicting the edge noise at the porous trailing edge. Finally, the utilized source model for trailing edge noise is specified in detail. \par
\subsection{Hybrid CFD/CAA approach} \label{sec:hybrid_cfd_caa_approach}

The approach pursued by DLR relies on a hybrid two-step procedure for a first-principle based prediction of broadband trailing-edge noise. The first step rests on a Computational Fluid Dynamics (CFD) simulation of the time-averaged turbulent flow around the airfoil. In the second Computational Aeroacoustics (CAA) step time dependent linear equations are solved on structured multi-block (SMB) meshes. For the preparation of unsteady vortex sound sources a synthetic turbulence method developed at DLR is adopted to force acoustic perturbation equations. The stochastic approach is especially well suited for aeroacoustics purposes, i.e. realizing a '4-D' time-space based prediction of fluctuating sources in a restricted volume (denominated subsequently as 'source patch') and including the convection of synthetic eddies with proper imposed time decay. The Random Particle-Mesh Method (RPM) by \citet{Ewert.2005} allows to synthetically realize time-dependent fluctuations from time-averaged turbulence statistics. It generates Gaussian correlated synthetic turbulence of local integral length scale $\Lambda=c_l/C_\mu \sqrt{k}/\omega$ ($c_l\simeq 0.5$, $C_\mu = 0.09$) and variance proportional to the turbulence kinetic energy distribution, refer to Section~\ref{sec:fRPM} for further details. A numerical revisited approach (refer to \citet{Ewert.2008, Ewert.2011}), termed Fast Random Particle-Mesh Method (FRPM) was applied for the simulations presented in this article.

\begin{figure}[hbt]
 \centering
 \includegraphics[width=0.9\textwidth]{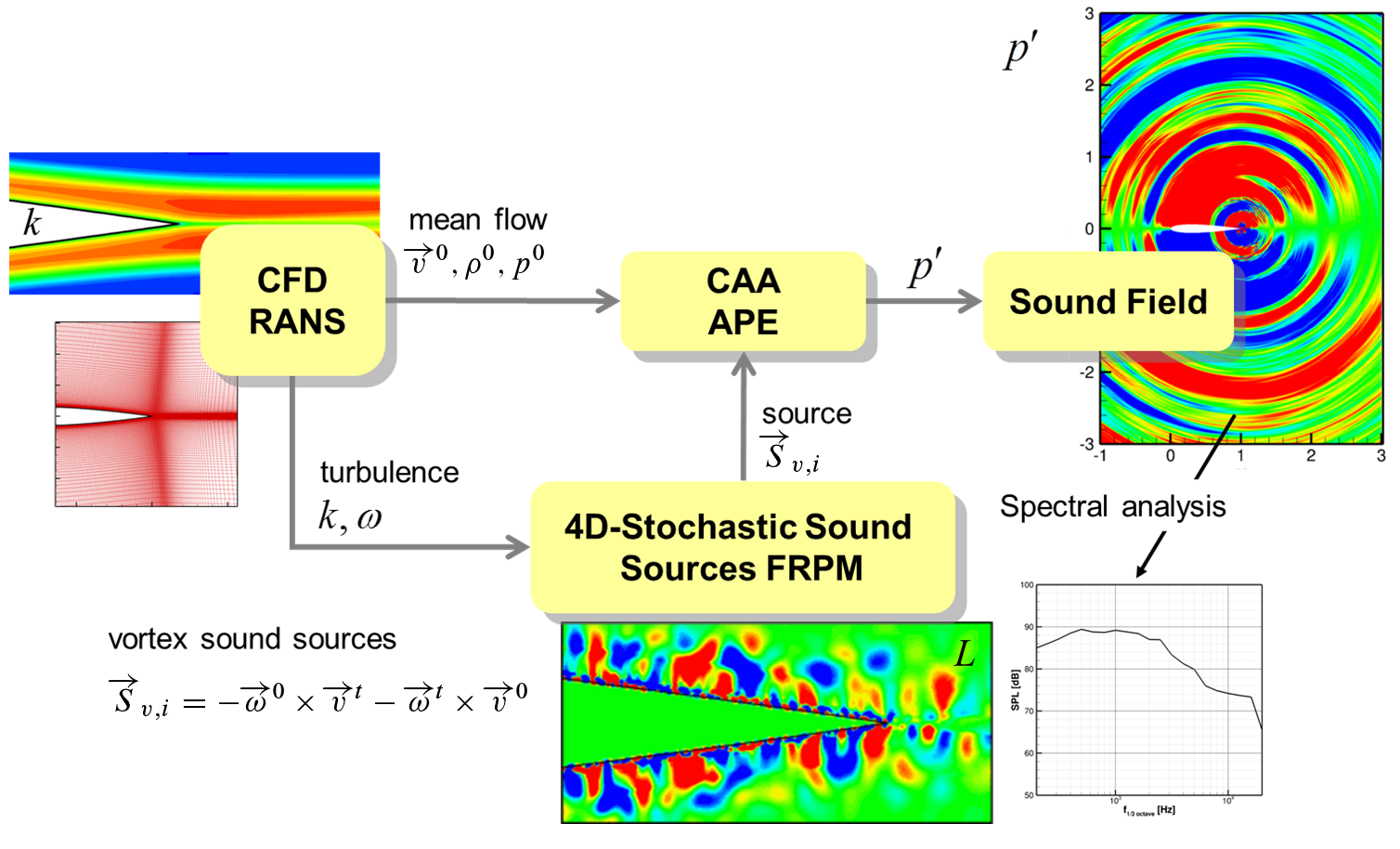}
 \caption{Schematic of CAA prediction method of DLR based on a two-step hybrid method using a steady RANS CFD step, followed by a CAA step solving Acoustic Perturbation Equations (APE) on structured multi block (SMB) meshes; the generation of fluctuating turbulent sound sources is realized with the 4-D FRPM synthetic turbulence method.} \label{f:frpm_flow_chart}
\end{figure}

\Fig{f:frpm_flow_chart} gives a general overview about the approach. The steady time-averaged RANS flow (which by means of Birkhoff's ergodic theorem is equivalent to an ensemble average of the flow) provides the mean-flow over which the time dependent aeroacoustic simulation is conducted. Furthermore, the turbulence statistics provided by RANS are utilized to generate the unsteady vortex sound sources that drive the governing equations.

In free field this turbulence is coupled with the CAA solver PIANO of DLR (refer to \citet{Delfs.2008}), which is based on the 4th order accurate DRP scheme proposed by \citet{Tam.1993}. The synthetic turbulence in conjunction with the RANS mean-flow defines the right-hand side fluctuating source terms of the Acoustic Perturbation Equations (APE) by \citet{Ewert.2003}, which are a modification of the Linearized Euler Equations (LEE) so that vorticity or entropy convection is entirely prescribed by the source term whereas acoustic generation and radiation is simulated dynamically. The APE realize a solution to the wave operator of irrotational flow. Together with proper right-hand side volume sources this becomes an acoustic analogy based on that wave operator. The source term mainly acts as a vorticity production term. Sound due to the interaction of vorticity with the trailing-edge is generated as part of the CAA simulation step. These vortex dynamics are dominated by the linear contributions to the source terms. Non-linear contributions to the source term (self noise term) mainly deemed responsible for sound generation of free turbulent flow are neglected. %
This is due to the short characteristic time scale related to vorticity passing by the trailing edge in comparison to the turbulent decay time scale in the vicinity of the edge. 
%
%
%
\subsection{Volume averaged perturbation equations} \label{sec:volume_averaging}
%
%
\paragraph{Volume averaging}%
The method of volume averaging is an important tool for multi-phase flows and since the late 60th, a considerable amount of work has been dedicated to the development of volume averaged conservation and transport equations, see e.g. \citet{Slattery.1969,Whitaker.1973,Gray.1975,Gray.1977,Hassanizadeh.1979a,Hassanizadeh.1979b,Drew.1983,Bear.1984,Ni.1991}. Basically, the averaging operation can be understood as spatial filtering of the flow variables. The superficial volume averaging is defined as follows:
\begin{equation}\label{eq:1.1}
	\left \langle \rho \right \rangle^s (\bfm x,t) := \frac{\int G \left ( \bfm x - \bfm x', \Delta \right ) \rho^*(\bfm x',t) d^3 x'}
	{\int G \left ( \bfm x - \bfm x', \Delta \right ) d^3 x'} .
\end{equation}
In this expression $G$ denotes the spatial filter applied for the volume averaging procedure. The filter is centered at $\bfm x$ and has a fixed extension defined by length scale $\Delta$, i.e. it decays to zero for $|\bfm x - \bfm x'| >> \Delta$. For example, the filter could be chosen to be a Gaussian with standard deviation~$\Delta$. 

The quantity $\rho^*$ denotes the generalized density variable which is well defined in the entire volume, i.e. in the porous volume as well as in the solid phase of a porous material. It is given by   
\begin{equation}\label{eq:1.2}
	\rho^*(\bfm x,t) = \rho(\bfm x,t) H(f(\bfm x)).
\end{equation} 
\begin{figure}[tbh]
  \centering
  \includegraphics[clip,scale=1,angle=0]{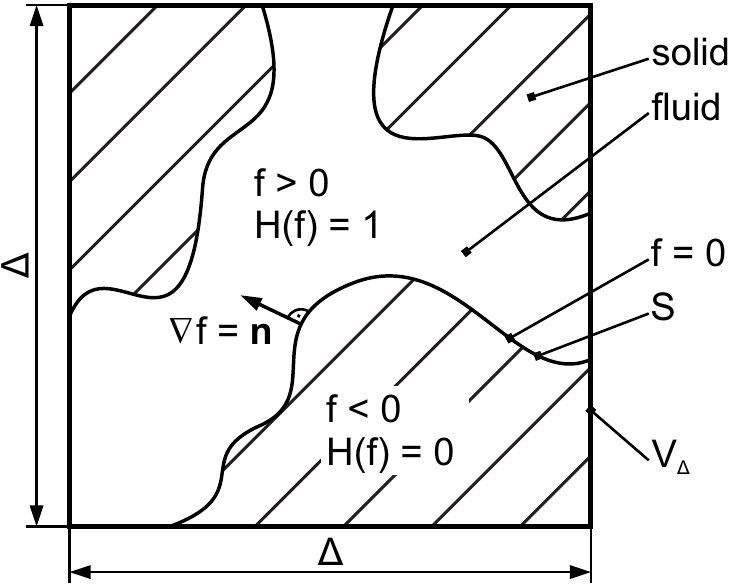}
  \caption{Schematic of porous material and definition of function $f(x)$.} \label{fig:Volume_Averaging_Porosity}
\end{figure}
Here, $H$ denotes the Heaviside-function and $f(\bfm x)$ is a function defined to be $f<0$ in the solid material and $f>0$ in the fluid, i.e. $f=0$ indicates the surface between solid and fluid in the porous medium, refer to \fig{fig:Volume_Averaging_Porosity}. The gradient of $f(0)$ is normal to the interface surfaces. Without losing generality we can define the scaling of $f$ such that the gradient is the wall normal unity vector and points into the fluid, i.e. $\nop f = \bfm n$. For further details regarding generalized variables and their application refer to \citet{Crighton.1996}. The integrals (without explicitly specified bounds) are taken over the entire space. In general, the integral over the filter kernel defines a characteristic filter volume, 
\begin{equation} \label{eq:filter_volume}
	V_\Delta = \int G \left ( \bfm x - \bfm x', \Delta \right ) d^3 x'.
\end{equation}

The intrinsic averaged density is defined by 
\begin{equation}\label{eq:1.3}
	\left \langle \rho \right \rangle^i (\bfm x,t)  := \frac{\int G \left ( \bfm x - \bfm x', \Delta \right ) \rho^*(\bfm x',t) d^3 x'}
	{\int G \left ( \bfm x - \bfm x', \Delta \right ) H(f(\bfm x')) d^3 x'}. 
\end{equation}
We can define a porosity factor $\bar \phi$ via 
\begin{equation}\label{eq:1.4}
	\bar \phi = \frac{\int G \left ( \bfm x - \bfm x', \Delta \right ) H(f(\bfm x')) d^3 x'}{\int G \left ( \bfm x - \bfm x', \Delta \right ) d^3 x'},
\end{equation}
which, based on the definitions \eqns{eq:1.1}{eq:1.3} for intrinsic and superficial averaged quantities, yields the following generally valid relationship between both volume averaged quantities:
\begin{equation}\label{eq:1.5}	
	\left \langle \rho \right \rangle^s = \bar \phi \;\left \langle \rho \right \rangle^i .
\end{equation}
It always is $0 \le \bar\phi \le 1$, where $\bar\phi = 1$ in free fluid and $\bar\phi = 0$ represents a solid body. \par
In the special case where the filter kernel is chosen to be a discontinuous top-hat function $G(\bfm x - \bfm x',\Delta) = g(x-x')g(y-y')g(z-z')$, where $g(x)$ is defined by 
\begin{equation}
  g(x) = 1-H\left( \left| x \right|-\Delta/2 \right),
\end{equation}
the superficial averaged density reads
\begin{equation}
	\left \langle \rho \right \rangle^s = \frac{1}{V_\Delta} \int_{V_F} \rho d^3 x'. 
\end{equation}
The top-hat function restricts the integration volume to a finite extension $V_\Delta=\Delta^3$ centered at the given position (window averaging). $V_F$ is the fluid volume of the porous material inside the actual window; it satisfies $V_F = \bar \phi V$. The intrinsic volume averaged variable in this case becomes
\begin{equation}
	\left \langle \rho \right \rangle^i = \frac{1}{{V_F}} \int_{V_F} \rho d^3 x'.
\end{equation} 
For a point inside the fluid phase the intrinsic density becomes in the limit $\Delta \to 0$ equal to the local density in the fluid, i.e.
\begin{equation}
	\left \langle \rho \right \rangle^i(\bfm x,t) \to \rho (\bfm x,t) \quad \mbox{for} \quad \Delta \to 0.	
\end{equation}
In order to smooth out geometrical details of the porous medium, such that the volume averaged variables become continuous over the porous material, a length scale $\Delta > D_p$ must be used, where $D_p$ denotes a length scale derived from a characteristic pore size. In this case, the intrinsic volume averaged density has the same order of magnitude as a local density, it is, however, a quantity defined over the entire space, which moreover can be spatially differentiated if a continuous filter function $G$ is applied. The superficial averaged density is smaller as defined by the porosity parameter $\bar \phi$. Hence, at interfaces between porous materials and free fluid the intrinsic density will only exhibit a gradual change over the interface, whereas the superficial averaged density will change rapidly over a scale $\Delta$. Inside a homogeneous porous material, the explicit value of $\bar \phi$ will be independent for sufficient large $\Delta$ from the explicit chosen length scale in \eqn{eq:1.4}. However, at an interface between the porous medium and a free fluid, $\bar \phi$ will change gradually over a length $\Delta$ from its value in the porous medium to one inside the fluid phase. 

Favre volume averaged velocities are defined via
\begin{equation}\label{eq:volav_Favre}
	[ v_i ] = \frac{\left \langle \rho v_i \right \rangle^{s,i}}{\left \langle \rho \right \rangle^{s,i}}.
\end{equation} 
As a consequence of definitions \eqns{eq:1.1}{eq:1.4} the definition is independent as to whether superficial or intrinsic averaging is applied. To derive volume averaged perturbation equations, the Navier-Stokes equations in conservative notation are volume averaged assuming the application of a spatial differentiable filter $G(|\bfm x - \bfm x'|)$.  In the fluid ($\bar \Phi = 1$) the resulting equations correspond to those used for large eddy simulation (LES), i.e. they formally correspond to the Navier-Stokes equations for volume averaged variables plus some extra subgrid scale stress terms on the right-hand side. In this derivation step, commutation of volume averaging and differentiation is applied, e.g. for the term $\rho v_i$ in the continuity equation it follows from the definition \eqn{eq:1.1}
\begin{equation}\label{eq:1.6}
	\left \langle \pp{\rho v_i}{x_i} \right \rangle^{s} = \pp{}{x_i} \left \{ \left \langle \rho \right \rangle^{s} \left [v_i \right ] \right \}.
\end{equation}
The derivation is shown in more detail in the Appendix~\ref{app:Volume_Averaging}.
\paragraph{Independent variables} \label{paragraph:independent_variables}
For numerical stability reasons the independent perturbation variables finally used for a reformulation of volume averaged APE are selected based on the prerequisite to be (almost) continuous across an interface between fluid and the porous medium. This way gradients that occur inevitably due to the sudden jump of porosity across the boundary can be lumped together in extra terms linear in the used independent variables which resemble a numerically resolved localized function with a distinct peak across the interface. If applied in conjunction with explicit time integration, these extra terms could trigger numerical instabilities. To circumvent this problem, their contribution can be treated implicitly in a mixed implicit-explicit time integration method (IMEX methods, refer to \citet{Ascher.1997,Boscarino.2007}), whereas all spatial gradient terms occurring in the governing equations can be treated further on by means of explicit time integration. Since the extra terms are localized, i.e. proportional to the fluctuating variables and do not involve information from neighbor nodes of the computational mesh, implicit treatment in the framework of SDIRK methods (see \citet{Ascher.1997}) demands only for inversion of an $n \times n$ matrix (depending on the dimension $n$ of the problem) composed out of steady mean-flow variables, which is computed and stored at the begin of an unsteady simulation cycle. Therefore, highly efficient treatment of these implicit gradient terms becomes feasible. 

\begin{figure}[tbh]
  \centering
  \includegraphics[clip,scale=1,angle=0]{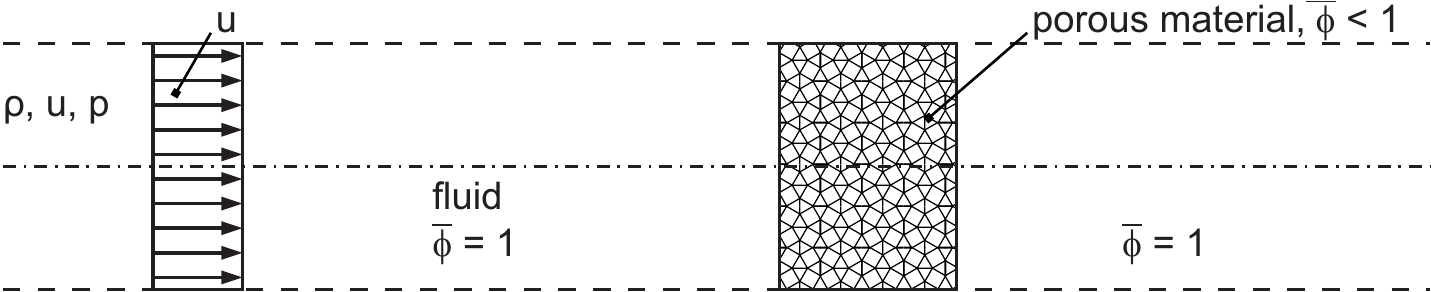}
  \caption{Sketch of channel flow with porous blocking.} \label{fig:Channel_Flow_Sketch}
\end{figure}
To discuss further the appropriate choice of independent variables, we consider a test set-up of an incompressible channel flow in which a zone of porosity is applied across the channel, refer to \fig{fig:Channel_Flow_Sketch}. For this case, the intrinsic density in the porous medium, where $0 < \bar \phi < 1$ holds, is a constant quantity, since
\begin{equation}\label{eq:volav_intrho}
	\left \langle \rho \right \rangle^i = \frac{\int G H \rho \mathrm{d}^3x}{\int G H  \mathrm{d}^3x} 
	= \frac{\rho \int G H  \mathrm{d}^3x}{\int G H  \mathrm{d}^3x} = \rho.
\end{equation} 
Mass conservation across the bulk and in the porosity implies $\rho u = \left \langle \rho \right \rangle ^s \left  [ u \right  ] = \const$. Using \eqn{eq:volav_intrho} and with the help of \eqn{eq:1.5} this yields   
\begin{equation}
	u = \bar{\phi} \left [ u \right ]. 
\end{equation}   
Hence, to accomplish almost continuous variables across the fluid-porous interface, we introduce a new velocity variable defined by  
\begin{equation}\label{eq:volav_velocity}
 \hat{v_i} :=  \bar \phi \left[ v_i \right] . 
\end{equation}
Furthermore, we chose the intrinsic volume-averaged fluctuating density $\left< \rho \right>^i$  and the intrinsic volume-averaged fluctuating pressure $\left< p \right>^i$ to close the set of independent variables. \par
\paragraph{Volume averaged perturbation equations} \label{paragraph:governing_equations}
For the derivation of volume averaged perturbation equations we decompose the volume averaged pressure and density into a mean, i.e. time-averaged, and a fluctuating part. Let $\vareps$ indicate density or pressure, i.e. $\vareps \in \{ \rho, p \}$, then the decomposition reads
\begin{equation}\label{eq:volav_eps_decomposition}
	\left \langle \vareps \right \rangle^{i,s} = \overline{\left \langle \vareps \right \rangle^{i,s}} + {\left \langle \vareps \right \rangle^{i,s}}^\prime,
\end{equation}
where the mean variable is defined by the time-average
\begin{equation}\label{eq:time_average}
	\overline{a} = \lim_{{\Delta t} \to \infty} \int_{t_0}^{t_0+\Delta t} a \mathrm{d}t.
\end{equation}

The Favre volume averaged velocity defined by \eqn{eq:volav_Favre} is decomposed into a Favre averaged mean-part plus a fluctuation, i.e. 
\begin{equation}\label{eq:volav_vi_decomposition}
	\left [v_i \right ] = \widetilde{\left [v_i \right ]} + \left [v_i \right ]'', 
\end{equation} 
where 
\begin{equation}
	\widetilde{\left [v_i \right ]} := \frac{\overline{\left \langle \rho \right \rangle^s \left [v_i \right ]}}{\overline{\left \langle \rho \right \rangle^s}},
\end{equation}
with the bar indicating a time average, \eqn{eq:time_average}. It is easy to prove that Favre averaging applied to volume averaged quantities still satisfies the usual relations 
\begin{equation}
	\overline{\left \langle \rho \right \rangle^s\left [v_i \right ]''}=0
\end{equation}
and 
\begin{equation}
	\overline{\left \langle \rho \right \rangle^s \left [v_i \right ] \left [v_j \right ]}= 
	\left \langle \rho \right \rangle^s \widetilde{\left [v_i \right ]} \widetilde{ \left [v_j \right ] } + 
	\overline{\left \langle \rho \right \rangle^s \left [v_i \right ]'' \left [v_j \right ]''}. 
\end{equation}	
Superficial volume averaging for density and pressure must be applied to the volume averaged Navier-Stokes equations to enable their reformulation in conservative notation that eventually can be transformed into a formulation based on primitive volume averaged variables, refer to Appendix~\ref{app:Volume_Averaging}. Introduction of the variable decomposition and linearization allows to derive volume averaged perturbation equations in terms of independent variables $\left ( {\left \langle \rho \right \rangle^s}^\prime, \left [ v_i \right ]'',{\left \langle p \right \rangle^s}^\prime \right )$. For the desired set of independent variables based on intrinsic volume averaged density and pressure as well as velocity defined by \eqn{eq:volav_velocity}, in a final step the variables have to be substituted accordingly. To simplify the syntax, subsequently we will use a notation without additional brackets to indicate intrinsic volume-averaged quantities, e.g. $\rho$, $\rho^0$, and $\rho'$ instead of $\left \langle \rho \right \rangle^i$,  $\overline{\left \langle \rho \right \rangle^i}$, and ${\left \langle \rho \right \rangle^i}^\prime$, respectively. Furthermore, a simplified notation is introduced for convenience by omitting the overbar on $\phi$, i.e. $\overline{\phi} \to \phi$. Then the substitutions 
\begin{equation} \label{eq:notation1}
	\phi \rho^\prime \to {\left \langle \rho \right \rangle^{s}}^\prime, 
	\quad \phi \rho^0 \to \overline{\left \langle \rho \right \rangle^{s}},
	\quad \phi p^\prime \to {\left \langle p \right \rangle^{s}}^\prime, 
	\quad p^0 \to \overline{\left \langle p \right \rangle^{s}}
\end{equation}
have to be applied to eventually reformulate the perturbation equations in the desired set of variables. Proper velocities according to the definition \eqn{eq:volav_velocity} are obtained by the replacements  
\begin{equation} \label{eq:notation2}
	\frac{\hat v_i^0 }{\phi} \to \widetilde{\left [v_i \right ]}, \quad \frac{ \hat v_i^\prime}{\phi} \to \left [v_i \right ]''. 
\end{equation}

The extended APE as resulting from consequent application of the volume averaging procedure as outlined before derive as follows: The continuity equation of the APE reads 
\begin{equation} \label{eqn:conti_APE}
  \pp{\rho^\prime}{t} + \phi^{-1} \left[ \hat{v}_i^0  \pp{\rho^\prime}{x_i} + \hat{v}_i^\prime \pp{\rho^0}{x_i} + \rho^0 \pp{\hat{v}_i^\prime}{x_i} + \rho^\prime \pp{\hat{v}_i^0}{x_i} \right] 
  = S_\rho \punkt
\end{equation} \par
The APE momentum equation becomes 
\begin{equation} \label{eqn:momentum_APE}
  \begin{split}
  \pp{\hat{v}_i^\prime}{t} &+ \phi^{-1} \pp{}{x_i} \left( \hat{v}_k^0 \hat{v}_k^\prime \right) + \phi\pp{}{x_i}  \left( \frac{p^\prime}{\rho^0} \right) \\ %
  &+ \underbrace{\frac{\phi \nu}{\kappa} \delta_{ij} \hat{v}_j^\prime}_{\text{Darcy terms}} + \underbrace{\frac{\phi c_\ind{F}}{\sqrt{\kappa}}\sqrt{\hat{v}_k^0\hat{v}_k^0} \left( e^0_i e^0_j + \delta_{ij} \right) \hat{v}_j^\prime}_{\text{Forchheimer terms}} \\ %
  &+ \underbrace{\hat{v}_i^0 \hat{v}_j^\prime \pp{}{x_j}\phi^{-1} + \delta_{ij} \hat{v}_j^\prime \hat{v}_k^0 \pp{}{x_k}\phi^{-1} - \phi^2 \frac{p^\prime}{\rho^0} \frac{\gamma - 1}{\gamma} \pp{}{x_i}\phi^{-1}}_{\text{gradient model terms}} = S_{v,i} \komma
  \end{split}
\end{equation}
where $\nu$ denotes the kinetic viscosity, $\kappa$ identifies the permeability and the Forchheimer coefficient is represented by $c_\ind{F}$.  It is $\gamma=1.4$ the isentropic exponent of ambient air. Further, $\delta_{ij}$ means the Kronecker delta and $e^0_{i}$ indicates the direction of the time averaged mean flow velocity. The inclusion of the Darcy and Forchheimer terms in the governing equations is shown in more detail in Appendix~\ref{app:Volume_Averaging}.\par
The right-hand source terms of this set of equations are indicated by $S_{\rho,v,p}$. All additional models terms of the momentum equation can be combined to a single matrix $\vM{\mu}$, such that these terms are simply added in the form $\mu_{ij} \hat{v}'_j$ to the perturbed momentum equation. The matrix $\vM{\mu}$ reads 
\begin{equation} \label{eqn:matrix_mu_ij}
   \footnotesize
   \vM{\mu} = \begin{pmatrix}
                 D\!+\!F\!\left( \frac{\hat{v}_1^0\hat{v}_1^0}{\left| {\vM{v}^0} \right|^2}\!+\!1\right)\!+\!\hat{v}_1^0\pp{\phi^{-1}}{x_1}\!+\!P & 
                 F\!\frac{\hat{v}_1^0\hat{v}_2^0}{\left| {\vM{v}^0} \right|^2}\!+\!\hat{v}_1^0\pp{\phi^{-1}}{x_2} & 
                 F\!\frac{\hat{v}_1^0\hat{v}_3^0}{\left| {\vM{v}^0} \right|^2}\!+\!\hat{v}_1^0\pp{\phi^{-1}}{x_3} \\ 
                 F\!\frac{\hat{v}_2^0\hat{v}_1^0}{\left| {\vM{v}^0} \right|^2}\!+\!\hat{v}_2^0\pp{\phi^{-1}}{x_1} & 
                 D\!+\!F\!\left( \frac{\hat{v}_2^0\hat{v}_2^0}{\left| {\vM{v}^0} \right|^2}\!+\!1\right)\!+\!\hat{v}_2^0\pp{\phi^{-1}}{x_2}\!+\!P & 
                 F\!\frac{\hat{v}_2^0\hat{v}_3^0}{\left| {\vM{v}^0} \right|^2}\!+\!\hat{v}_2^0\pp{\phi^{-1}}{x_3} \\ 
                 F\!\frac{\hat{v}_3^0\hat{v}_1^0}{\left| {\vM{v}^0} \right|^2}\!+\!\hat{v}_3^0\pp{\phi^{-1}}{x_1} & 
                 F\!\frac{\hat{v}_3^0\hat{v}_2^0}{\left| {\vM{v}^0} \right|^2}\!+\!\hat{v}_3^0\pp{\phi^{-1}}{x_2} & 
                 D\!+\!F\!\left( \frac{\hat{v}_3^0\hat{v}_3^0}{\left| {\vM{v}^0} \right|^2}\!+\!1\right)\!+\!\hat{v}_3^0\pp{\phi^{-1}}{x_3}\!+\!P \\ 
               \end{pmatrix}
\end{equation}
The following abbreviations have been used therein:
\begin{eqnarray*}
   \footnotesize
   D &:=& \frac{\phi \nu}{\kappa} \\
   F &:=& \frac{\phi c_\ind{F}}{\sqrt{\kappa}} \left| {\vM{v}^0} \right| \\
   \left| {\vM{v}^0} \right| &:=& \sqrt{\left( \hat{v}_1^0 \right) ^2 + \left( \hat{v}_2^0 \right) ^2 + \left( \hat{v}_3^0\right) ^2} \\
   P &:=& \left( \hat{v}_1^0 \pp{\phi^{-1}}{x_1} + \hat{v}_2^0 \pp{\phi^{-1}}{x_2} + \hat{v}_3^0 \pp{\phi^{-1}}{x_3} + \right) \punkt
\end{eqnarray*}
The resulting energy equation for the APE in terms of pressure is identical to that one of the LEE:
\begin{equation} \label{eqn:energy_APE}
  \pp{p'}{t} + \phi^{-1} \left( \hat{v}_i^0\pp{p'}{x_i} + \hat{v}_i^\prime\pp{p^0}{x_i} \right)  + \gamma \phi^{-1} \left( p^0\pp{\hat{v}_i^\prime}{x_i} + p^\prime \pp{\hat{v}_i^0}{x_i} \right) + (\gamma - 1)\, \left(p^0 \hat{v}_i^\prime + p^\prime \hat{v}_i^0 \right) \pp{}{x_i} \phi^{-1} = S_p \punkt
\end{equation} \par
It was derived by formal application of volume averaging and reformulating the equation into perturbation form. Further, a thermally and calorically ideal gas was assumed and viscous as well as entropy modes were neglected. \par
%
%
%
\subsection{Fast Random Particle Method (FRPM)} \label{sec:fRPM}
For the simulation of broadband sound generation the APE are excited by stochastically generated right-hand side sources. As the dominating source of vortex sound the fluctuating (linearized) Lamb vector is modeled on the right-hand side of the APE momentum equation, i.e.
\begin{equation}\label{eq:ape4src}
  S_{v,i} = -\epsilon_{ijk} \omega^0_j v^t_k -  \epsilon_{ijk}\omega^t_j v^0_k,
\end{equation}  
where $\epsilon_{ijk}$ is the Levi-Civita-symbol. Altogether the system
\eqns[to]{eqn:conti_APE}{eqn:matrix_mu_ij} with source~(\ref{eq:ape4src}) constitutes an acoustic analogy based on the wave-operator of irrotational flow, taking into account only the vortex sound source contributions. For the CAA simulations the steady RANS solution is used to prescribe the mean-flow. The mean flow vorticity $\omega^0_i = \epsilon_{ijk} \partial v^0_k/\partial x_j$ needed to specify the source is computed from the RANS mean-flow velocity. The fluctuating turbulent velocities $v^t_i$ are modeled stochastically with the FRPM method, refer to the next section, from which the fluctuating vorticity is derived as $\omega^t_i = \epsilon_{ijk} \partial v^t_k/\partial x_j$.

The APE suppress the vorticity mode otherwise present in the LEE and as such convective and absolute hydrodynamic instabilities that can plague the LEE are removed, see \citet{Ewert.2003}. Convecting vorticity can be present in the APE perturbation velocity, but is entirely prescribed by the right-hand side source term. To be precise, vortex sound sources serve on the one hand as a direct sound source, describing the sound generation in free turbulence. On the other hand it acts as a pure vorticity source in the APE.  Airframe noise generation is due to the interaction of unsteady convecting turbulence (vorticity) with sharp edges. In previous work with PIANO it was extensively demonstrated, e.g. by injecting test vortices into the linearized Euler equations, that the linear CAA equations are capable of predicting the essential noise sound generation at trailing edges, i.e. the physical conversion process of the vorticity mode into an acoustic mode that take place in the vicinity of geometrical inhomogeneities such as sharp trailing edges, see \citet{lummer03}.

\paragraph{Numerical realization of the FRPM approach}

In the framework of Random Particle-Mesh methods synthetic turbulence is generated by means of Lagrangian particles distributed over a limited spatial domain where sound sources ought to be realized. The Lagrangian particles are convected by a given steady (or unsteady) background flow. Each particle that crosses an outflow boundary of the source domain will be re-injected at an inflow boundary into the source domain such that the local particle density in the source domain is conserved. 

The number of particles necessary to obtain a numerically converged stochastical realization of turbulence was studied in \citet{ewert07} and \citet{Siefert.2009}. It was found that for locally evaluated particle densities convergence can be achieved for particle numbers exceeding 5 particles per cell (\unit{ppc}) in 2-D and \unit[10]{ppc}, respectively. Using a globally specified particle density, minimally \unit[2]{ppc} are needed for a 2-D problem and \unit[1]{ppc} in 3-D, refer to \citet{Ewert2010}. To be precise, for particle densities higher than the previous given values, a solution independent of particle number can be achieved. Unlike stochastic approaches in Fourier space, which aim at realizing a specific (isotropic) turbulence spectra with a given local energy and length scale, FRPM realizes local 2 point correlations of the synthetically generated fluctuating velocity components. In the standard approach, a Gaussian correlation function is realized, thus providing Gaussian turbulence spectra. Via superposition of Gaussian spectra (typically of the order of 10 are used) other turbulence spectra can be realized, e.g. of Liepmann type (see  \citet{Rautmann.2014}) or von-K\'arm\'an type (see \citet{Wohlbrandt.2015}), but are not considered in the current publication.

Each particle has associated a set of $i=1 \ldots n$ random variables; the actual value depends on the dimension of the considered problem. The actual random values attached to each particle will be modified over time by means of a stochastic partial differential equation. The cloud of random particles represents a numerical approximation to spatially delta-correlated white-noise. A fluctuating source component $\psi_i$ is obtained by distributing the $i\text{th}$ random deviate $r_{ik}$ of particle $k$ at (variable) position $x\;^c_k(t)$ with a mollifier kernel onto a computational mesh. The numerical approach realizes a weighted sum over all $N$ random particles, \citet[see][]{Ewert.2008}, defined by 
\begin{equation}\label{eq:qi_sum}
  \psi_i\left( \bfm{x},t \right) =   
  \sum_{k=1}^N \hat A^{n} \mathcal G
  \left(\bfm{x}-\bfm{x}\,^c_k(t)\right) 
  \frac{r_{ik}(t)}{\rho^0(\bfm{x}\,^c_k)}.
\end{equation}
The quantity $\rho^0$ denotes the mean-flow density ($\nop \cdot \rho^0 \bfm u^0=0$). In general, the approach is applicable to compressible fluids of considerable density variation. The amplitude scaling function $\hat A$ can be either a function of $\bfm x$ (i.e. depending on the receiver position) or a function of the particle position $\bfm{x}_k^c$. In this work it is chosen to be a function of  $\bfm x$. The Gaussian filter kernel is defined by  
\begin{equation}
 \mathcal G(\bfm x - \bfm{x}\,^c_k) = 
  \exp\left (-\frac{\pi}{2}\frac{|\bfm x - \bfm{x}\,^c_k|^2}{\Lambda^2} \right ),
\end{equation}
where $\Lambda$ is the (local) integral length scale of turbulence, corresponding to the amplitude scaling function taken either at position $\bfm x$ or $\bfm{x}_k^c$. The properties of the random variables $r_{ik}$ satisfy for frozen turbulence, i.e. in the absence of turbulence decay, 
\begin{eqnarray}
  \label{eq:xri1.2}
  \left \langle r_{ik}(t) \right \rangle &=& 0 \\
  \label{eq:xri2.2}
  \left \langle r_{ik}(t) r_{jl}(t) \right \rangle &=& \delta_{ij} \delta_{kl}
  \; \delta m_k\\
  \label{eq:xri2.3}
  \dot{r}_{ik} &=& 0 .
\end{eqnarray}
To summarize the meaning of these equations, $r_{ik}$ represent mutually un-correlated random variables, \eqn{eq:xri2.2}, with vanishing mean, \eqn{eq:xri1.2}, and a constant variance proportional to magnitude $\delta m_k$, which is an average fluid mass fraction related to each particle, defined by the fluid mass in the source domain divided by the number of particles used in that domain. Furthermore, the particles move with a convection velocity related to the convection velocity field value (typically, the time-averaged mean flow is used to prescribe the convection velocity field) at the actual particle position, i.e. 
\begin{equation}\label{eq:xri2.4}
  \dot{\bfm{x}}^c_k = \bfm{v}^0 \left (\bfm{x}\,^c_k \right ).
\end{equation}

For frozen turbulence the random variables of each particle remain constant. In case of an exponential decay the properties of the random variable $r_{ik}$ are prescribed by an individual Langevin equation for each random variable and particle, i.e. 
\begin{equation}\label{eq:xdot_ridef_v3.2}
  \dot{r}_{ik} = -\frac{1}{\tau_s} r_{ik}
  +
  \sqrt{\frac{2}{\tau_s}} s_{ik}.
\end{equation}
A (temporal) white-noise source term $s_{ik}$ appears on the right-hand side, with properties 
\begin{eqnarray}
\label{eq:xsik_def1}
\left \langle s_{ik}(t)\right \rangle &=& 0 \\
\label{eq:xsik_def2}
\left \langle s_{ik}(t)s_{jl}(t+\tau )\right \rangle &=& 
\delta m_k \delta(\tau) \delta_{ij}\delta_{kl}. 
\end{eqnarray} 
In other words, $s_{ik}$ represents (temporal) white-noise scaled with a factor of magnitude $\delta m_k$. The solution of the Langevin equation~(\ref{eq:xdot_ridef_v3.2}) with a source having correlation \eqn{eq:xsik_def2} yields a correlation of variable $r_{ik}$ 
\begin{equation}\label{eq:xrik_langevin01}
\left \langle r_{ik}(t)r_{jl}(t+\tau )\right \rangle = 
\delta m_k \delta_{ij}\delta_{kl} \exp\left(-\frac{|\tau|}{\tau_s}\right).
\end{equation}    

The Langevin equation~(\ref{eq:xdot_ridef_v3.2}) can be solved numerically by the finite-difference equation (see \citet[pp. 484]{pope00}),
\begin{equation}
  r_{ik}(t+\Delta t) = \left ( 1-\frac{\Delta t} {\tau_s} \right )  r_{ik}(t)
  + \left ( \frac{2\delta m_k \Delta t}{\tau_s}
 \right )^ {1/2} \sigma_{ik}(t),  
\end{equation}
where $\sigma_{ik}(t)$ are $m \times N_p$ mutually uncorrelated standardized Gaussian random variables ($\left \langle \sigma_{ik}(t) \right \rangle =0$, $\left \langle \sigma_{ik}(t)\sigma_{jl}(t) \right \rangle = \delta_{ij}\delta_{kl}$) which are independent of themselves at different times ($\left \langle \sigma_{ik}(t)\sigma_{ik}(t') \right \rangle =0$, for $t'\ne t$), and which are independent of $r_{ik}(t)$ at past times (e.g., $\left \langle \sigma_{ik}(t) r_{ik}(t') \right \rangle =0$ for $t'\le t$.

Alternative turbulence decay models using second order Langevin equations have been also proposed, refer to \citet{Siefert.2009, Ewert2012, dieste12, Neifeld2013}.
A correlation function $\mathcal{C}(\bfm r,\tau)=\langle \psi_i(\bfm
x,t)\psi_i(\bfm x+\bfm r,t+\tau)\rangle/\langle \psi_i(\bfm x,t)^2\rangle$ is
related to the filter-kernel function $\mathcal G$ via
\begin{equation}\label{r0}
  \mathcal{C}(\bfm r,\tau ) = 
  \int \mathcal G( \bfm r -\bfm v_0^c\tau - \bfm \xi) 
  \mathcal G(\bfm \xi) \mathrm{d}^n
  \bfm \xi .
\end{equation}
A Gaussian filter kernel
\begin{equation}\label{eq:3d_kernel}
  \mathcal G(x) = 
  \exp\left [ -\frac{\pi}{2}\frac{x^2}{\Lambda^2}\right ]
\end{equation}
realizes Gaussian correlations $\mathcal{C}(\bfm r,\tau)$ with a width $\sqrt{2}$ times larger than that of the filter kernel. Furthermore Taylor's hypothesis at convection velocity $\bfm v_c^0$ is built in: \begin{equation}\label{eq:ro-2d}
 \mathcal{C}(\bfm r,\tau) = 
  \exp \left [ -\frac{\pi |\bfm r -\bfm{v}_c^0   \tau|^2}
       {4 \Lambda^2}
       \right ]. 
\end{equation}
The generation of 3-D isotropic turbulence is achieved by taking fluctuating velocities from the curl of a fluctuating 3-D vector function $\psi_k(x_i,t)$,
$k \in \{1,2,3\}$,
\begin{equation}\label{eq:3d-fluctuations}
  v^t_{i}=\epsilon_{ijk}\pp{\psi_k}{x_j}
\end{equation}
For a homogeneous problem the solenoidal velocities $v^t_i$ realize the two-point velocity cross-correlations of homogeneous isotropic turbulence
\begin{equation}
  R_{ij}(\bfm r,\tau) = 
  \frac{\left \langle v^t_i(\bfm x,t)v^t_j(\bfm x+\bfm r,t+\tau) \right
    \rangle}
  {n^{-1}\left \langle {v^t_i}(\bfm x,t)^2 {v^t_i}(\bfm x,t)^2\right \rangle 
    }  
  = \frac{f(r)-g(r)}{r^2} r_i r_j + g(r)\hat \delta_{ij},
\end{equation}  
where $r=|\bfm r|$. In the denominator summation over the equal index product is assumed. Application of the Gaussian kernel yields for the longitudinal correlation function a Gaussian
\begin{equation}\label{eq:longcorr-gauss-2d}
  f(r) = \exp\left (-\frac{\pi}{4}\frac{r^2}{\Lambda^2}\right ), 
\end{equation} 
where the integral length scale is identical to the parameter $\Lambda$,
\begin{displaymath}
  L=\int_0^\infty 
  f(r) \mathrm{d} r = \Lambda.
\end{displaymath}
The theoretical relation between the longitudinal and lateral correlation functions $f(r)$ and $g(r)$ is given for 3-D homogeneous isotropic turbulence
by 
\begin{equation}
  g(r) = f(r) + \frac{r}{2} \frac{\mathrm{d}f(r)}{\mathrm{d}r},
\end{equation} 
which is achieved for the stochastic realization, applied to a homogenous problem, in which case the lateral correlation becomes 
\begin{equation}\label{eq:latcorr-gauss-2d}
  g(r) = \left(1- \frac{\pi r^2}{4\Lambda^2} \right)
    \exp\left (-\frac{\pi}{4}\frac{r^2}{\Lambda^2}\right )
\end{equation} 
with lateral length scale $\Lambda/2$.  The amplitude $\hat A^{(n)}$ in \eqn{eq:qi_sum} must be chosen such that the fluctuating velocities achieve a local turbulence kinetic energy $k_T=1/2\left \langle {v^t_i}(\bfm  x,t)\langle {v^t_i}(\bfm x,t)\right \rangle$. In 3-D this is accomplished by
\begin{equation}\label{eq:amp3d}
  \hat A^{(3)} = \sqrt{\frac{2}{3\pi}\frac{k_T}{\Lambda}}.
\end{equation}
In a 2-D realization of synthetic turbulence \eqn{eq:qi_sum} is applied for $n=2$ to generate a scalar streamfunction $\psi$, which defines the two-component velocity field 
\begin{equation}
  v_i^t = \epsilon_{ij}\pp{\psi}{x_j}
\end{equation}
with turbulent kinetic energy $k_T= 1/2 \left ( ((v^t_1)^2 + (v^t_2)^2 \right)$. In this case the amplitude function must be scaled according to \citet{Ewert.2008},
\begin{equation}\label{eq:amp2d}
   \hat A^{(2)} = \sqrt{\frac{4k_T}{3\pi}}.
\end{equation}

\section{COMPUTATIONAL SETUP}  \label{sec:Comp_Setup}
This chapter presents the grids used for the aerodynamic and the aeroacoustic simulations for both, the solid generic airfoils and that one with a porous trailing edge. Further, the most important parameters necessary for the computions are presented. \par 
\subsection{CFD simulations}
\begin{figure}[tbh]
\centering
\includegraphics[width=0.475\textwidth]{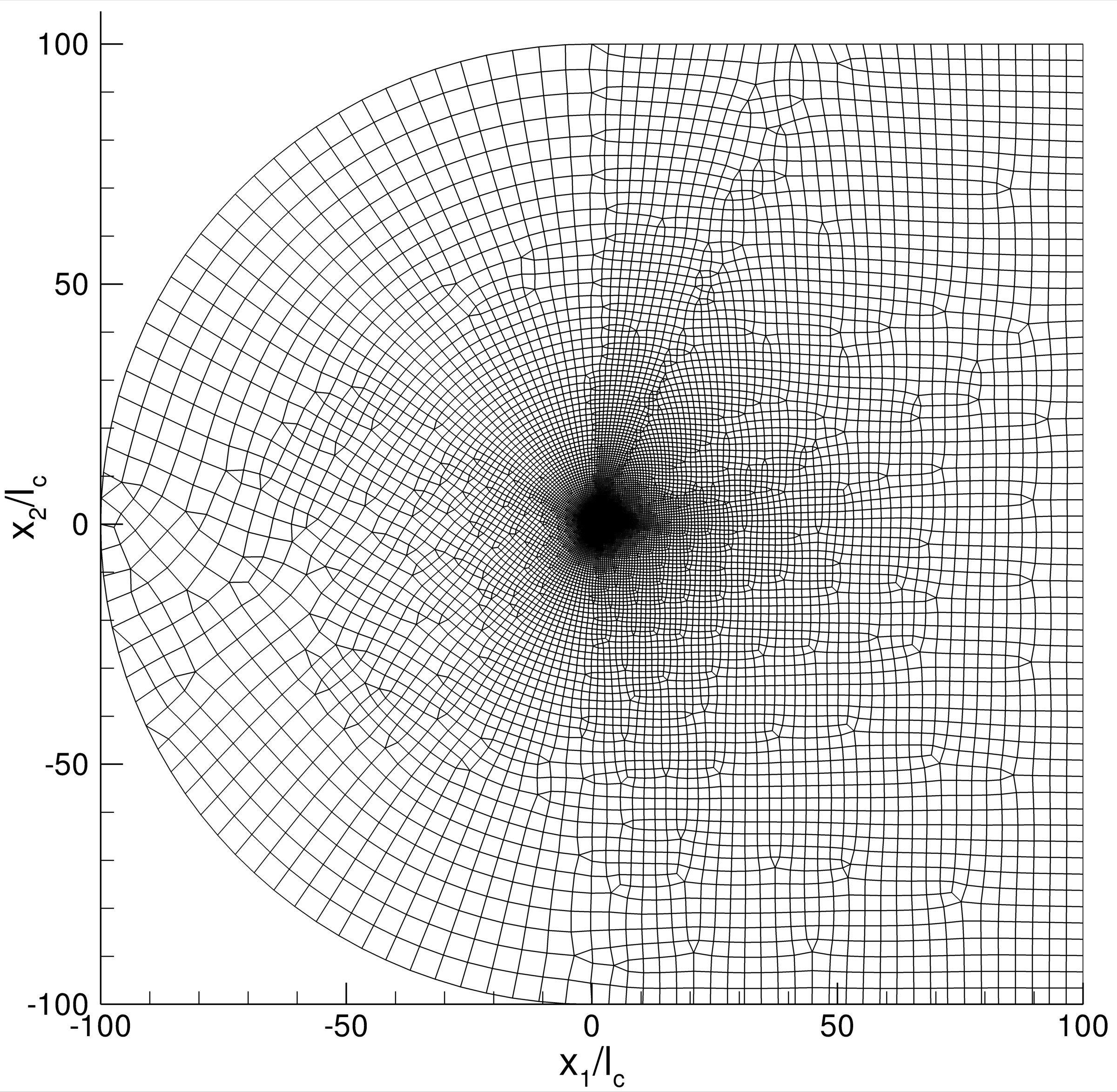} \hfill
\includegraphics[width=0.475\textwidth]{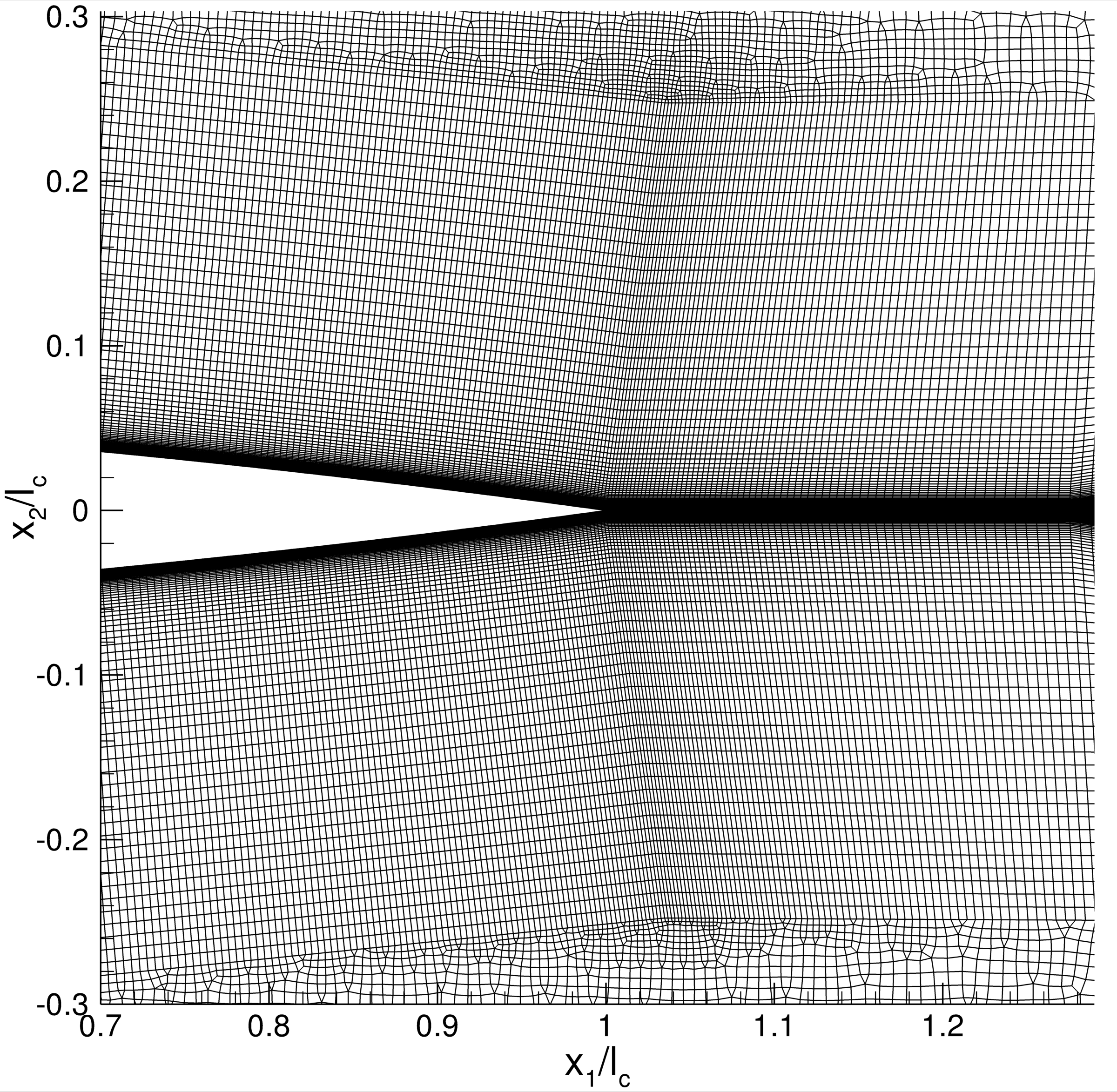}
\caption{(left) Computational mesh for CFD simulations; (right) detailed view of the structured mesh region near the airfoil.}
\label{f:cfd_mesh}
\end{figure}

\paragraph{Generic airfoils} %
RANS CFD simulations as the basis of the CAA simulations were carried out using the unstructured DLR in-house CFD solver TAU, refer to \citet{Gerhold.1997, Schwamborn.2006}. \comment{@ Roland: Revier 1 wuenscht: A short description about the "`TAU"' code is necessary.} For the NACA0012 airfoil a two-dimensional computational domain is generated, with the outer boundaries extending about 100 chord lengths away from the airfoil. A hybrid grid is chosen, where the viscous sub layer is resolved by a structured region and the parts far away from the airfoil are resolved by a coarser quad-dominated unstructured grid. Along the airfoils surface 225 nodes are distributed along each the upper and lower side. The height of the first cell layer is determined so, that a dimensionless wall distance of $y^{+}<1$ is achieved. The dimensionless tangential distance is $x_\text{nose}^{+}\approx 250$ at the nose, $x_\text{thick}^{+}\approx 1322$ where the airfoil has its greatest thickness and $x_\text{te}^{+}\approx 50$ at the trailing edge. The structured near-airfoil grid extends about 110 layers from the wall with an exponential growth rate. Thus, a good resolution of the viscous sub-layer is realized and so reducing the numerical error for flow variable and turbulence statistic calculations. The hybrid structured/unstructured meshing approach reduces cells in less interesting areas far away from the airfoil and cuts the total number of grid cells to $10^{5}$. The grid, normalized by the chord length $l_c$ is shown in \fig{f:cfd_mesh}. The finer resolution in the vicinity of the airfoil is depicted on the right-hand side of \fig{f:cfd_mesh}.

The airfoils suction and pressure side are defined as viscous walls with a fixed laminar turbulent transition location at $x_{tr}=x_1/l_c=0.065$ on both sides. The transition is fixed at this position to ensure the same conditions as given by the test case parameters form the experiments, see \citet{Herr.2013}. A far-field boundary condition with the flow values ($U_\infty$ and $\alpha$) was applied at the outside boundaries. The two-equation SST-k-$\omega$ turbulence model as proposed by \citet{Menter.1994} was used for the simulation of viscous effects and turbulence statistics. \par
\begin{figure}[tbh]
  \centering
  \includegraphics[clip,scale=1,angle=0]{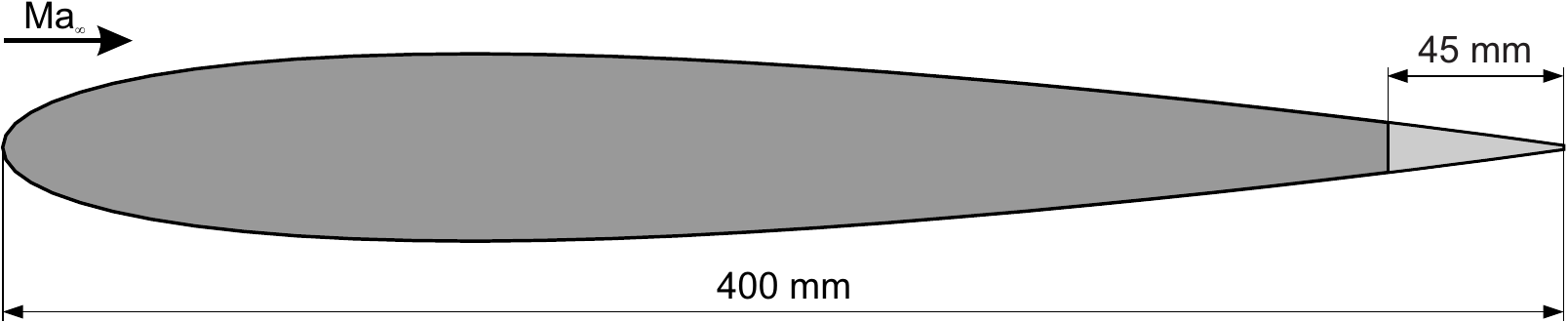}
  \caption{Set up of the porous trailing edge at a NACA0012 geometry at $\Ma = 0.118$ and  $\Rey = 1\!\cdot\!10^{6}$.} \label{fig:up_Slits}
\end{figure}

\paragraph{Porous trailing edges} %
Hybrid RANS based CFD/CAA prediction of aeroacoustic sound generation demands for a suitable RANS solution that also takes into account the effect of porosity on the mean flow and turbulence statistics. A model for simulating porous regions of a computational area was added to DLR in-house CFD solver TAU (see \citet{Moessner.2013}). With this, a 2-D RANS computation of the flow field around and through a NACA0012 airfoil with a realistic porosity was computed. The porosity was located at the trailing edge and covers the last $11.25\%$ of the airfoil. The numerical setup is depicted in \fig{fig:up_Slits} for the NACA0012 airfoil. \par
\begin{figure}[tbh]
\centering
\includegraphics[width=0.475\textwidth]{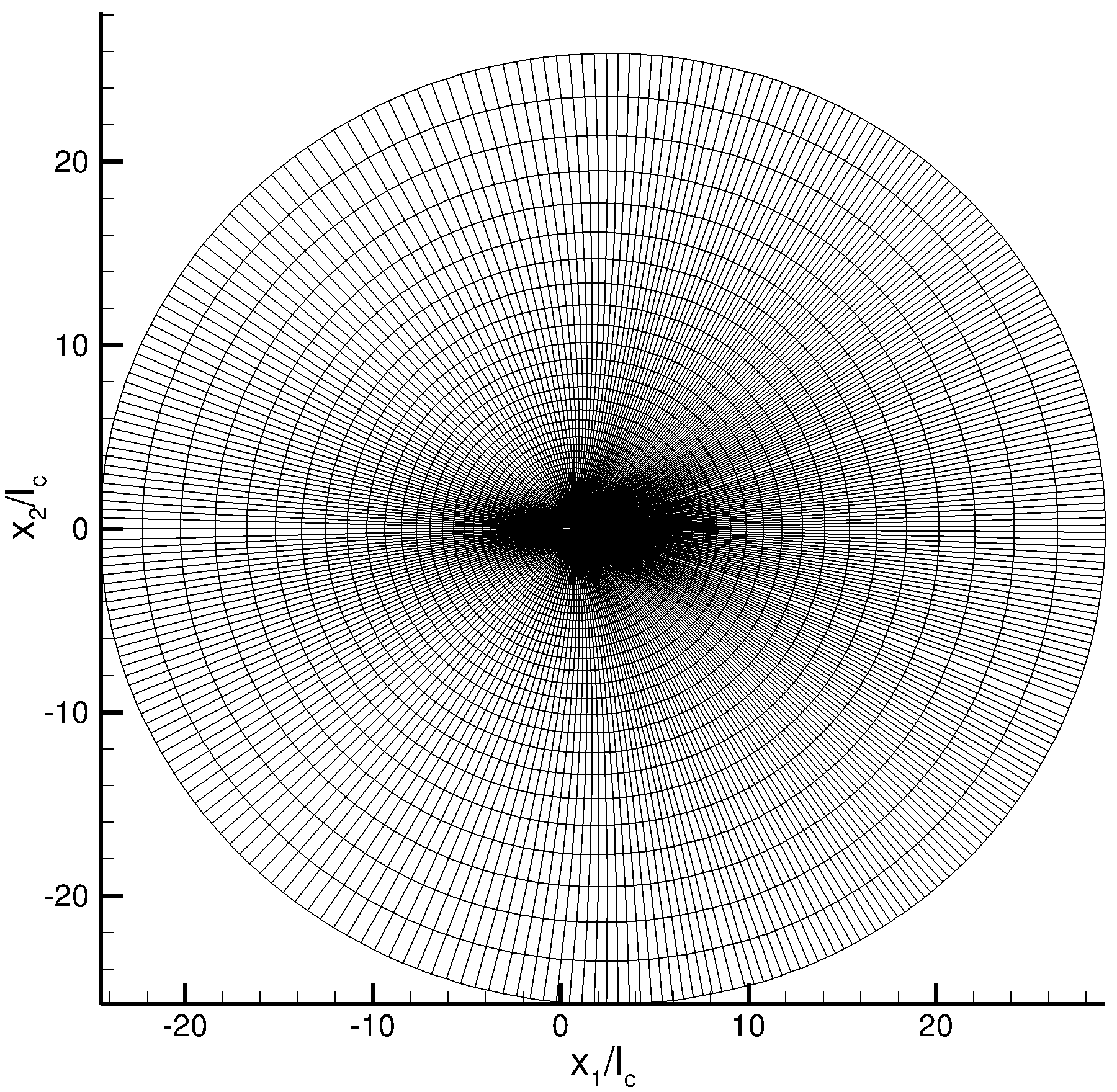} \hfill
\includegraphics[width=0.475\textwidth]{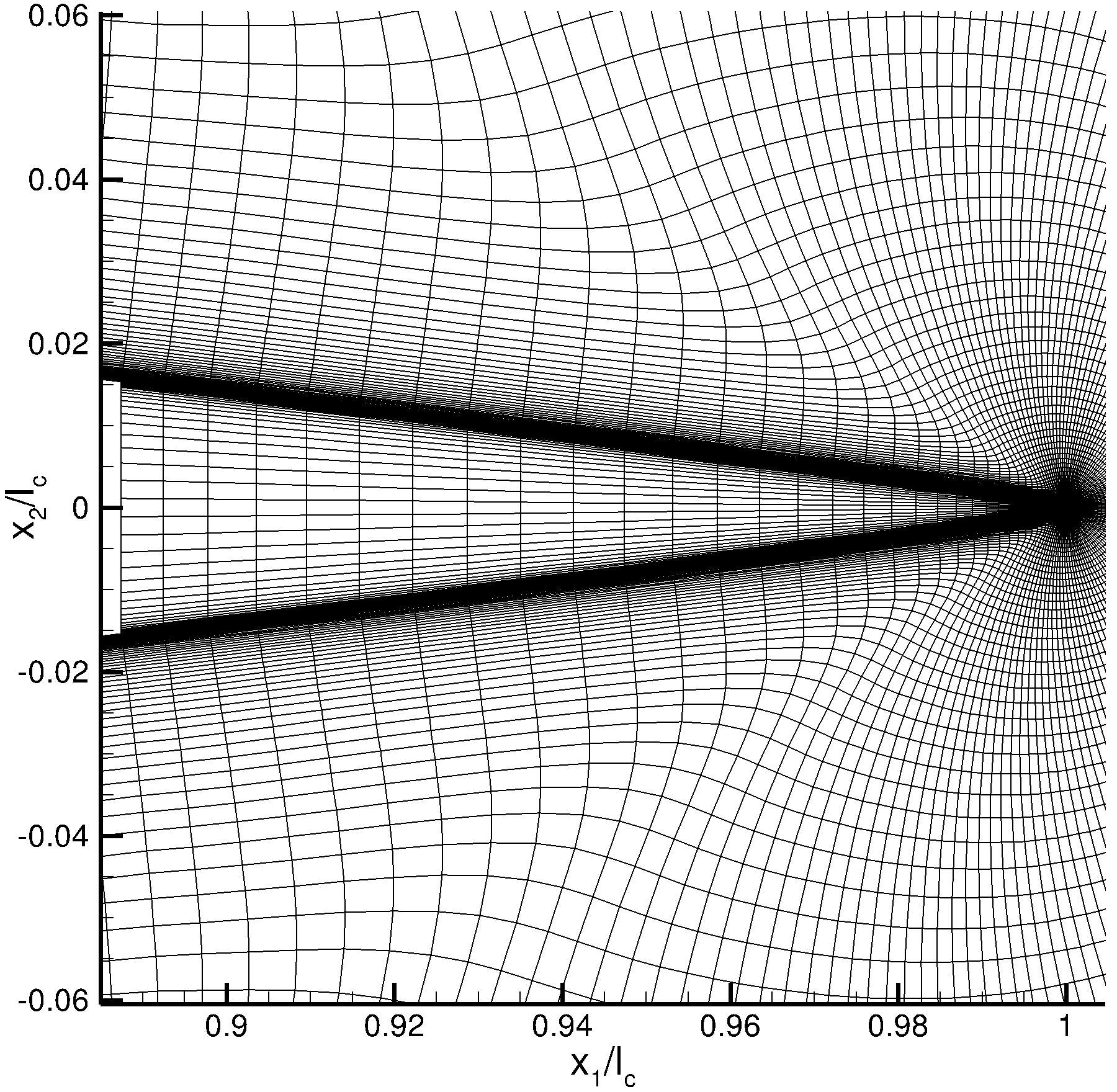}
\caption{(left) computational mesh for CFD simulations of the porous NACA0012 airfoil; (right) detailed view of the porous region at the trailing edge.}
\label{fig:cfd_mesh_porous}
\end{figure}

For the 2-D porous trailing edge, a fully structured mesh was built. The computational domain has an extension of $-24.5\cdot l_c$ in upstream, $29\cdot l_c$ in downstream and $\pm26\cdot l_c$ in vertical direction. The complete mesh consists of $\approx2.3\!\cdot\!10^{6}$ grid points and $\approx2.15\!\cdot\!10^{6}$ elements, respectively. The boundary layer is resolved by about $50$ grid points, such that a wall distance of $y^{+}<1$ is achieved. The computational domain for the solid reference case is mainly the same, except for the porous region at the trailing edge. \Fig{fig:cfd_mesh_porous} shows the grid, normalized by the chord length $l_c$, in an overview and the magnification of the porous resolved region. \par
\subsection{CAA simulations}
\paragraph{Generic airfoils} %
For saving computational time, the 2-D CAA grid has a reduced extension. This is $6\cdot l_c$ in both streamwise and vertical direction. The structured CAA mesh consists of $\approx1.1\!\cdot\!10^{6}$ grid points, distributed to $64$ blocks. The mesh is designed for a field frequency resolution of \unit[10]{kHz} using $7$ points per wavelength resolution. The smallest mesh cells are located in the FRPM area covered by the patch. A timestep of $\Delta t \approx \unit[4.7 \cdot 10^{-7}]{s}$ is used. In total $4\!\cdot\!10^{5}$ time steps was simulated resulting in a total real time sampling length of $t \approx \unit[0.19]{s}$. \par
\begin{figure}[tbh]
\centering
\includegraphics[width=0.475\textwidth]{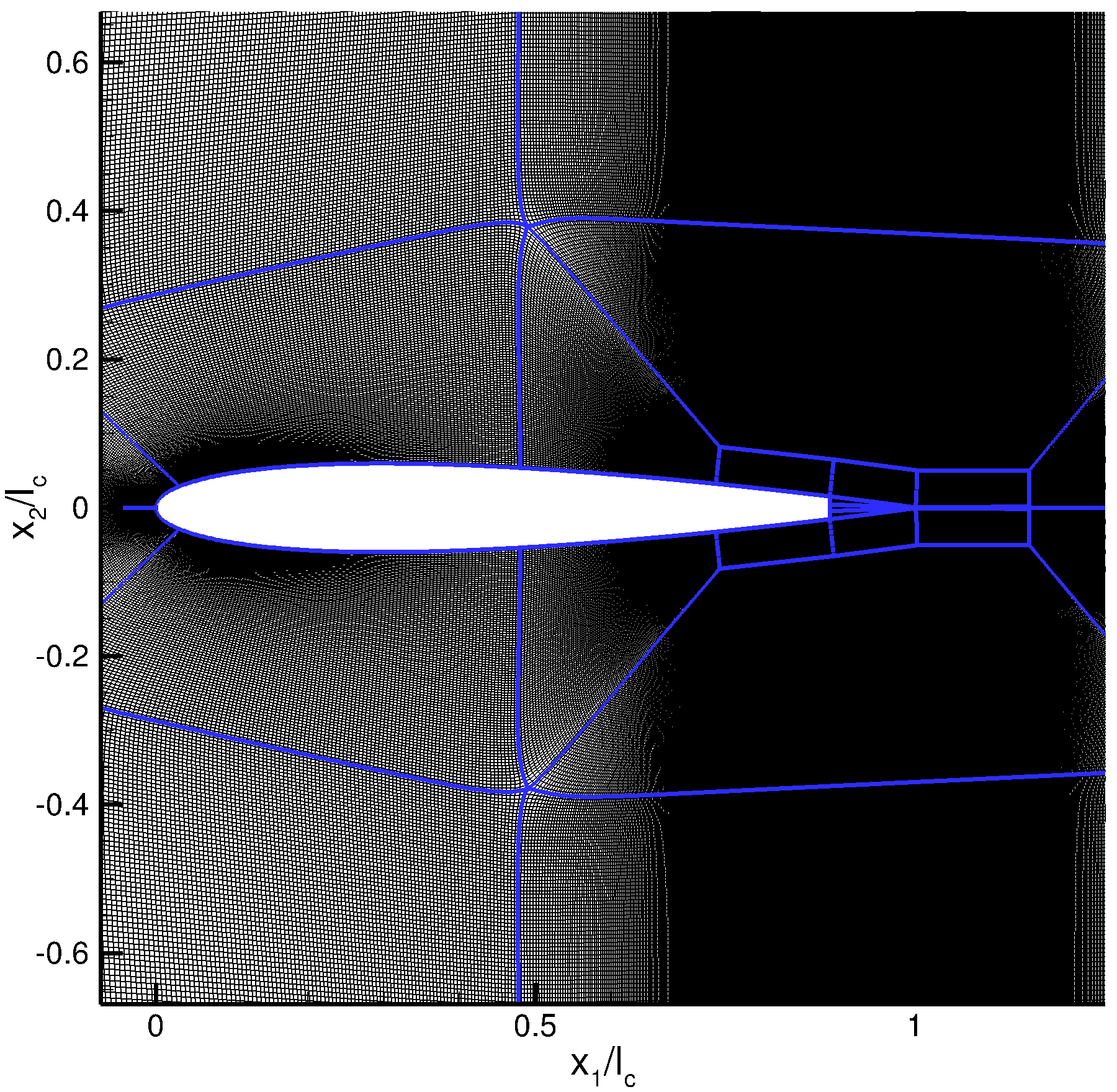} \hfill
\includegraphics[width=0.475\textwidth]{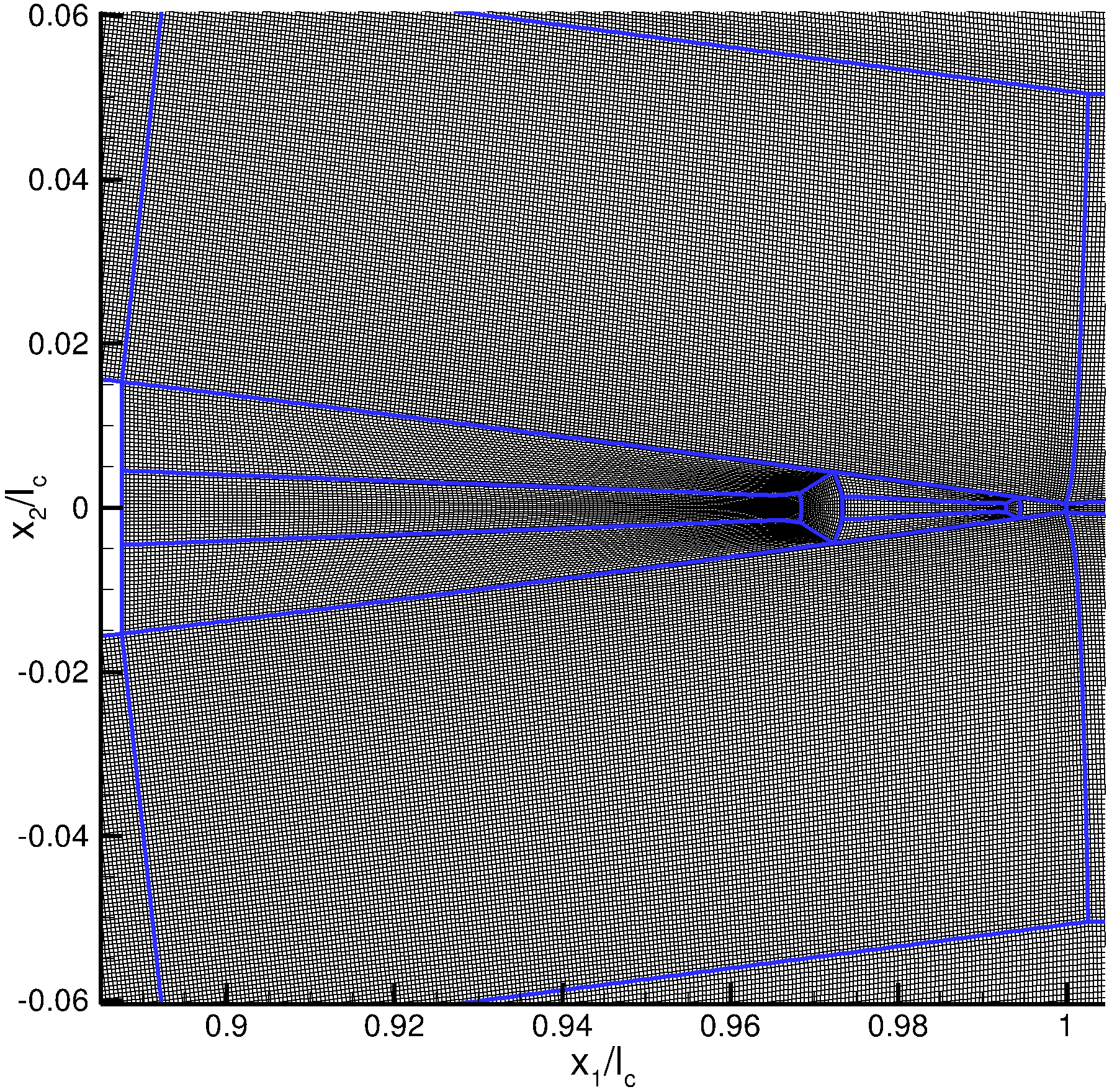}
\caption{(left) computational mesh for CAA simulations of the porous 2-D airfoil; (right) detailed view of the porous region at the trailing edge.}
\label{fig:caa_mesh_porous}
\end{figure}

\paragraph{Porous trailing edges} %
For the porous trailing-edge test cases, a similar CAA grid was chosen. Only the size of the higher resolved area in the vicinity of the trailing-edge was chosen larger as the porous region at the trailing edge had to be resolved in that case. Thus, an overall cell number of $\approx 1.5\!\cdot\!10^{6}$ grid points was achieved at equal resolution quality. This computational domain is divided into $78$ blocks, of which $9$ resolve the porous region with about $12.0\!\cdot\!10^{3}$ grid points. \Fig{fig:caa_mesh_porous} shows the grid, normalized by the chord length $l_c$, in an overview and the magnification of the porous resolved region. The minimal time step allowed is $\Delta t \approx \unit[1.3 \cdot 10^{-7}]{s}$ due to the small cells within the porous region. But, to prevent any numerical instability caused by the porosity model, a smaller time step of $\Delta t \approx \unit[1.2 \cdot 10^{-7}]{s}$ was used. This leads to a roughly $15\%$ higher computational time in total. The computation of $4\!\cdot\!10^{5}$ time steps was performed within less than \unit[48]{hours} on a state-of-the-art cluster system, using $12$ CPUs in parallel. \par
\section{SIMULATION RESULTS}  \label{sec:Results}
In this chapter, the result of a generic verification of the modified APE is presented. Then, the CFD/CAA approach is discussed in detail. Attention is drawn to the reconstruction of the turbulent statistics with the FRPM model. Finally, the experience from solid airfoils is transfered to the edge noise reduction problem at the trailing edge of a NACA0012 airfoil. The computational results will be compared to measurements, as long as they are available. \par
\subsection{Generic test problems}
As a simple verification test case for the implementation of the porous terms, the wave propagation in a quiescent medium is studied for different porous materials. Two different isotropic porosities were chosen: A generic one with artificially chosen porosity parameters and a realistic one with parameters deduced from measurements. The realistic case is based on sintered metals fiber felt. A second test case is wave propagation in an anisotropic porous material without mean-flow. For both types of verification, the 2-D-computational domain has an extension of $x = \unit[-1]{m}\dots \unit[+1]{m}$ and $y = \unit[-0.5]{m} \dots \unit[0.5]{m}$. The fluid phase is ambient air considered as an ideal gas with properties $c_0 = \unit[343]{m/s}$, $p^0 = \unit[1.01325\cdot10^5]{Pa}$ and $\rho^0 = \unit[1.205]{kg/m^3}$. The grid is designed to resolve frequencies up to $f = \unit[20]{kHz}$. No additional background damping was used.

\begin{figure}[tbh]
  \centering
  \includegraphics[clip,width=0.32\linewidth,angle=0]{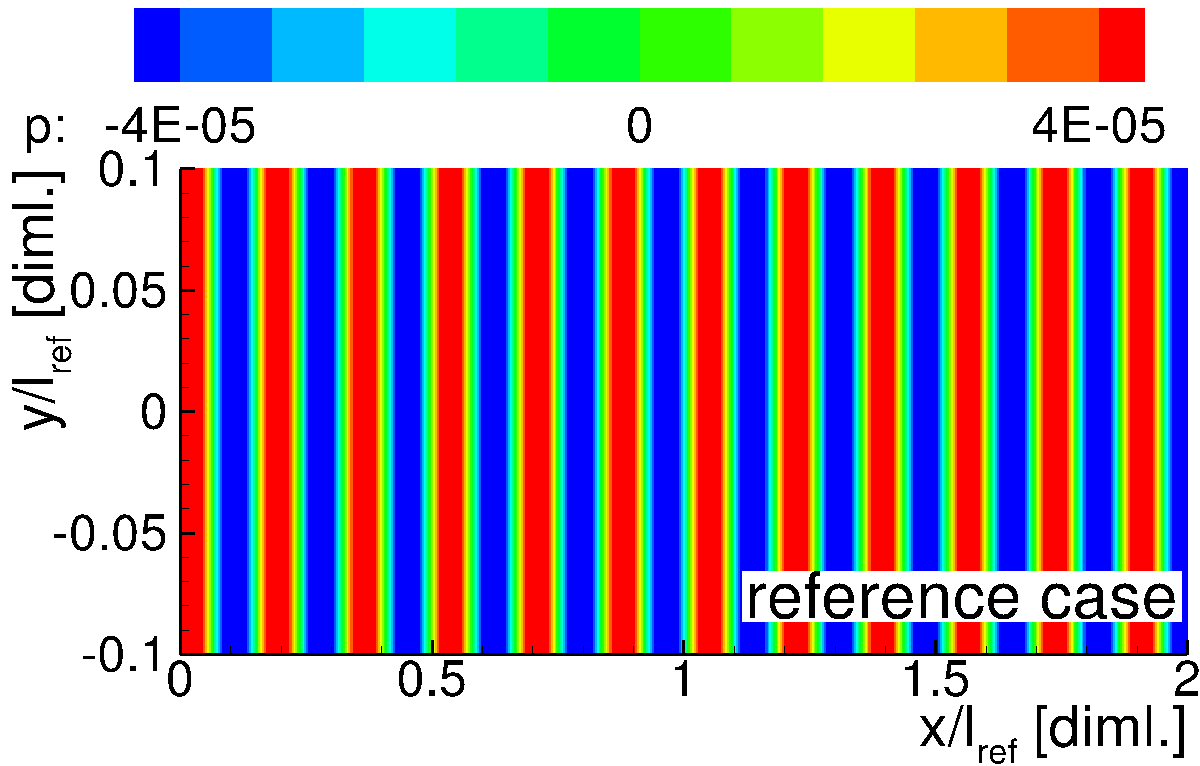} \hfill%
  \includegraphics[clip,width=0.32\linewidth,angle=0]{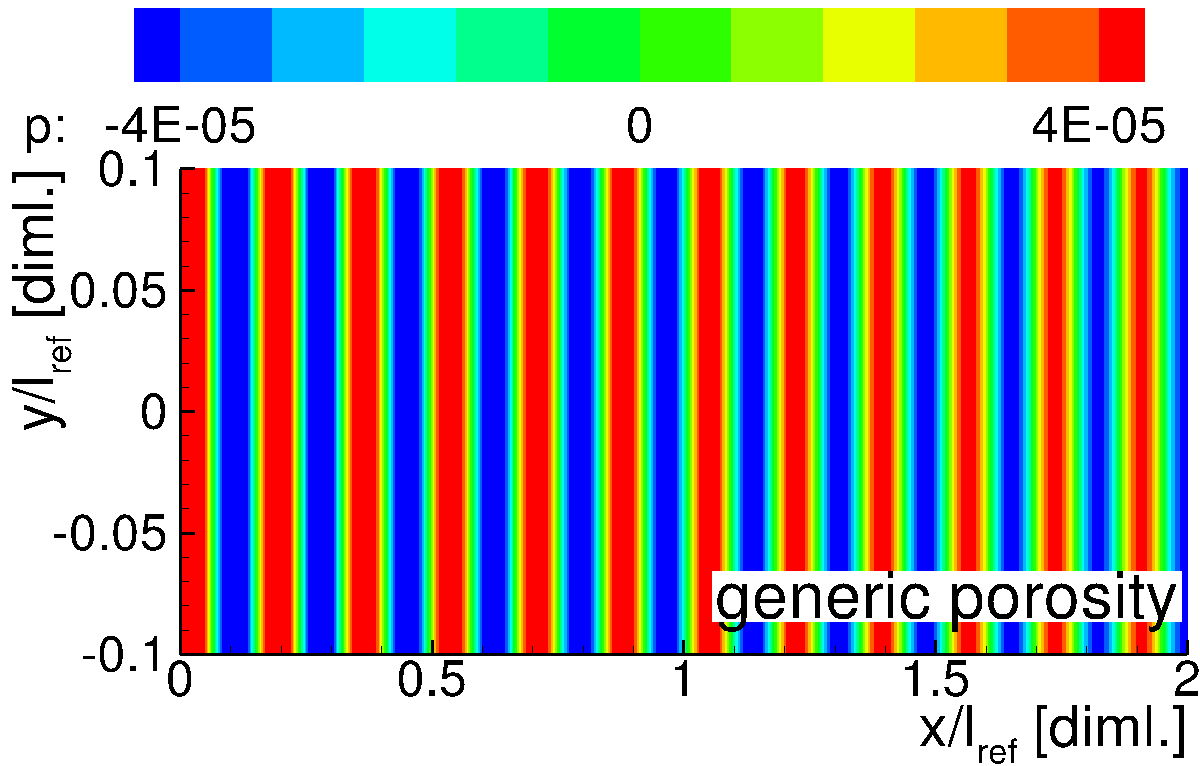} \hfill%
  \includegraphics[clip,width=0.32\linewidth,angle=0]{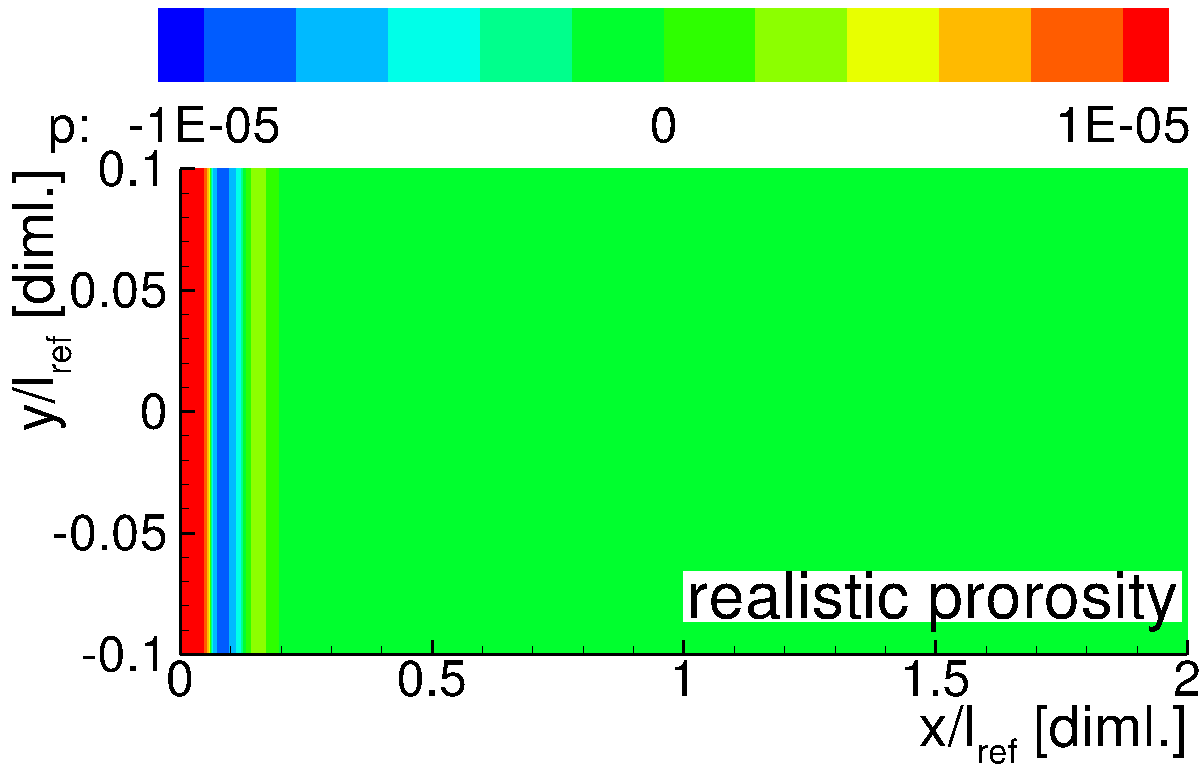} \hfill%
  \caption{Snapshot of the acoustic response to a harmonic plane wave front at \unit[2]{kHz} for three different isotropic materials; (left) free medium as a reference,  (middle) generic porosity and (right) realistic porosity.}\label{fig:snap_Shot_Thompson}
\end{figure}

For the first test case with isotropic porosity and a quiescent medium, monofrequent incoming plane waves with different frequencies were specified at the left boundary through a Thompson boundary condition, refer to  \citet{Thompson.1987,Thompson.1990}. In \fig{fig:snap_Shot_Thompson}, a snap shot of the acoustic response is shown for a harmonic signal at \unit[2]{kHz}. For quantitative evaluation, signals were picked from a horizontal line at $y=\unit[0.1]{m}$. \par
\begin{figure}[tbh]
 \centering
 \includegraphics[clip,scale=1,angle=0]{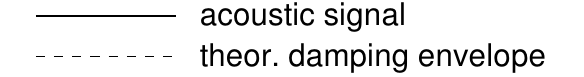}\\[0.35\baselineskip]%
 \parbox{0.31\linewidth}{\includegraphics[clip,width=\linewidth,angle=0]{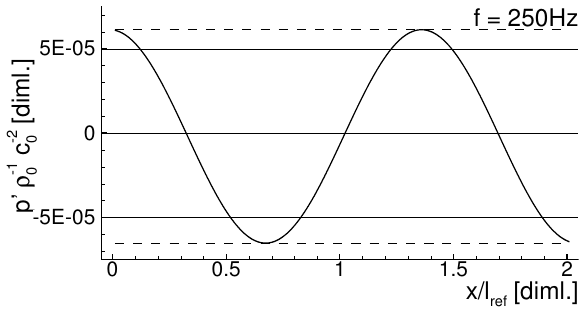}\\[0.3\baselineskip]%
                         \includegraphics[clip,width=\linewidth,angle=0]{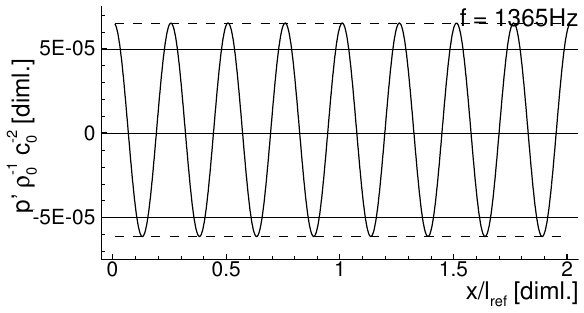}\\[0.3\baselineskip]%
                         \includegraphics[clip,width=\linewidth,angle=0]{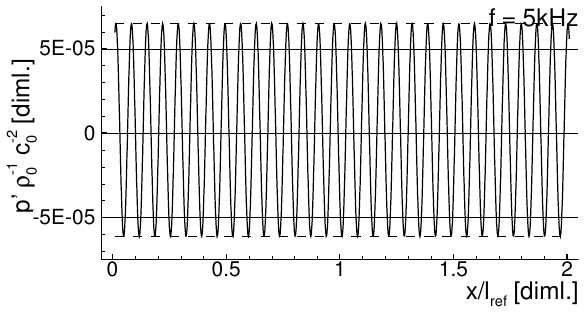}\\[0.3\baselineskip]%
                         \includegraphics[clip,width=\linewidth,angle=0]{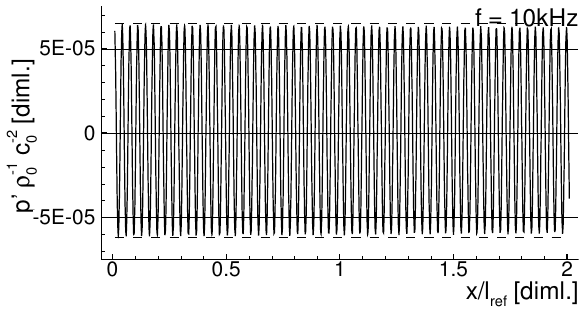}\\[0.3\baselineskip]%
                         {\small reference case: \\ $\phi = 1, \frac{\nu}{\kappa} = \unit[0]{\frac{1}{s}} $}}\hfill
 \parbox{0.31\linewidth}{\includegraphics[clip,width=\linewidth,angle=0]{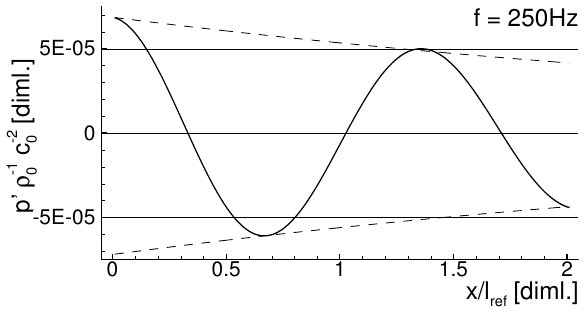}\\[0.3\baselineskip]%
                         \includegraphics[clip,width=\linewidth,angle=0]{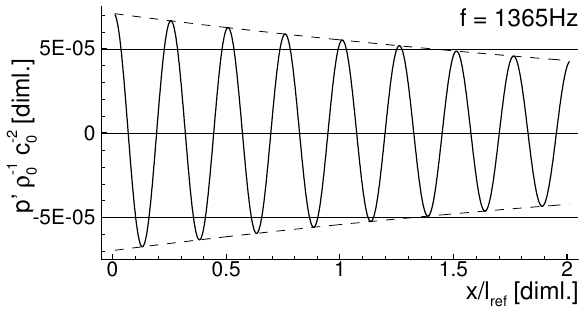}\\[0.3\baselineskip]%
                         \includegraphics[clip,width=\linewidth,angle=0]{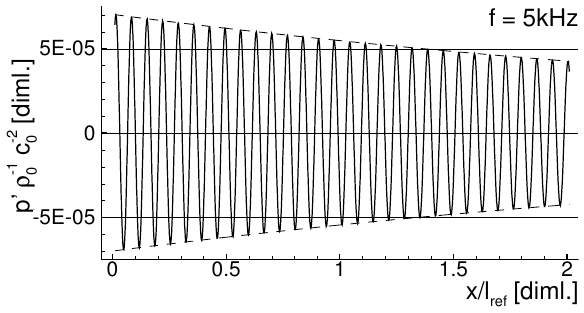}\\[0.3\baselineskip]%
                         \includegraphics[clip,width=\linewidth,angle=0]{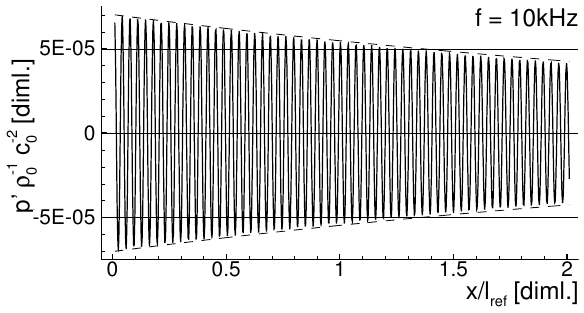}\\[0.3\baselineskip]%
                         {\small generic porosity: \\ $\phi = 0.8, \frac{\nu}{\kappa} = \unit[2.144]{\!\cdot10^{2}\frac{1}{s}} $}}\hfill
 \parbox{0.31\linewidth}{\includegraphics[clip,width=\linewidth,angle=0]{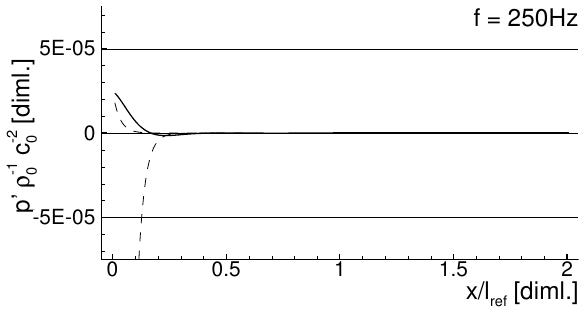}\\[0.3\baselineskip]%
                         \includegraphics[clip,width=\linewidth,angle=0]{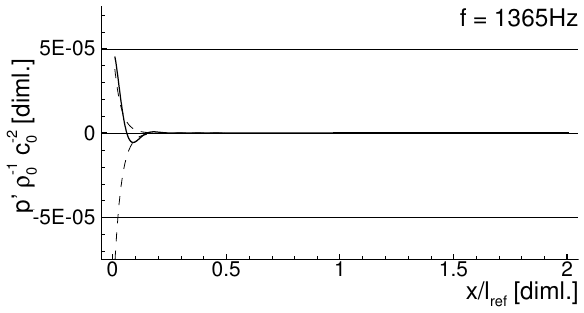}\\[0.3\baselineskip]%
                         \includegraphics[clip,width=\linewidth,angle=0]{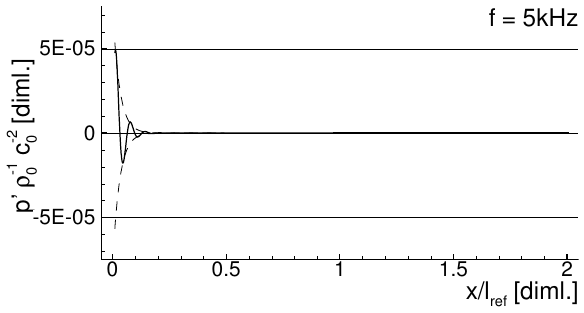}\\[0.3\baselineskip]%
                         \includegraphics[clip,width=\linewidth,angle=0]{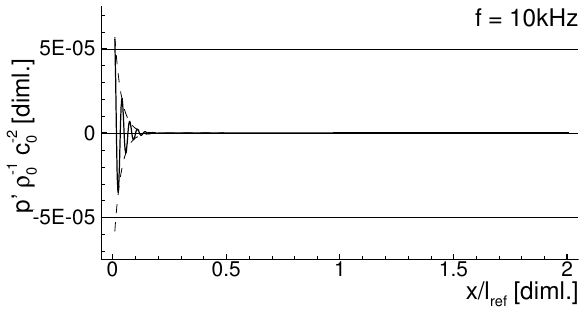}\\[0.3\baselineskip]%
                         {\small realistic porosity: \\ $\phi = 0.86, \frac{\nu}{\kappa} = \unit[26.336]{\!\cdot10^{3}\frac{1}{s}} $}}\hfill
 \caption{Comparison of CAA results with Acoustic Pertubation Equations (APE) with theoretical damping envelope of the 1-D wave equation in a medium at rest from \eqn{eqn:theoretical_damping} for isotropic porosity.} \label{fig:1D_WavePropagation_porosity}
\end{figure}

For a medium at rest, the theoretical damping envelope can be derived from the 1-D wave equation. This is shown in Appendix~\ref{app:Wave_Equation}, based on the acoustic potential. The damping can be reduced to the ratio of the local root mean square value $\tilde p$ of the sound pressure $p^\prime$ to its maximal value at the position $x_s$ of the source, i.e. $\tilde p^2 = \overline{\left( p^\prime\right)^2}$. Considering a homogeneous, isotropic porosity, the damping envelope with respect to $x$ results from \eqns{eqn:app_acoustic_pressure}{eqn:app_acoustic_potential_expanded}:
\begin{eqnarray}
  \frac{\tilde p (x)}{\tilde p_\ind{max}} &=&\exp\left( - \frac{\omega}{c_0}\sqrt{\frac{1}{2}\sqrt{\frac{(\phi\nicefrac{\nu}{\kappa})^2}{\omega^2} + 1} - \frac{1}{2}} \; x \right) = \exp(\lambda^\ind{(D,sim)} x) \label{eqn:theoretical_damping} \\
   && \text{with } \tilde p_\ind{max} = \tilde p (x_s) \Punkt \nonumber
\end{eqnarray}
Further, it is $L_\ind{p}$ the sound pressure level of $\tilde p$. Namely it is
\begin{equation}
  L_\ind{p}^\ind{(D,sim)}(x) = 20 \log \left( \frac{\tilde p (x)}{\tilde p_\ind{max}} \right) = \frac{20}{\ln(10)} \lambda^\ind{(D,sim)} x \Punkt
\end{equation}
While the exponent $(\ind{D})$ denotes the analytical damping behavior, the exponent $(\ind{sim})$ indicates the damping behavior predicted by the simulation. Finally, the degree of agreement between the analytical result and the computiation can be expressed by
\begin{equation} \label{eq:Comparison_Damping}
  L^\ind{(sim)}_\ind{p} = f(\lambda^\ind{(D)} x) \Punkt
\end{equation}
In \fig{fig:1D_WavePropagation_porosity}, the results from simulation are juxtaposed to the theoretical decay envelope. The acoustic signal from CAA matches the theoretical damping envelope from \eqn{eqn:theoretical_damping} very well. \par
\begin{figure}[tbh]
  \centering
  \includegraphics[clip,width=\linewidth,angle=0]{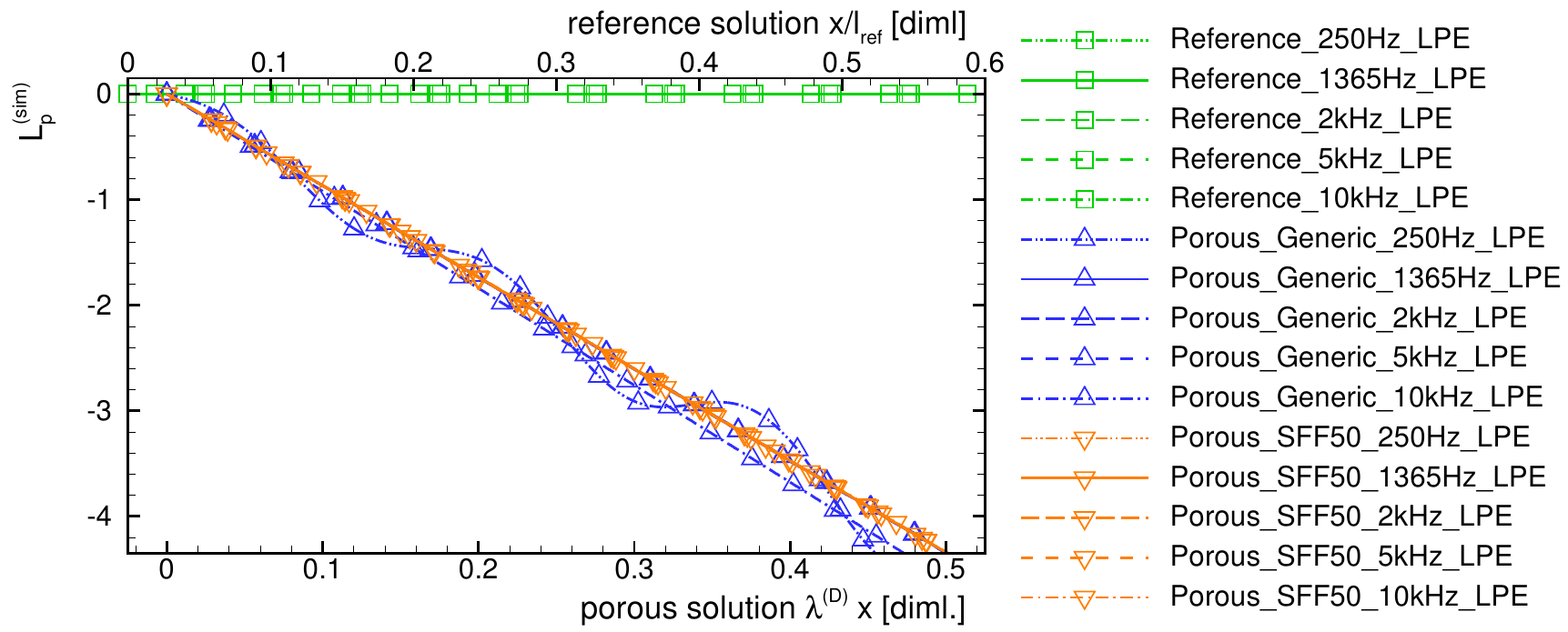}
  \caption[]{Comparison of CAA results with Acoustic Pertubation Equations (APE) with theoretical damping envelope of the 1-D wave equation in a medium at rest from \eqn{eqn:theoretical_damping} for isotropic porosity. For the reference case, the computational prediction is directly given, without comparison with the theoretical damping coefficient, what is $\lambda^\ind{(D)} = 0$ for the free medium.} \label{fig:1D_WavePropagationRMS_porosity}
\end{figure}

The plot of the level decay of the rms-values  within the computational domain is shown in \fig{fig:1D_WavePropagationRMS_porosity} in the way introduced by \eqn{eq:Comparison_Damping}. The computational results fit convincingly well the theoretical envelope along the range of \unit[0.25]{m}. With respect to technical application, this is a good result, as a typical installation depth is less. The prediction accuracy is comparable for a large frequency range. \par
 \par
\begin{figure}[tbh]
  \centering
  \includegraphics[clip,width=.5\linewidth,angle=0]{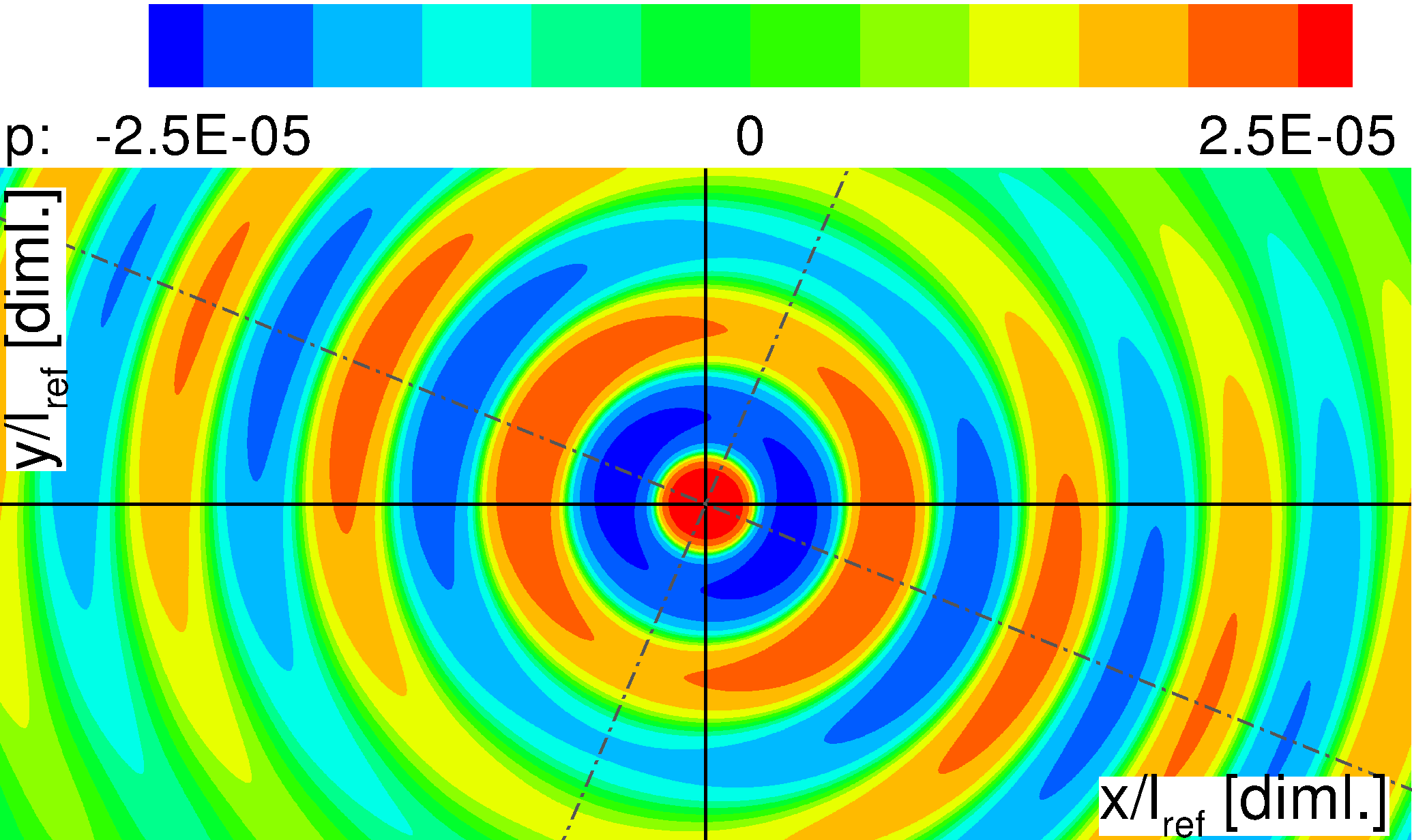}
  \caption{Snap shot of the acoustic response of a anisotropic homogeneous material to an harmonic pressure pulse at a frequency of \unit[1365]{Hz}; the main axis system is indicated by dashed lines.}\label{fig:Hauptachsen_anisotropic_porosity}
\end{figure}

For the anisotropic porous material, a qualitative evaluation of the computational results is done. The matrix $\mu_{ij}$ describing the damping properties was arbitrarily chosen to
\begin{equation}
 \mu_{ij} = \begin{bmatrix}
             1.0&2.0  \\
             2.0&5.0 
           \end{bmatrix}\unit{\frac{1}{s}}
\end{equation}
The resulting eigenvalues $\lambda_{1,2}$ and the related eigenvectors $\vM{x}_{1,2}$ are:
\begin{eqnarray*}
    \lambda_1 = 3 - \sqrt{8}\unit{\frac{1}{s}} \Komma & \vM{x}_1= \left( 0.924, -0.383 \right)^T \unit{m} \\%
    \lambda_2 = 3 + \sqrt{8}\unit{\frac{1}{s}} \Komma & \vM{x}_2= \left( 0.383,  0.924 \right)^T \unit{m}%
\end{eqnarray*}
In \fig{fig:Hauptachsen_anisotropic_porosity}, the acoustic response to an harmonic pressure pulse is shown for a material with anisotropic damping properties. The plot illustrates the orientation of the two eigenvectors $\vM{x}_{1,2}$ and emphasizes that the signal decay along the direction of the first eigenvector $\vM{x}_1$ is less than the signal decay along $\vM{x}_2$. This corresponds to the magnitude of the related eigenvalues. \par
 \par
%
%
The results verify proper implementation of the porous model. \par
\subsection{Generic airfoils}
Noise simulations for generic airfoils with the strategy described in the previous part were carried out for the trailing-edge problem statement of the BANC-II workshop, refer to \citet{Herr.2012}. Herein flow and noise characteristics of 2-D airfoil sections under nominally uniform flow, defined by the flow velocity $U_\infty$ and the angle of attack $\alpha$, were evaluated from an experimental and a numerical perspective. Measurement data for the validation of numerical codes was made available in the BANC-II problem statement. The computations in this article are mainly focused on test cases \#1, \#2 and \#3, where a NACA0012 airfoil (chord length $l_c=0.4m$) with different angles of attack under a flow Mach number of $\Ma=0.16$ and a Reynolds number of $\Rey_c=1.5\cdot10^6$ is considered. The angles of attack were ranging from $\alpha=\unit[0]{^\circ}$ (case \#1) over $\alpha = \unit[4]{^\circ}$ (case \#2) to $\alpha=\unit[6]{^\circ}$ (case \#3).
\begin{figure}[tbh]
\centering
  \includegraphics[width=0.3\textwidth]{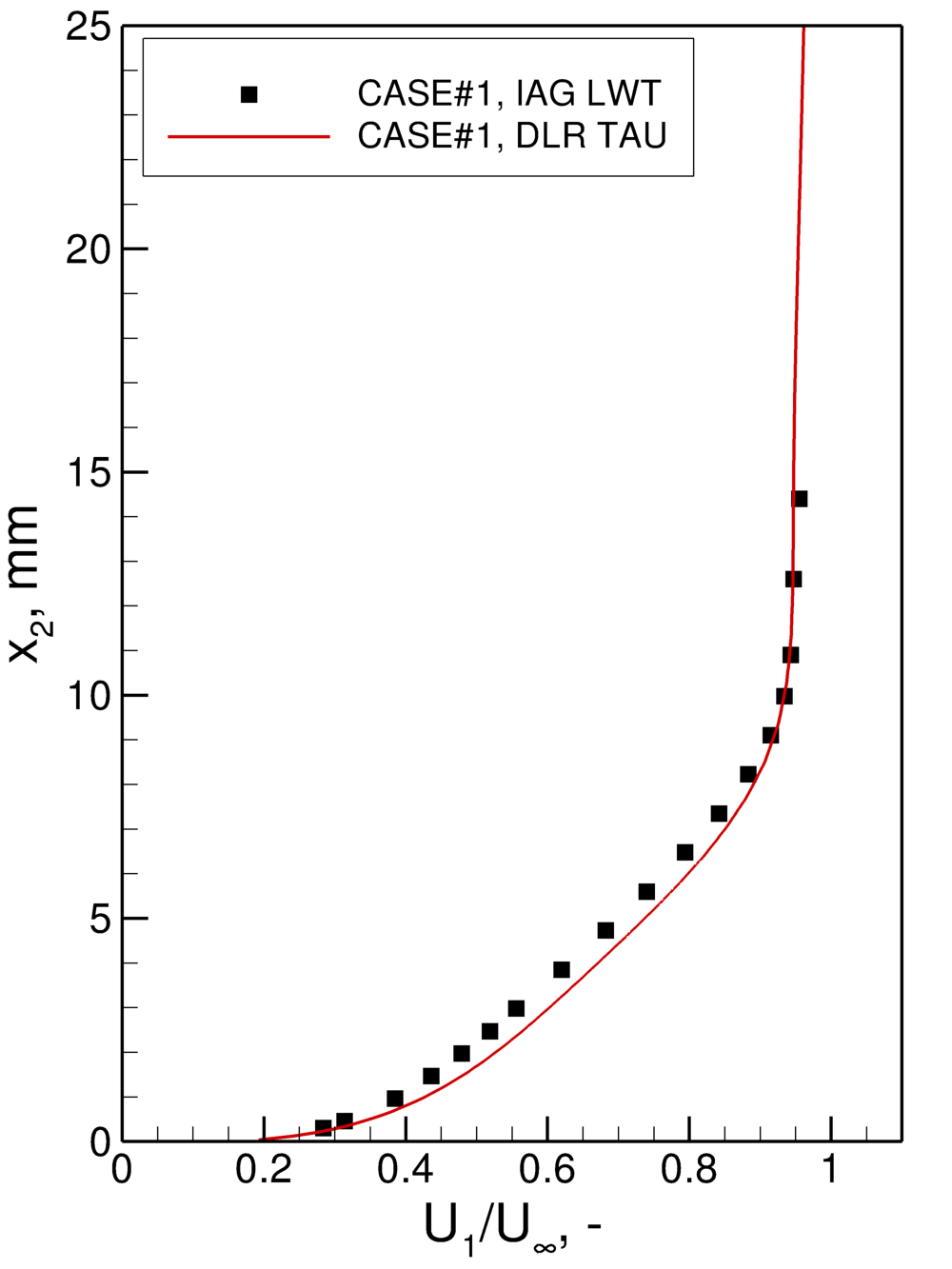}\hfill
  \includegraphics[width=0.3\textwidth]{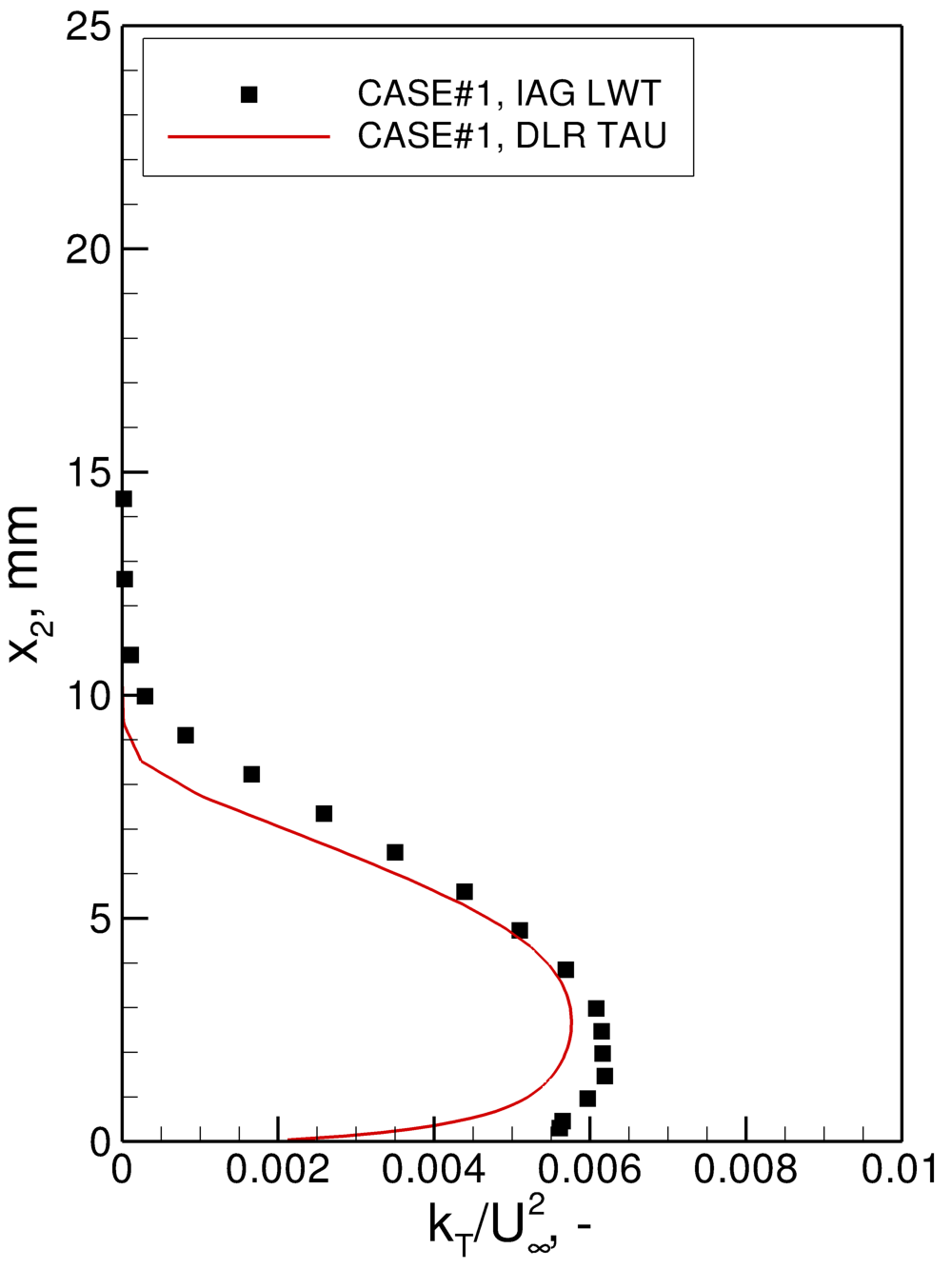}\hfill
  \includegraphics[width=0.3\textwidth]{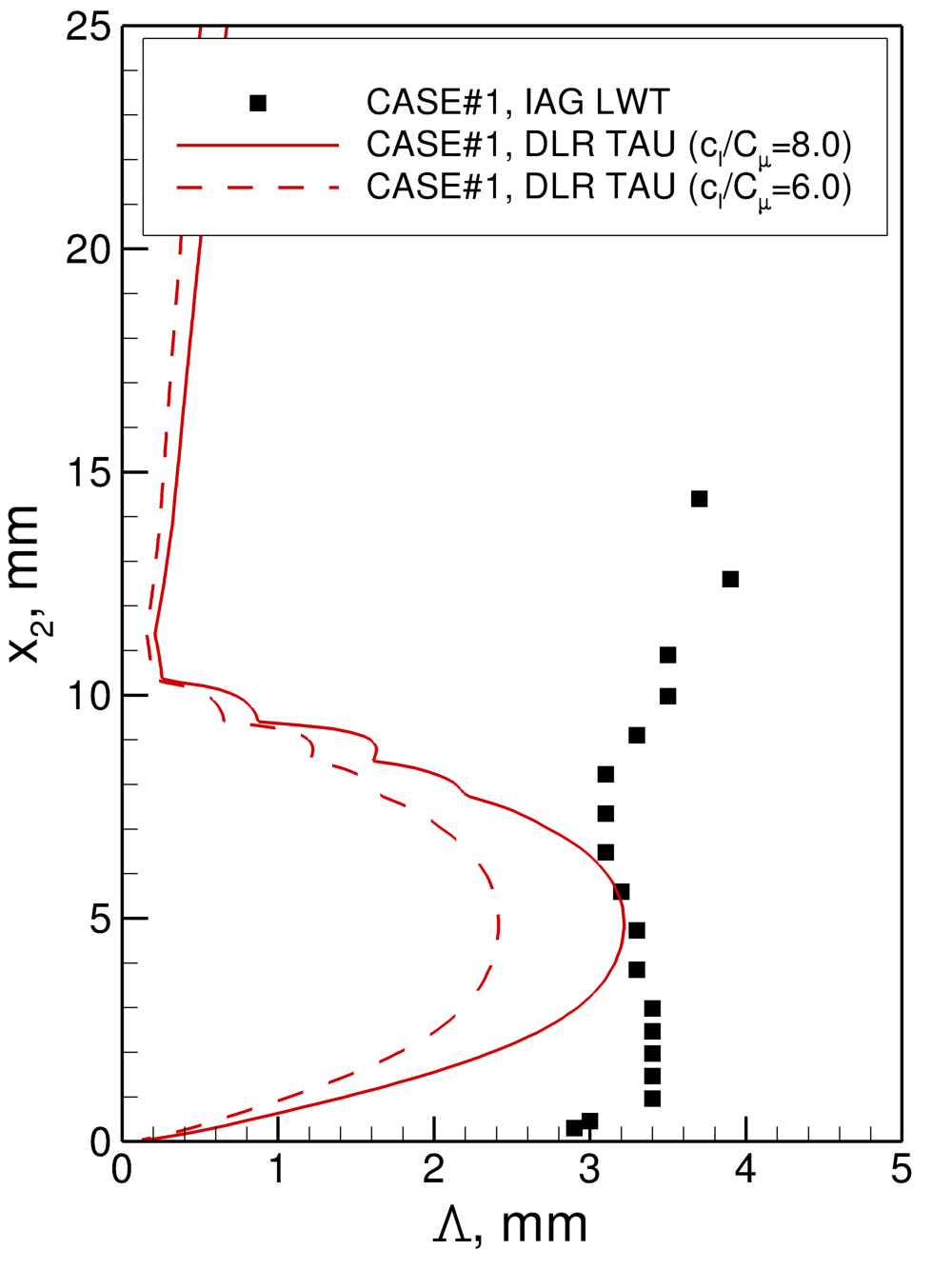}
\caption[]{NACA0012 test case \#1 ($Ma=0.16$, $\alpha=\unit[0]{^\circ}$) trailing-edge boundary layer profiles; (left) normalized velocity profile in streamwise direction; (middle) normalized turbulent kinetic energy; (right) longitudinal integral length scale for different $c_l/C_\mu$ ratios.}
\label{fig:cfd_bl}
\end{figure}
\Fig{fig:cfd_bl} shows boundary layer values simulate by TAU and data from the experiments conducted by IAG Stuttgart, \citet{Herrig.2011}. For the evaluation, the profiles are compared along a vertical line (perpendicular to the airfoil chord) positioned $0.38\%$ main chord behind the airfoils trailing-edge. The NACA0012 zero degree angle of attack test case is chosen for the presentation of the CFD results. The results are in good agreement with the experimental data. The values for the normalized velocity profile in streamwise direction $x_1$ (\fig{fig:cfd_bl}~(left)) are in line with the RANS computations. The velocity profile of the TAU simulation shows slightly higher values. However, the general shape and boundary layer thickness (location where $0.99\,U_\infty$ is reached) match quite closely, implying that the used k-$\omega$-SST turbulence model is suitable for this case. Moreover, the non-dimensionalized turbulent kinetic energy profile $k_T/U_\infty^2$ (\fig{fig:cfd_bl}~(middle)) matches the experimental data. For the used two-equation model with its isotropic turbulence ($\langle u'_1u'_1 \rangle=\langle u'_2u'_2 \rangle=\langle u'_3u'_3 \rangle = \langle u'u' \rangle$) the longitudinal turbulent length scale $\Lambda$ (see \fig{fig:cfd_bl}~(right)) can be calculated from the turbulent kinetic energy $k_T$ and the specific dissipation $\omega$ :
\begin{equation}
  \Lambda=\frac{c_l}{C_\mu} \frac{\sqrt{k_t}}{\omega}.
  \label{eq:length_scale}
\end{equation} 
The quantities $c_l$ and $C_\mu$ are constants from the turbulence model. The fraction\footnote{for $C_\mu=0.09$ this implies a value of $c_l=0.54$} $\nicefrac{c_l}{C_\mu}$ is now referred to as $c_\Lambda$. It can be observed that, for a $c_\Lambda = 6.0$ the measured length scales (black squares) are higher than the computed ones (dashed line). If the ratio is adapted to $c_\Lambda = 8.0$ (solid line), the fitted computational results match the measured values concerning the absolute level. Despite the absolute values of the length scale $\Lambda$, the simulated flow field and its turbulence statistics are in good agreement with the experimental values. Concerning the details of the comparison between the measuremts and the prediction of the length scale, a deviation is obvious. Nonetheless, in the region of maximum turbulent kinetic energy, the accordance of the numerical result with the measurement is sufficiently good. This is the region of the most relevance for the reconstruction of the final sound spectra. Thus, a validated and reliable input data set is provided for the following CAA simulations. In the case of non-zero angles of attack, the CFD is underpredicting the profiles of the turbulence statistics. For details refer to \citet{Herr.2015}. \par
\begin{figure}[tbh]
\centering
  \includegraphics[width=0.475\textwidth]{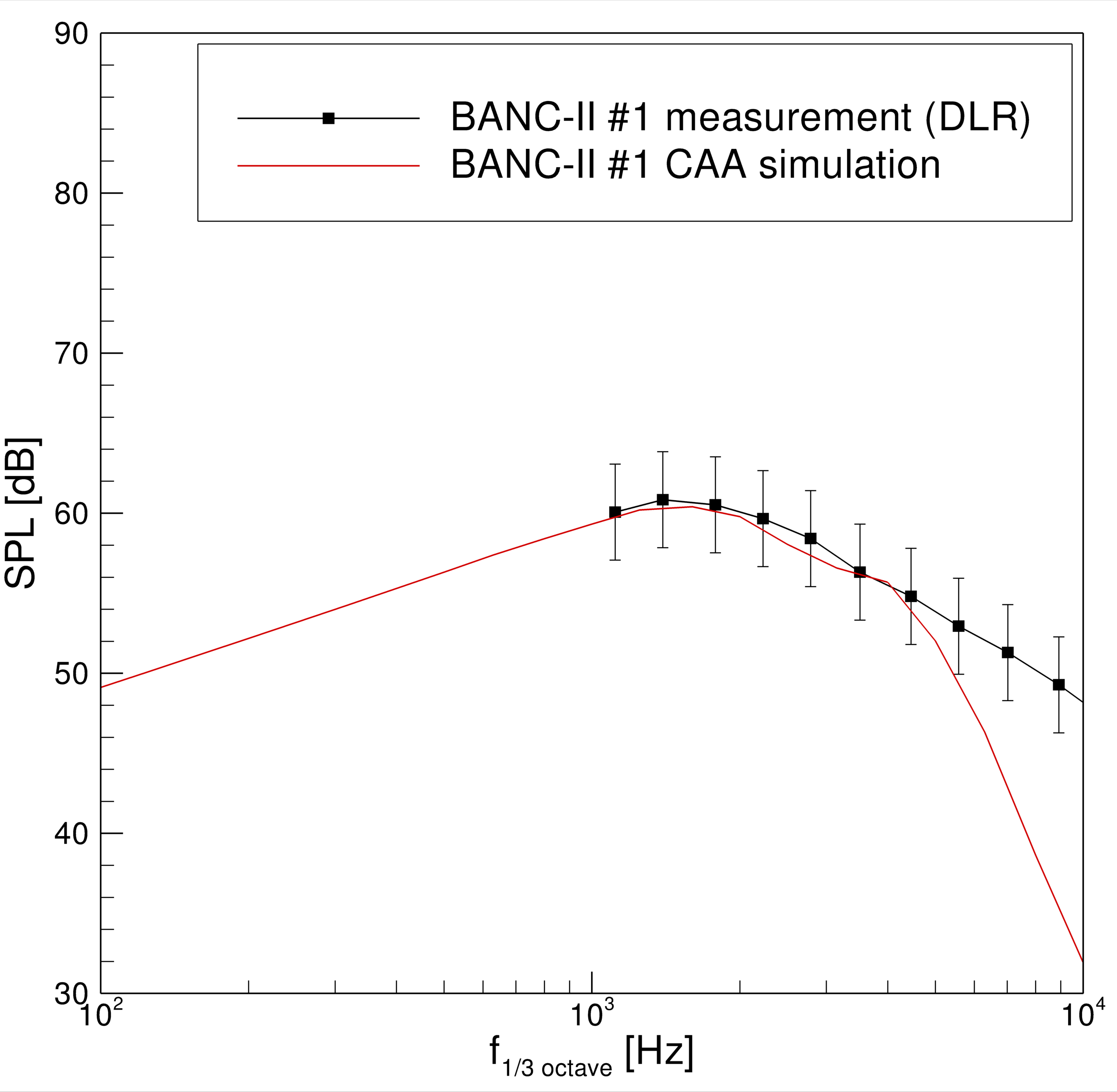}\hfill
  \includegraphics[width=0.475\textwidth]{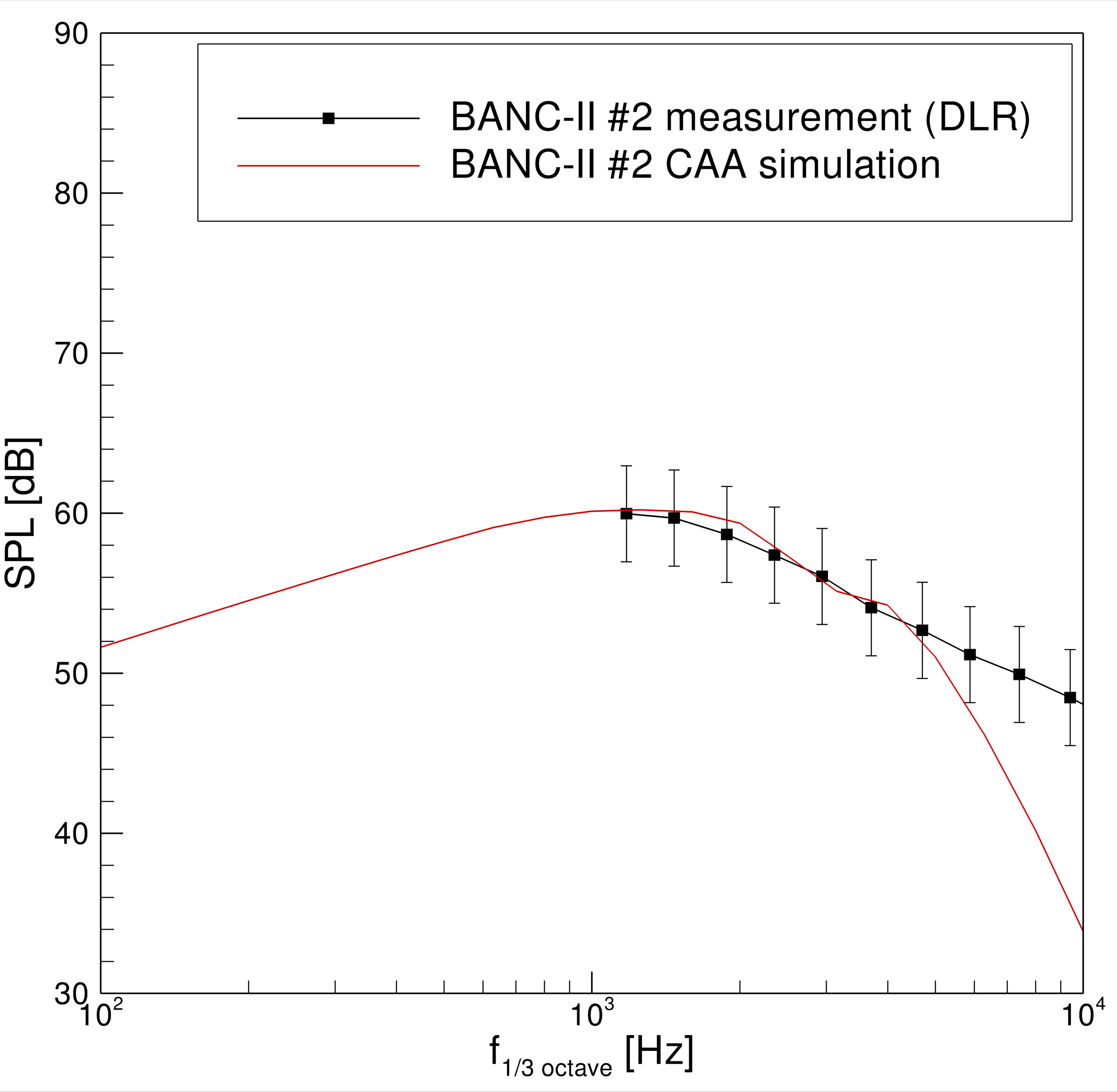}
\caption[]{One-third-octave band spectra from CAA simulation compared to available measurement data from BANC-II; (left) test case \#1 ($\alpha=\unit[0]{^\circ}$); (right) test case \#2 ($\alpha=\unit[4]{^\circ}$).}
\label{fig:caa_gen_airfoils1}
\end{figure}
\begin{figure}[tbh]
\centering
  \includegraphics[width=0.475\textwidth]{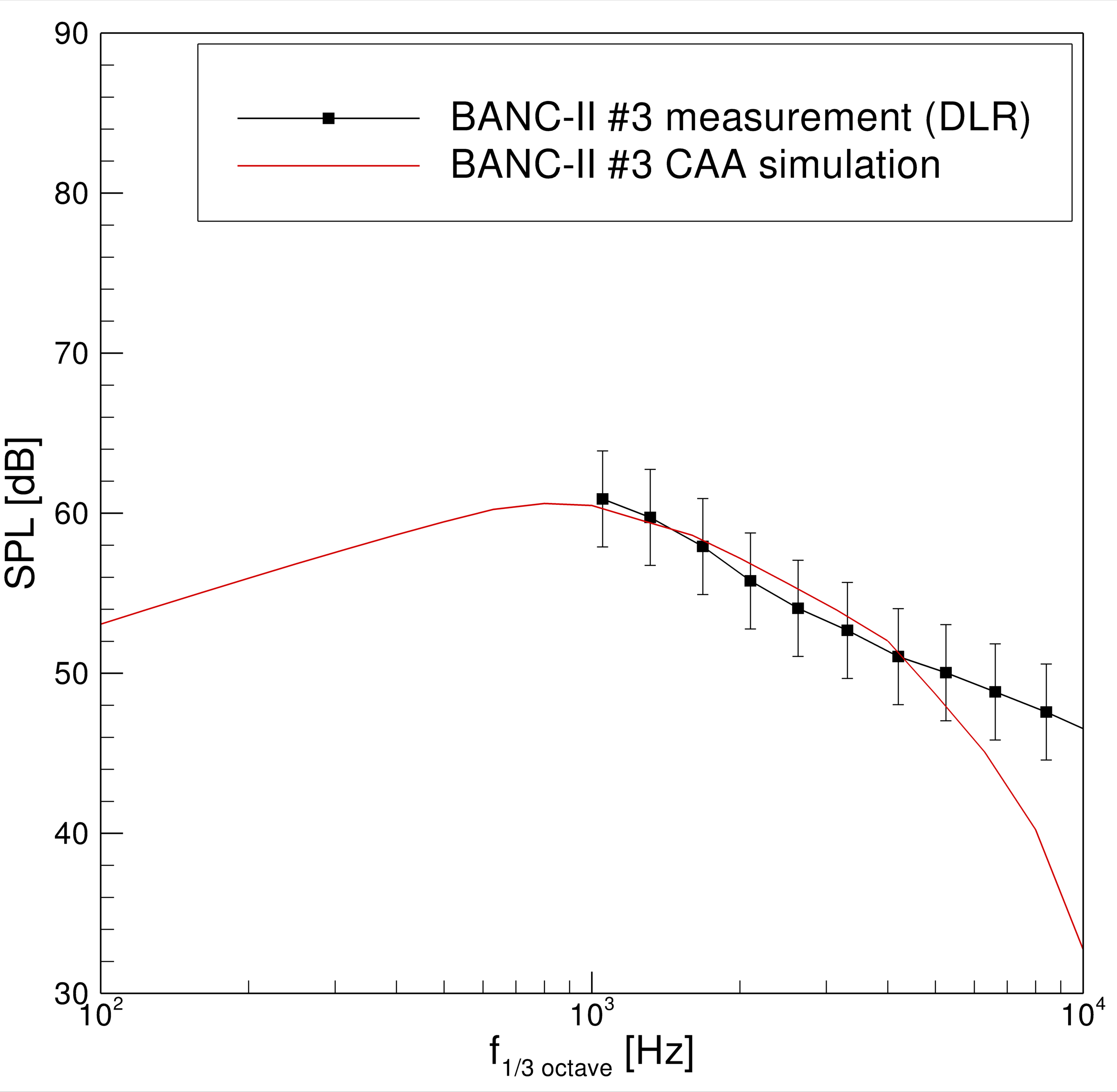}
\caption[]{One-third-octave band spectrum from CAA simulation compared to available measurement data from BANC-II; test case \#3 ($\alpha=\unit[6]{^\circ}$).}
\label{fig:caa_gen_airfoils2}
\end{figure}
Simulated and measured broadband one-third-octave band spectra for the three test cases are shown in \figs{fig:caa_gen_airfoils1}{fig:caa_gen_airfoils2}. For the evaluation a microphone position perpendicular to the chord in a distance of $2.5\,l_c$ below the trailing-edge was chosen. Sampling data was recorded every \unit[$2.5 \cdot 10^{-5}$]{s}. For the spectral evaluation a fast Fourier transform with a Hanning window was used. An average of 20 samples was taken to achieve a smoother shape of the curve. The \unit[3]{dB} uncertainty given by the measurements is indicated by the error bars. For the simulated sound pressure level spectra shown in this section, the 2-D to 3-D correction as proposed by \citet{Ewert.2009} has been used. Thus, the sound radiation is corrected to account for the 3-D case, as measured in the BANC-II test cases. Untapered, unswept airfoil sections with a wetted span of \unit[1]{m} were considered. A common off-set calibration of \unit[-2.5]{dB} was used on all simulated CAA spectra. With the chosen best practice setup this additional calibration on top of the 2-D to 3-D correction is needed to match the experimental results in terms of absolute levels. \par
It can be seen that the CAA spectrum (red line) matches the peak frequency of the experimental data and is in the uncertainty range until approximately \unit[5]{kHz} for all test cases. Until that point the decay behavior towards high frequencies matches the experimental observations. Beyond this value the simulated SPL values are lower than in the experiment. A procedure to successfully remedy this problem is currently under investigation, see \citet{Rautmann.2014}. However, for the low and medium frequency range the procedure is able to correctly predict the broadband noise emitted by the specific airfoil under different operational parameters. For the high frequency range a relative comparison is possible. \par
For the increasing angle of attack the peak frequency of the spectrum is shifted further towards lower values. Lying at approximately \unit[1.5]{kHz} for the zero degree angle of attack test case (\#1) the peak frequency moves to \unit[0.8]{kHz} for test case \#3 with $\alpha=\unit[6]{^\circ}$. The general trend of the angle of attack increase is reproduced. The peak frequency shifts towards lower values as for the higher frequencies the SPL values are going up. This is due to the fact that the spectrum is a combination of two single spectra. One low-frequent part coming from the suction side with its relatively wide boundary layer and a second high-frequent part coming from the pressure side with the thin boundary layer. With higher angles of attack, these two spectra become more separated. Hence, the anticipated physical behavior is reproduced by the combined CFD-CAA approach. The computational time for the full sound field around the airfoil was about \unit[15]{hours} on a 16-CPU machine. \par
\subsubsection{Reconstruction of source statistics}
As mentioned in Section~\ref{sec:hybrid_cfd_caa_approach} FRPM realizes turbulent eddies with a local integral length scale $\Lambda$. Referring to \eqn{eq:length_scale}, $\Lambda$ can be calculated from the turbulence statistics given by the preliminary conducted CFD simulation. The FRPM turbulence reconstruction has to be limited to a lower length scale $l_{min}$. The reason for that lies in the spatial resolution of the source patch. The minimum resolved length scale should be greater than four patch cells to avoid numerical errors due to under resolution of the turbulent eddies. Note, values coming from the CFD solution lower than the chosen $l_{min}$ will be set to the value of $l_{min}$. It was found that the length scale distribution over the patch has to be smoothed out after cutting it at the minimum value to avoid spurious peaks in the reconstructed TKE spectrum. Results from test case \#1 were chosen for illustration.
\begin{figure}[tbh!]
\centering
 \includegraphics[width=0.475\textwidth]{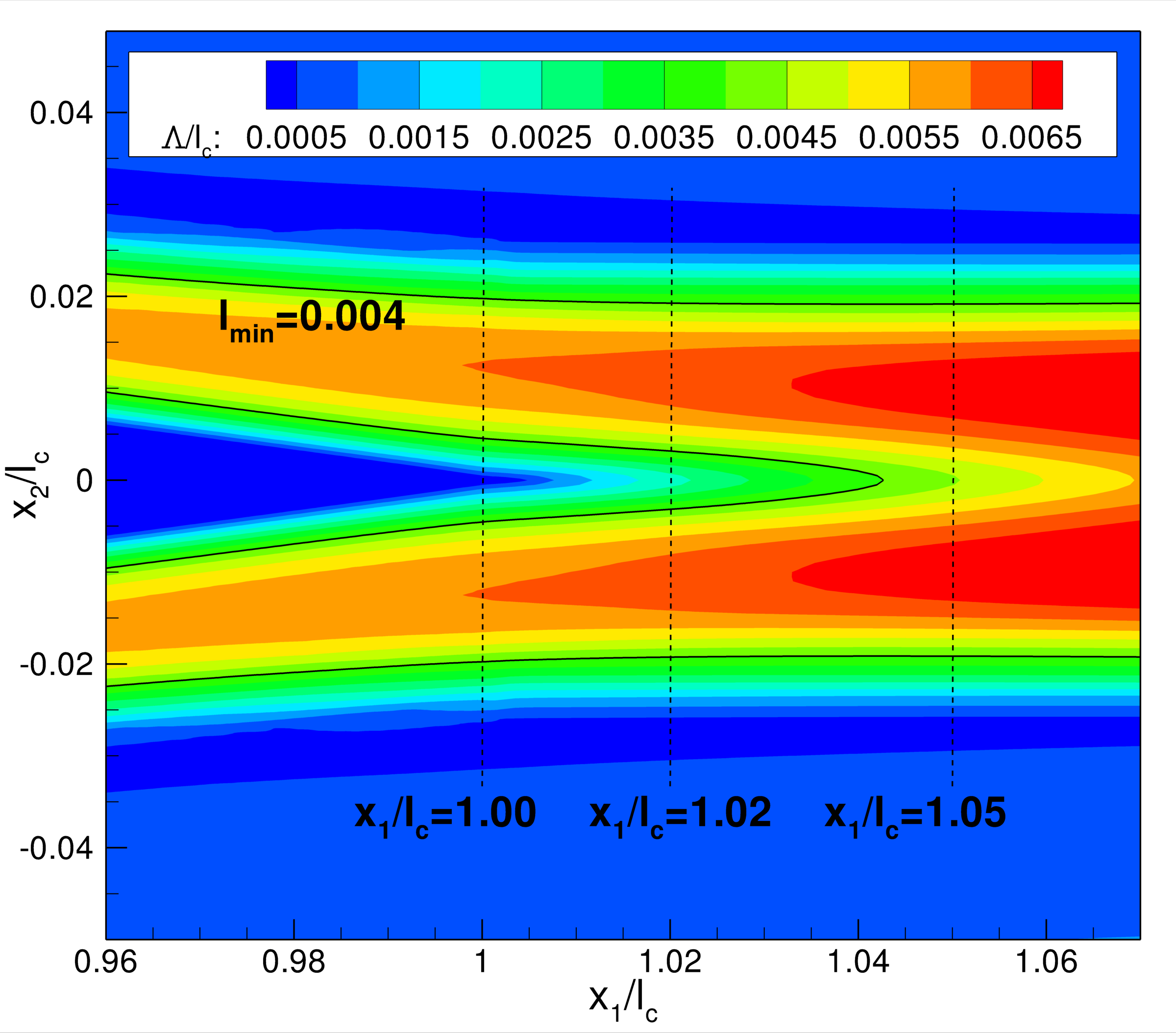}\hfill
 \includegraphics[width=0.475\textwidth]{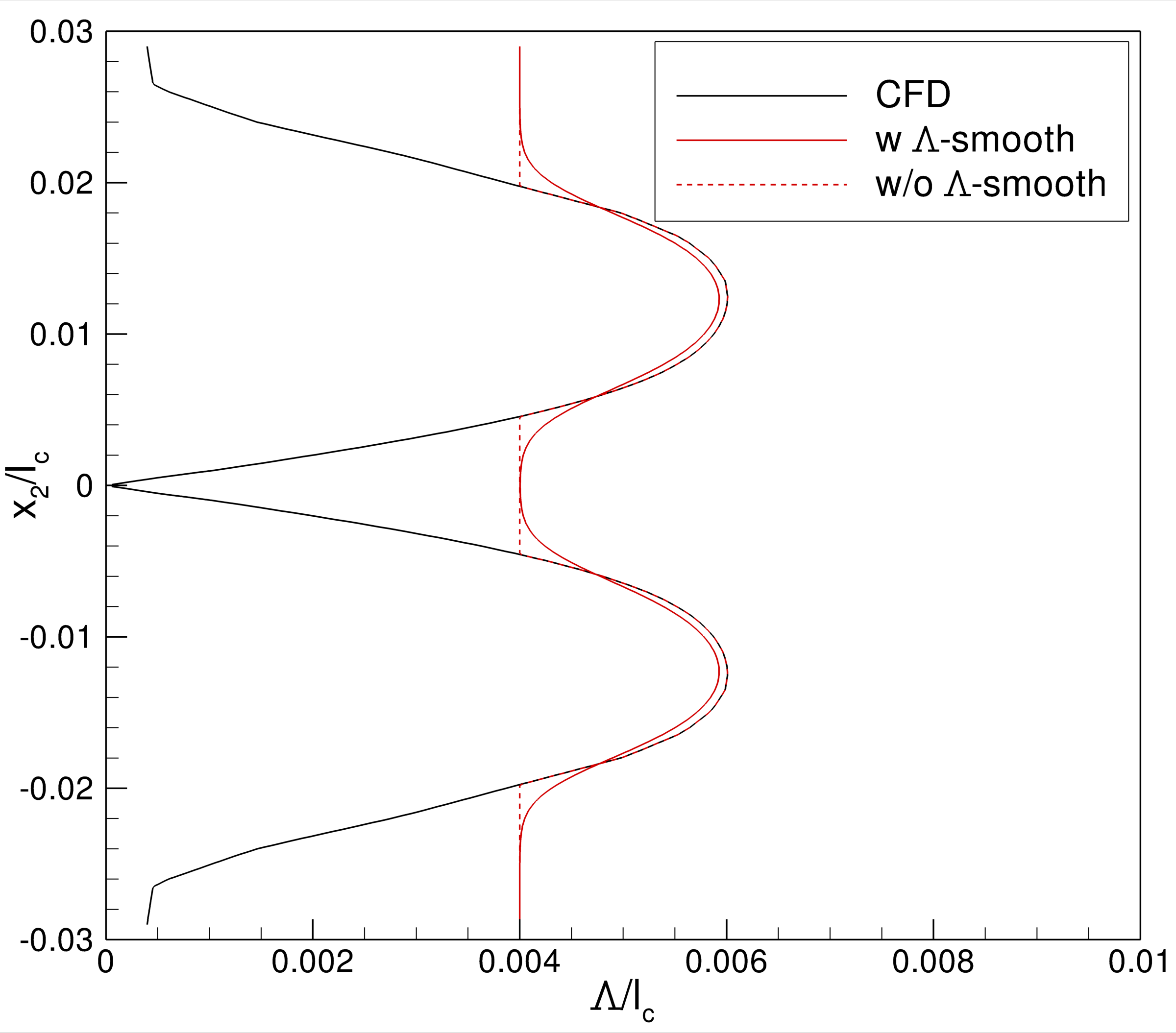}
\caption{(left) Contour plot of $\Lambda$ in the vicinity of the TE; (right) different $\Lambda$ treatments evaluated a $x_1/l_c=1$.}
\label{fig:lmin1}
\end{figure}
\Fig{fig:lmin1} (left) reveals the $\Lambda$ field around the trailing edge. The solid line indicates the chosen minimum length scale (in this case $l_{min}=0.004$). The transition between the variable length scale and the lower threshold $l_{min}$ has to be smoothed in order to avoid a discontinuity in the variable field. The smoothing is shown in \fig{fig:lmin1} (right). The dashed line shows the unsmoothed $\Lambda$-values along a line in $x_2$ direction directly at the trailing-edge. The effect of the smoothing is shown by the solid line.
\begin{figure}[tbh!]
\centering
 \includegraphics[width=0.3\textwidth]{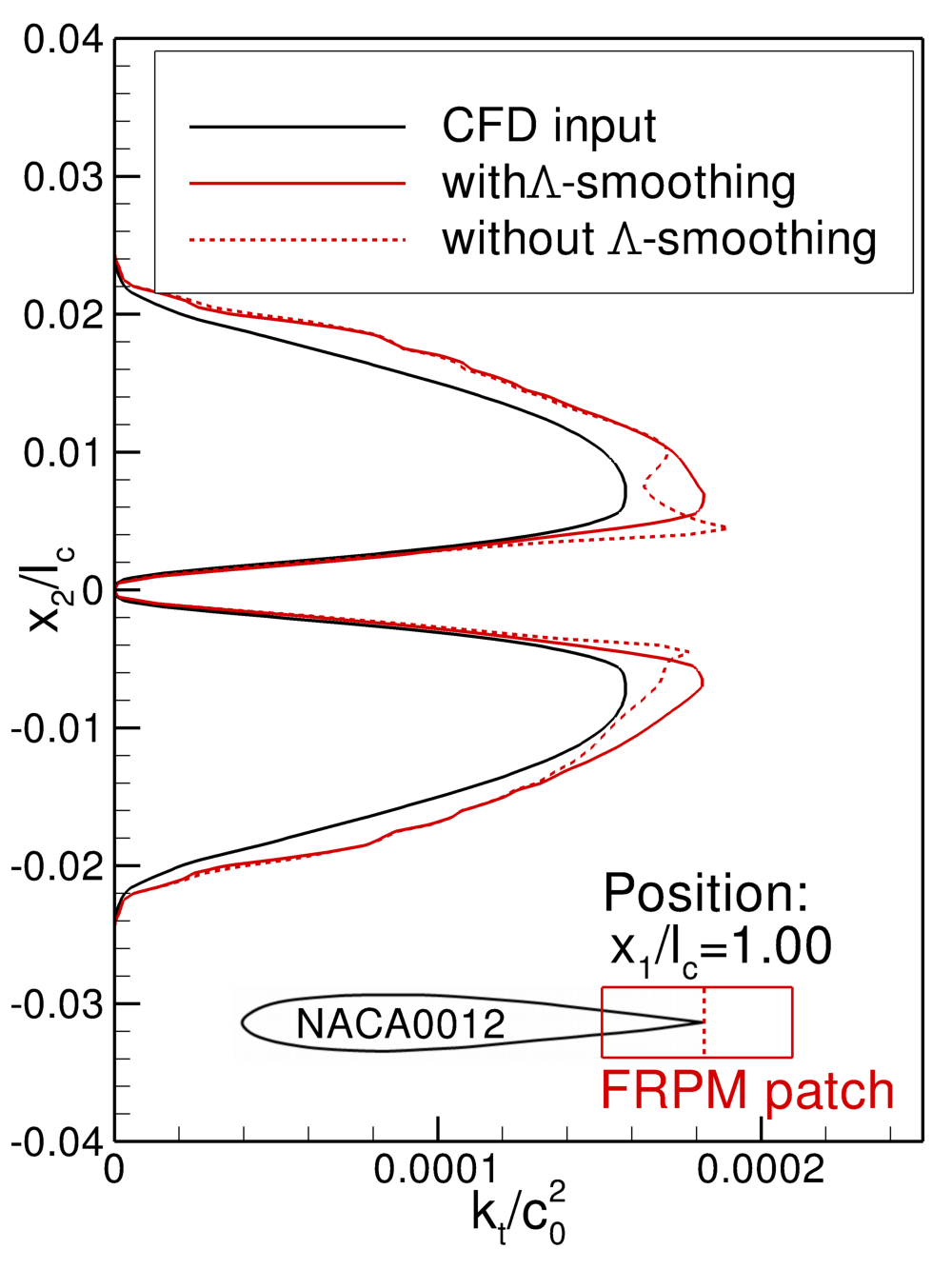}\hfill
 \includegraphics[width=0.3\textwidth]{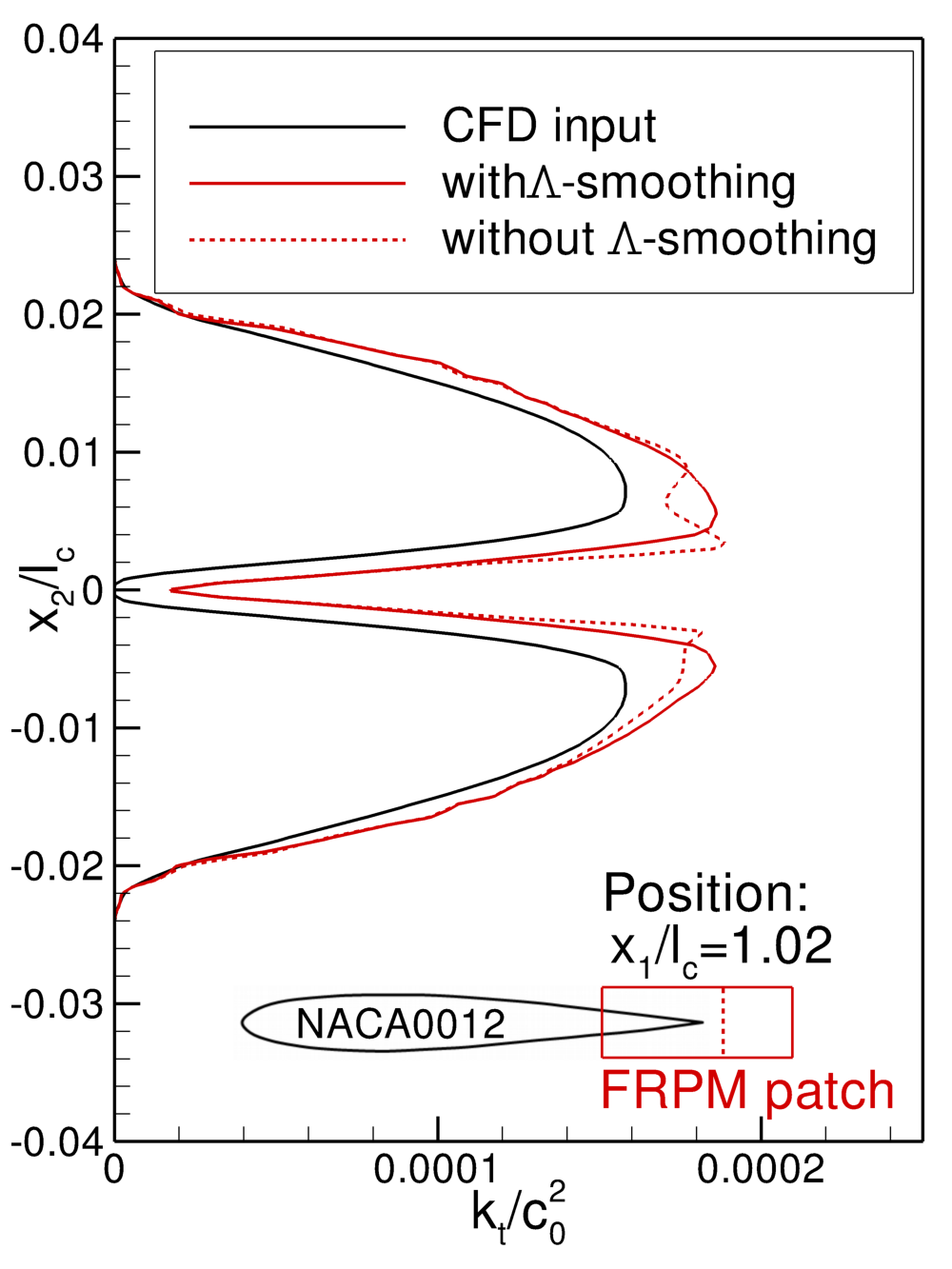}\hfill
 \includegraphics[width=0.3\textwidth]{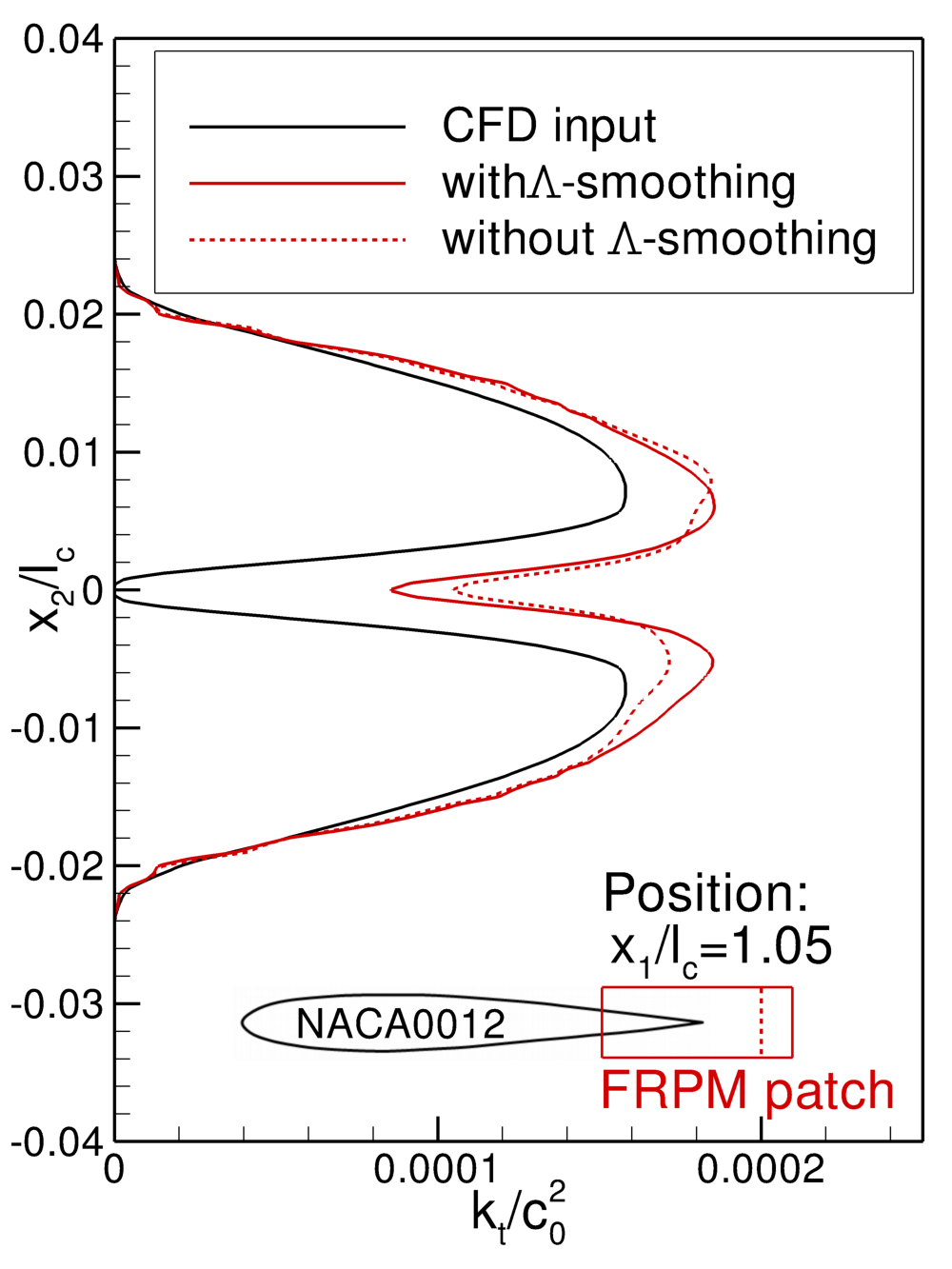}
\caption[FRPM smooth]{Reconstruction of TKE for NACA0012 ($\alpha=\unit[0]{^\circ}$, $Ma=0.1664$) with different $\Lambda$-treatments; (left) normalized TKE at $x_1/l_c=1.00$; (middle) normalized TKE at $x_1/l_c=1.02$; (right) normalized TKE at $x_1/l_c=1.05$.}
\label{fig:lmin2}
\end{figure}
The influence on the reconstructed TKE distribution is shown in \fig{fig:lmin2}. Compared are similar FRPM simulations once with the length scale smoothing (solid lines) and one without (dashed lines). The evaluation positions refer to the dashed black lines in \fig{fig:lmin1}~(left). It can be seen that  the peak in the TKE distribution at the position where the length scale is cut to the lower threshold is reduced by the smoothing of $\Lambda$. Also the symmetrical shape for the TKE profile is better reproduced. Only for the last downwind position at $x_1/l_c=1.05$ the influence is much smaller. This is due to the fact, that there is no cut down of $\Lambda$ in the inner region of the patch as the values here are all above the threshold. In the outer regions the position were $\Lambda<l_{min}$ is in a region with low turbulent kinetic energy. Thus, no influence is visible. It can be concluded that the new smoothing procedure achieves a better reconstruction of the turbulent field in terms of statistics. This will directly lead to a more precise simulation of the sound sources and the related sound field. Generally speaking, a good reproduction of the turbulent field around the trailing-edge and its statistics was achieved with the FRPM code.
\begin{figure}[tbh!]
\centering
\includegraphics[width=0.475\textwidth]{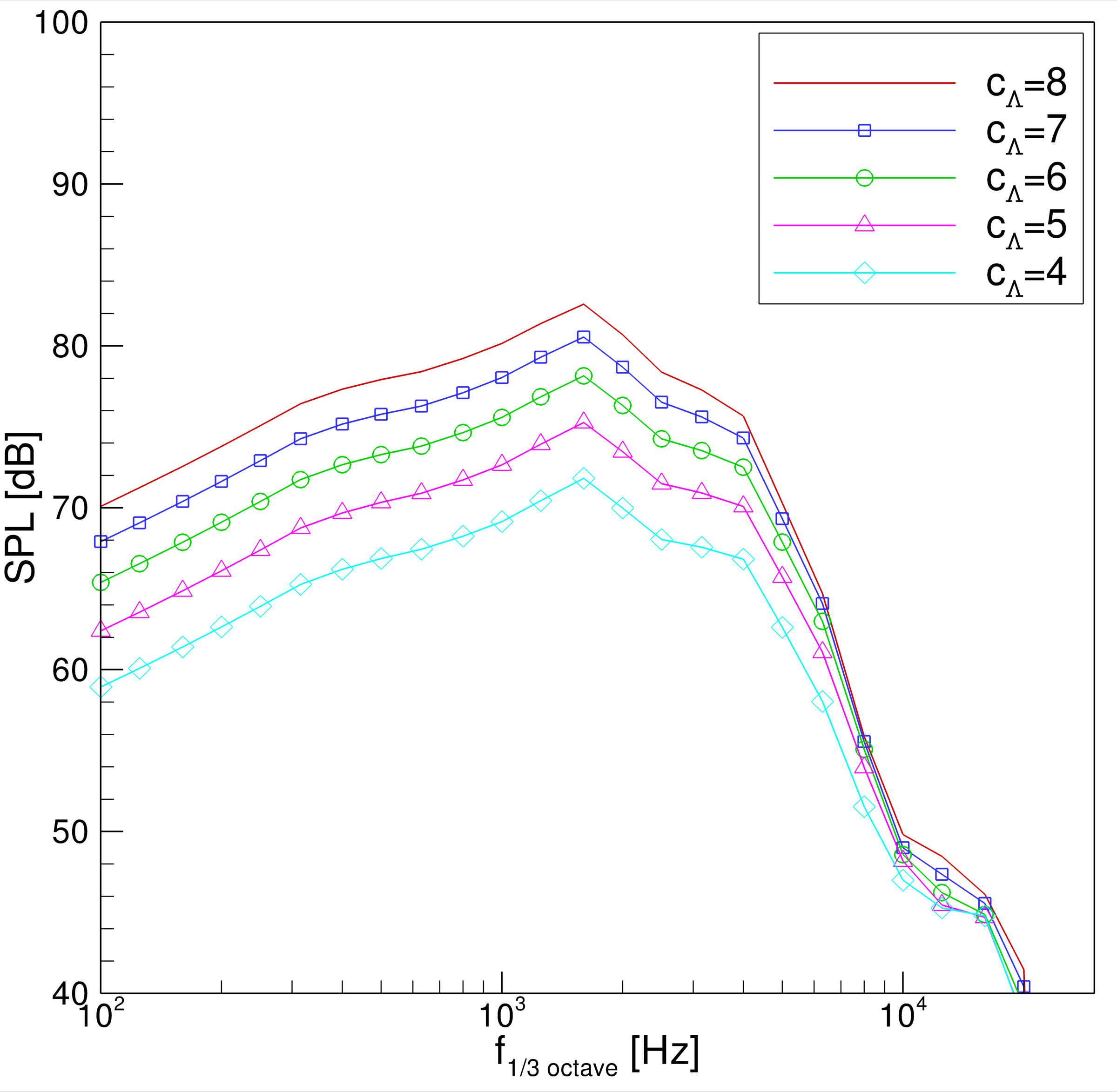}
\caption[Lfac]{One-third octave sound pressure level spectra for the NACA0012 test case with different values for $c_\Lambda = {c_l}/{C_\mu}$, \eqn{eq:length_scale}.}
\label{fig:lfac}
\end{figure}
Another important thing to consider concerning the length scale is the so called length scale factor $c_\Lambda$. \Fig{fig:lfac} shows the NACA0012 test case calculated with different values of $c_\Lambda$.  It can be observed, that for the variation of $c_\Lambda$ the maximum of the spectrum stays at the same frequency of about \unit[1.5]{kHz}, while there is a severe influence on the slope of the decaying spectra and the SPL values. The higher the value of $c_\Lambda$ is chosen, the steeper is the slope and the higher are the absolute SPL values. For the CAA simulations a $c_\Lambda$ value of 8 was chosen, as this value matches the measured length scales (see \fig{fig:cfd_bl}~(right)) and shows good results in turbulence statistic reconstruction. Also, the FRPM grid is able to realize vortices with this length scale. \par
\subsection{Porous trailing edges}
\begin{figure}[tbh]
\centering
\includegraphics[width=0.475\textwidth]{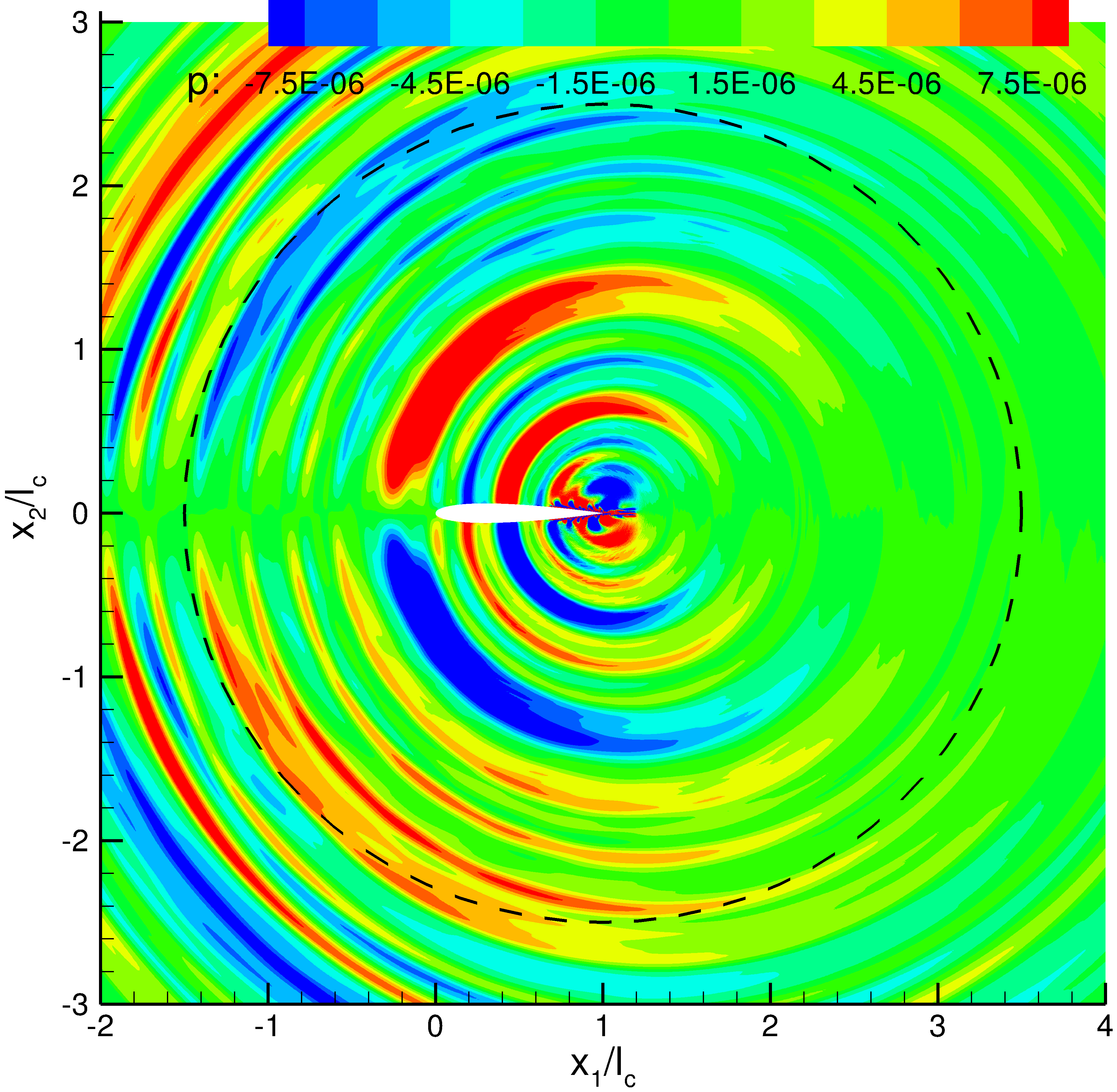} \hfill
\includegraphics[width=0.475\textwidth]{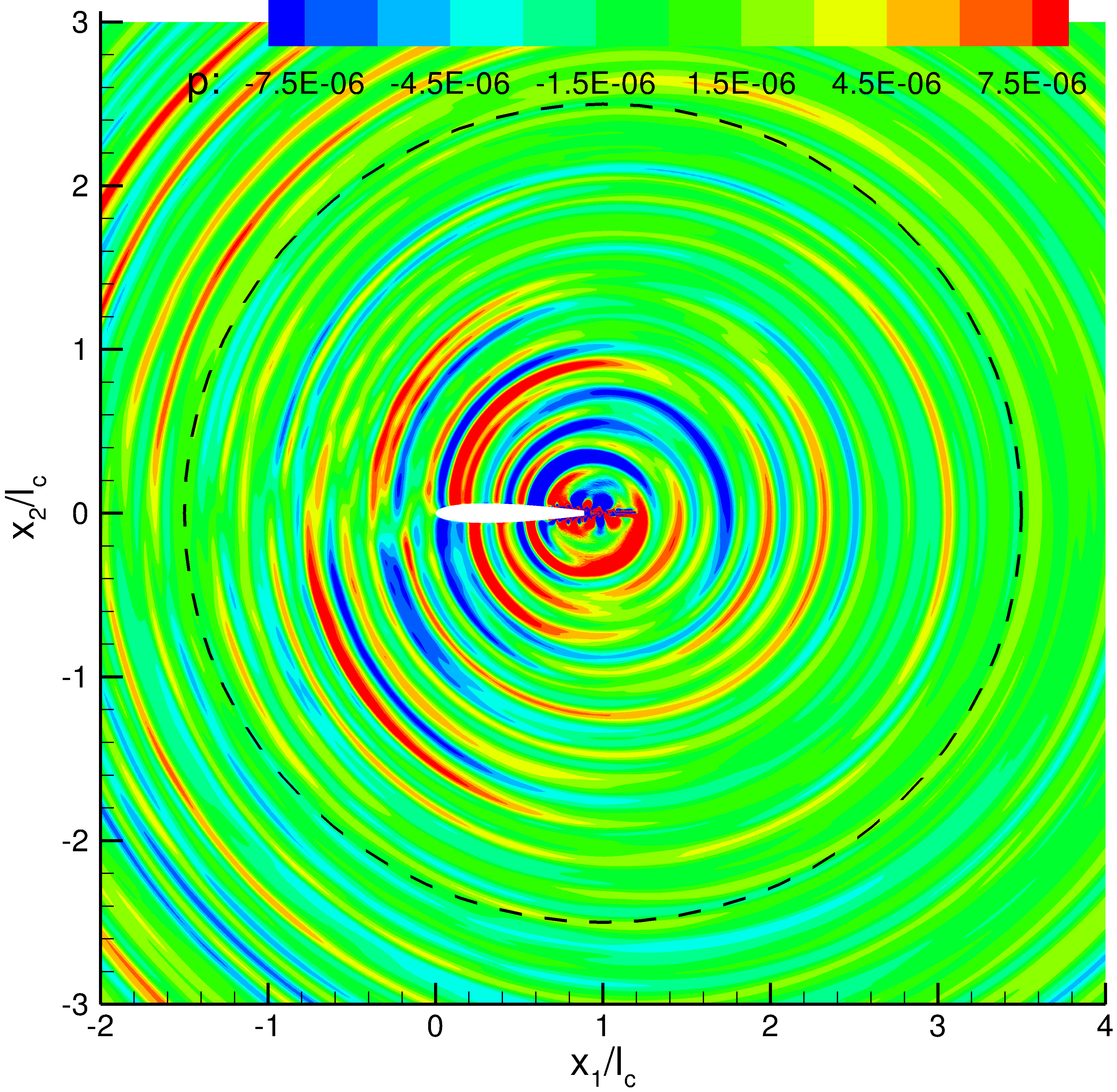}
\caption{Snap shot of the sound field calculated with CAA by means of FRPM-sources of (left) the solid reference case and (right) a trailing edge with the realistic porosity SFF50. The evaluation microphones are indicated by a dashed circle.} \label{fig:caa_contourplot_solid_porous}
\end{figure}

The above mentioned technique of porous inlays was then applied to the prediction of the reduction of TBL-TEN due to a porous trailing edge. The realistic properties of the porous material SFF50 were used. In parallel, the reference set-up with solid trailing edge was computed. The predictions were qualitatively and quantitatively evaluated. The snap shot of the sound field comparing the solid reference case and the airfoil with the porous trailing edge in \fig{fig:caa_contourplot_solid_porous} reveals two details: the porous trailing edge radiates sound at higher frequencies and sound is additionally radiated into direction downstream. For the reference case, the directivity fits the expected behavior of TBL-TEN. \par
\begin{figure}[tbh]
\centering
\includegraphics[width=0.475\textwidth]{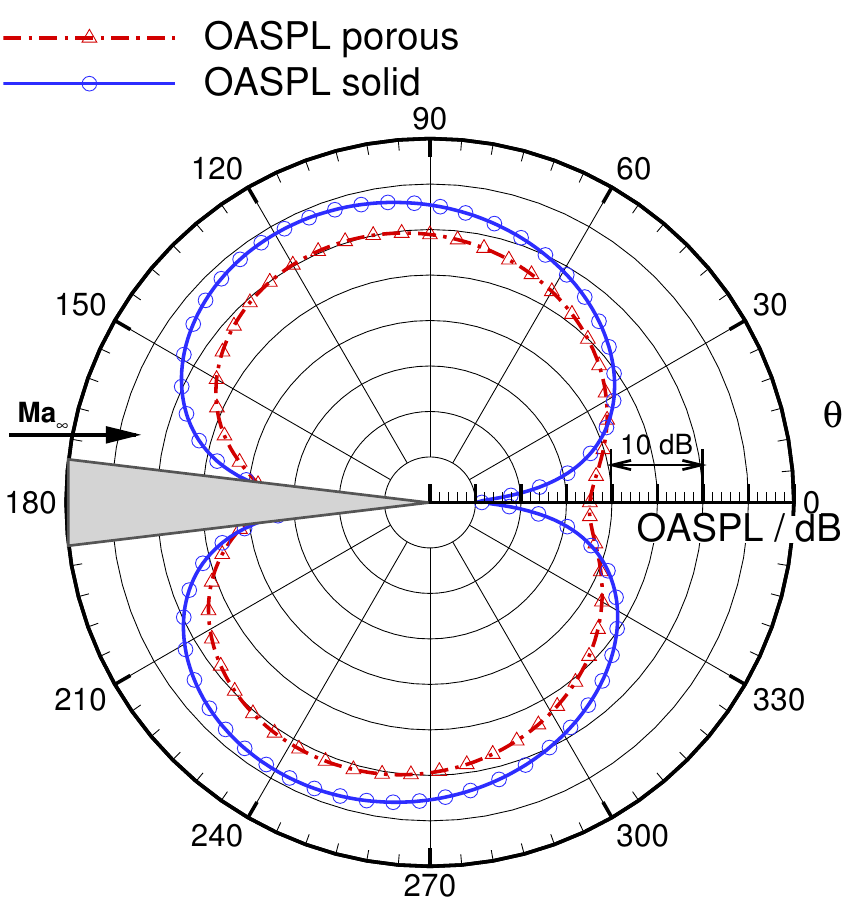} \hfill
\includegraphics[width=0.475\textwidth]{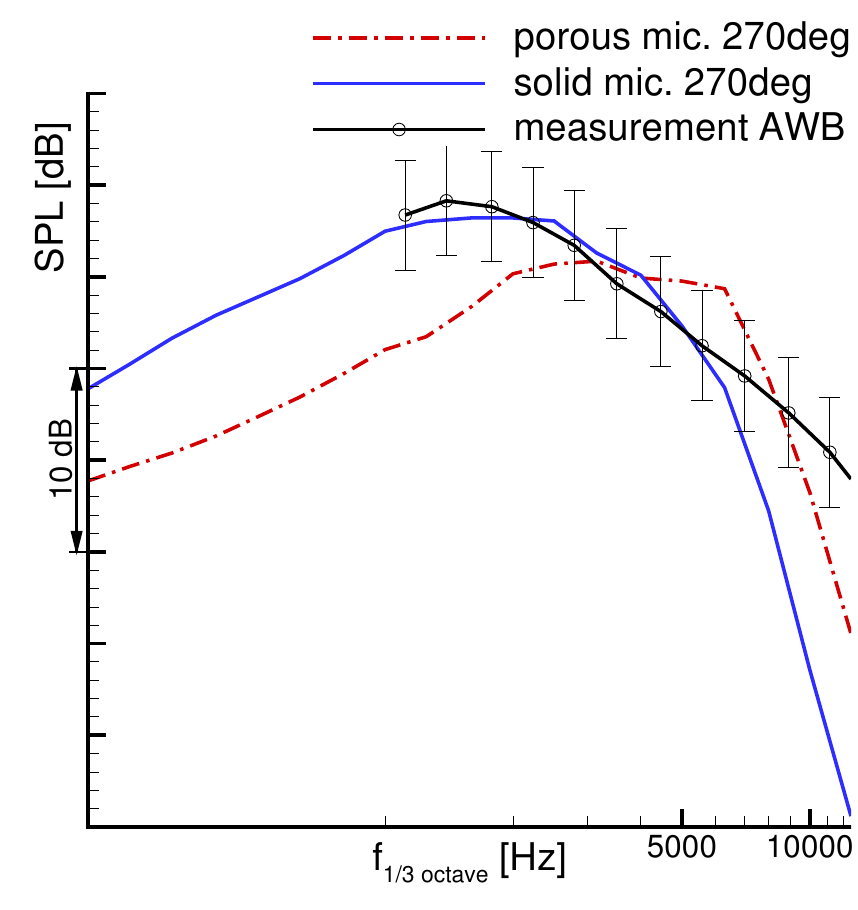}
\caption{(left) directivity of the overall sound pressure level (OASPL) of the solid reference case compared to the case with porous trailing edge; only frequencies up to \unit[2.5]{kHz} were considered; (right) resulting one-third-octave band spectra for the two cases at a microphone position \unit[90]{deg} below the trailing edge compared to a measurement with solid trailing edge, see \citet{Herr.2012}.} \label{fig:caa_DirectivitySprectrum_solid_porous}
\end{figure}

The corresponding directivity is shown in \fig{fig:caa_DirectivitySprectrum_solid_porous}. It gives evidence that radiation of sound in direction downstream occurs. The present data reveal a reduction in direction vertical and upstream of the overall sound pressure level (OASPL) in the range \unit[3]{dB}. Note, for the evaluation of the shown OASPL only frequencies up to \unit[2.5]{kHz} were considered. \par
The spectra presented in \fig{fig:caa_DirectivitySprectrum_solid_porous} exhibits a sound reduction up to \unit[6]{dB} in the low frequency range. The sound pressure level for the porous trailing edge exceeds that of the solid trailing edge above \unit[4]{kHz}. In \citet{Herr.2014}, the experimental results of a cambered F16 airfoil are presented. There, an intersection of the sprectra of the sound pressure levels occurs at frequencies above \unit[10]{kHz}. The maximum resulting sound reduction during the measurements with the F16 airfoil is about \unit[6]{dB} to \unit[8]{dB}. Similar measurements by \citet{Herr.2007} with a NACA0012 airfoil with narrow lengthwise slits at the trailing edge give a similar result of about \unit[6]{dB}. \par

\section{SUMMARY AND CONCLUSION}  \label{sec:Conclusion}
A hybrid CFD-CAA prediction method for the simulation of broadband trailing-edge noise was presented. Details of the underlying theory and the implementation of porous media treatment were given. In a first step, the procedure was used to calculate the acoustic field around a NACA0012 airfoil with solid trailing-edge. The mean flow characteristics of the flow field were determined by CFD simulations. It could be shown that acoustic one-third-octave band spectra were in good agreement with experimental data and thus provide a good basis for the following CAA simulations. A closer look was taken on the influence of simulation parameters on the CAA result. A length scale treatment which reduces spurious peaks in the reconstructed turbulence field was presented. Together with this treatment a best practice procedure for trailing-edge noise was achieved. Sound pressure levels calculated by this procedure were in good agreement with measurements from acoustic wind tunnels and thus prove the validity of the method. Beside the capability of solid trailing-edge the simulation the CFD-CAA method was also applied to a porous trailing-edge. For this purpose extended Acoustic Perturbation Equations from volume averaging have been introduced \par
The method to model porous inlays was tested. To verify the implementation against analytical results, tests of the damping of acoustic waves in homogeneous and anisotropic porous medium were performed first. Further, the porous model was used to predict the sound reduction of an airfoil with porous trailing edge treatment in comparison to a solid airfoil. A decrease of the overall sound pressure level was found for frequencies up to \unit[2.5]{kHz}. Finally, it was demonstrated, that the present approach is able to calculate physically plausible results within reasonable simulation time.
\par
The only drawback of the actual implementation is the need for a smaller timestep if porous treatment is applied with explicit time integration. To overcome this, an IMEX approach will be implemented as proposed by \citet{Ascher.1997} and \citet{Boscarino.2007}. By this, the 2-D porous model can be applied without time step limitation. \par
\section{ACKNOWLEDGMENT}
Financial support has been provided by the German Research Foundation (Deutsche Forschungsgemeinschaft, DFG) in the framework of the Sonderforschungsbereich~880. Computational resources have been provided by German Aerospace Center (Deutsches Zentrum f\"ur Luft- und Raumfahrt e.V., DLR), Institute of Aerodynamics and Flow Technology. \par
The authors thank Michael M\"o{\ss}ner and Kumar Pradeep from Institute of Fluid Mechanics of Technische Universit\"at Braunschweig for providing the CFD of the porous NACA0012 use case. \par

\bibliographystyle{plainnat}
\bibliography{01_fRPM_TE_porous}
\appendix
\section{PROPERTIES OF VOLUME AVERAGING} \label{app:Volume_Averaging}

One property of volume averaging is that it is interchangeable with summation, i.e.
\begin{equation}
	\left \langle a + b \right \rangle^{i,s} 
	= \left \langle a \right \rangle^{i,s} + \left \langle b \right \rangle^{i,s}.
\end{equation}
It follows directly from the definition \eqn{eq:1.1} (just shown for the superficial component)
\begin{equation}
	\left \langle a+b \right \rangle^s  
	= \frac{\int G \left ( a + b \right ) d^3 x'}
	{V_\Delta}  = \frac{\int G a d^3 x'}
	{V_\Delta} + \frac{\int G b d^3 x'}
	{V_\Delta} = \left \langle a \right \rangle^{s} + \left \langle b \right \rangle^{s}.
\end{equation}

Commutation of volume averaging and time differentiation follows immediately from the definitions \eqns{eq:1.1}{eq:filter_volume}, i.e. 
\begin{equation}
	\left \langle \pp{\rho}{t} \right \rangle^s 
	= \frac{1}{V_\Delta}\int \pp{\,G H \rho}{t} d^3 x'  = \pp{}{t}\frac{1}{V_\Delta}\int G H \rho\, d^3 x'
	= \pp{}{t}\left \langle \rho \right \rangle^s.
\end{equation}
 
Commutativity of volume and averaging with respect to spatial differentiation can be also derived from definition \eqn{eq:1.1}, 
\begin{eqnarray} \label{eq:app1.1}
	\left \langle \pp{ \rho  v_i }{x_i} \right \rangle^{s}&=& 
	 \frac{\int G \left ( \bfm x - \bfm x', \Delta \right ) \pp{}{x_i'} \left \{ \rho v_i(\bfm x',t) \right \} H(f(\bfm x'))  d^3 x'}
	{\int G \left ( \bfm x - \bfm x', \Delta \right ) d^3 x'} \\ 
	&= &
	\nonumber
	\frac{\int {G}  
	\pp{}{x_i'} \left \{ \rho v_i(\bfm x',t) H(f(\bfm x'))\right \}  d^3 x'}
	{\int G  d^3 x'} 
	- \underbrace{\frac{\int G \rho v_i \pp{f}{x'_i}  \delta(f(\bfm x')) d^3 x'}
	{ \int G d^3 x' }}_{(i)} 
	.
\end{eqnarray}  
The Dirac delta function in the second of the finally resulting terms results from the spatial derivation of the Heaviside function, i.e. $\partial/\partial x_i (H(f)) = \delta(f) \partial f/\partial x_i$. The volume integration over the delta function reduces the volume integral to a surface integral over all boundaries between solid and fluid phase, where $f=0$. Since the gradient of $f$ is normal to the surface, $\partial f/\partial x_i = n_i$---refer to \fig{fig:Volume_Averaging_Porosity}---the wall normal velocity, i.e. the scalar product of velocity with the wall normal, vanishes ($v_n = v_i n_i = 0$) due to the no-slip condition of the velocity and thus the entire second term $(i)$ vanishes. 

The remaining terms can be manipulated further, i.e.
\begin{eqnarray}
	\nonumber
	\int {G}  
	\pp{\rho v_i H(f)}{x_i'}   d^3 x' &=&
	\underbrace{\int \pp{}{x_i'} \left \{ {G}  
	 \rho v_i H(f)\right \}  d^3 x'}_{Ia}
	 - \underbrace{\int \pp{{G}}{x_i'}  \rho v_i H(f)  d^3 x'}_{Ib} \\
	 &=& \underbrace{\int \pp{{G}}{x_i}  \rho v_i H(f)  d^3 x'}_{II} 
	 = \underbrace{\pp{}{x_i} \left [ \int {G}  \rho v_i H(f)  d^3 x' \right ]}_{III} 
\end{eqnarray}
In a first step the differentiation has been shifted to the filter kernel. By means of Gauss theorem the first integral ($Ia$) will vanish, since the filter function decays for large magnitudes of its argument towards zero. Due to the dependence of the filter function on argument $\bfm x- \bfm x'$, a derivative with respect to $x_i'$ on $G$ can be changed into a derivative with respect to $x_i$. This has been applied to proceed from $Ib$ to $II$. In step $III$, the derivative is shifted out of the integral. Taking into account the last integral, and utilizing that a Gaussian filter function (which is even) satisfies\footnote{The property \eqn{eq:app_denominator} would not hold in general for intrinsic volume averaged quantities, since for the denominator $\partial/\partial x_i \int G H(f) \mathrm{d}^3x'$ it cannot be guaranteed that it is always identical zero.}
\begin{equation} \label{eq:app_denominator}
	\pp{}{x_i} \int {G}  d^3 x' = \int \pp{{G}}{x_i}  d^3 x' = -\int \pp{{G}}{x_i'}  d^3 x' = 0,
\end{equation}
it follows 
\begin{equation}
	\left \langle \pp{ \rho  v_i }{x_i} \right \rangle^{s} = 
	 \pp{}{x_i} \left \{ \left \langle \rho \right \rangle^{s} \left [v_i \right ] \right \} 
	 = \pp{}{x_i} \left \langle \rho v_i \right \rangle^{s} = \pp{}{x_i} \left \{ \frac{\int {G}  \rho v_i H(f)  d^3 x'}{\int {G}  d^3 x'} \right \}.
\end{equation}
This result proves commutativity of superficial volume averaging and spatial differentiation for the term $\rho v_i$, \eqn{eq:1.6}. Note, strict satisfaction of commutativity is a result of the no-slip condition that causes the vanishing of term $(i)$ in \eqn{eq:app1.1}. For other terms this is not in general the case. For example, the previous steps applied to variable pressure $p$ instead of $\rho v_i$ yields for this term~$(i)$
\begin{equation} \label{eq:app1.2}
	f^p_i = \frac{\int G \; p n_i \;   \delta(f(\bfm x'))  d^3 x'}
	{ V_\Delta } = \frac{1} {V_\Delta} \int_{S} G \; p n_i \;   dS.
\end{equation}
This represents the pressure force from the solid onto the fluid. Consequent application of the volume averaging procedure to the continuity and momentum equation yields 
\begin{equation}
	\left \langle \pp{\rho }{t} +  \pp{\rho v_i }{x_i} \right \rangle^s =
	\pp{\left \langle \rho \right \rangle^s}{t} 
	+  \pp{}{x_i} \left \{ \left \langle \rho \right \rangle^s \left [v_i \right ]\right \}
	=0
\end{equation}
and  
\begin{eqnarray}
	&&\left \langle \pp{\rho v_i}{t} +  \pp{\rho v_i v_j}{x_j} + \pp{p}{x_i} - \pp{\tau_{ij}}{x_j} \right \rangle^s =\\ 
	\nonumber
	&& \pp{\left \langle \rho \right \rangle^s \left [ v_i \right ]}{t} 
	+  \pp{\left \langle \rho \right \rangle^s \left [v_i \right ]\left [v_j \right ]}{x_j} 
	+ \pp{\left \langle p \right \rangle^s}{x_i} - \pp{\left \langle \tau_{ij}\right \rangle^s}{x_j} 
    - \mathcal{F}_i=0
\end{eqnarray}
with 
\begin{equation}
	\mathcal{F}_i=f_i - \underbrace{ \pp{\left ( \rho  v_i v_j  - \left \langle \rho \right \rangle^s \left [v_i \right ]\left [v_j \right ]\right )}{x_j}}_{SFS}.
\end{equation}
Here the interchangeability of volume averaging and summation has been applied. Furthermore, for all gradient terms commutativity of volume averaging and differentiation has been used.

The volume force $f_i$ represents the effect of term $(i)$ from pressure $p$ as shown by \eqn{eq:app1.2} and a similar shear stress contribution from $\tau_{ij}$. Due to the no-slip condition satisfied by velocity, the term $\rho v_i v_j$ will provide no contribution to $f_i$. Note, volume averaging in general creates the sub-filter stress ($SFS$) indicated above from the sub-filter velocity fluctuations $v_i'' = v_i - \left [ v_i \right ]$. The sub-filter velocity component must be distinguished from the Favre fluctuations of the volume averaged velocity defined in \eqn{eq:volav_vi_decomposition}. To be precise, the complete velocity is given by 
\begin{equation}
	v_i = \widetilde{\left [ v_i \right ]}  + \left [ v_i \right ]'' + v_i''.
\end{equation} 
The volume force term and sub-filter contributions inside the porous medium are modeled by the ansatz proposed by Darcy and Forchheimer. The Darcy terms describes a velocity proportional volume force, whereas the Forchheimer term explicitly addresses a quadratic contribution to the force,
\begin{equation}
    - \mathcal{F}_i = \underbrace{\phi \frac{\left \langle \mu \right \rangle^s}{\kappa} [v_i]}_{\text{Darcy terms}} + \underbrace{\left\langle \rho \right\rangle^{s}\phi^2\frac{c_\ind{F}}{\sqrt{\kappa}}\sqrt{[v_k][v_k]} [v_i]}_{\text{Forchheimer terms}} \komma
\end{equation}
where $\phi$ denotes the porosity, $\mu$ is the dynamic viscosity, $\kappa$ identifies the permeability and the Forchheimer coefficient is indicated by $c_\ind{F}$.

Despite the additional terms, the resulting momentum equation corresponds formally with the Navier-Stokes equations in conservative form. They can be reformulated in primitive form. For example, the momentum equation is rewritten into an equation for the time derivative of velocity by removing with the help of the continuity equation the time derivative of density. The independent variables can be split further into resolved Favre averaged and fluctuating components for the velocity, i.e. \eqn{eq:volav_vi_decomposition}, and mean and fluctuating components for pressure and density as defined by \eqn{eq:volav_eps_decomposition}. Introducing the decomposition into the equations in primitive notation yields non-linear equations in disturbance form. Next, mean flow terms can be neglected by subtracting the mean of a disturbance equation from itself. Finally, non-linear terms might be neglected. This way, the governing equations can be reformulated into perturbation form as shown in Section~\ref{paragraph:governing_equations}. For example, the volume averaged continuity equations is rewritten as 
\begin{equation}
	\pp{{\left \langle \rho' \right \rangle^s}^\prime}{t} 
	+  \pp{}{x_i} \left \{ {\left \langle \rho \right \rangle^s}^\prime \widetilde{\left [v_i \right ]} + 
		\overline{\left \langle \rho \right \rangle^s} \left [v_i \right ]''\right \}
	= 0
\end{equation}
Utilizing the notation as introduced by \eqns{eq:notation1}{eq:notation2}, the previous equation agrees with \eqn{eqn:conti_APE}. For the derivation of the momentum equation in the from used in the APE (with some left-hand side terms shifted to the right-hand side to define appropriate vortex sound sources), the Darcy and Forchheimer terms have to be linearized as well, finally yielding the equation as outlined in \eqn{eqn:momentum_APE}. The APE pressure equation is obtained by substituting $\rho'$ with $p'/c_0$, where $c_0$ denotes the speed of sound in the free fluid. 

It may be emphasized (without explicit proof) that for a Gaussian filter function
\begin{equation}
	G_\Delta \left (\bfm x - \bfm x', \Delta \right ) 
	= \frac{1}{\left ( 2\pi \right )^{3/2} \Delta^3}\exp \left [ -\frac{\left | \bfm x - \bfm x' \right |^2}{2\Delta^2} \right ],
\end{equation}
the consecutive application of volume averaging to variable $a$ with different length scales for each filter step, i.e. 
\begin{equation}
	\left \langle \left \langle a \right \rangle^s_{\Delta_1} \right \rangle^s_{\Delta_2} = 
	\frac{1}{V_{\Delta_2}} \int G_{\Delta_2} \left \{ 
		\frac{1}{V_{\Delta_1}}\int G_{\Delta_1} a d^3 x'' \right \} ,
	d^3 x'
\end{equation}
is equal to one filter step with length scale $\Delta_1+\Delta_2$, i.e. 
\begin{equation}
	\left \langle \left \langle a \right \rangle^s_{\Delta_1} \right \rangle^s_{\Delta_2} = 
	\left \langle a \right \rangle^s_{\Delta_1+\Delta_2}.
\end{equation} 
For $\Delta_1 \to 0$ the first filter would represent a Dirac delta function, that is applicable for a pore size tending to zero, i.e. $D_p \to 0$. Consequently, volume averaged (mean and fluctuating) quantities based on length scale $\Delta_2$ can be obtained by consecutively volume average quantities initially obtained from volume averaging with $\Delta_1 \to 0$ with a filter of length scale~$\Delta_2$.

\section{MODIFIED WAVE EQUATION} \label{app:Wave_Equation}
We will verify the proper implementation of the extra terms describing the effect of porosity by comparing simulated results for a plane wave test problem with the analytical solution. For this we consider the simplified case of a medium at rest and a homogeneous porous material that fills out the computational domain. The effect of an anisotropic material is taken into account by considering the matrix $\mu_{ij}$ to represent a general symmetric matrix. Based on these prerequisites (i.e. $v^0_i=0$, $p^0=const.$, $\rho^0=const.$, $\phi=const.$), the pressure and momentum equation of the linearized Euler equations (as well as Acoustic Perturbation Equation) reduce to
\begin{eqnarray}
	\label{eq:A.1}
 	&& \pp{v'_i}{t} + \frac{\phi}{\rho^0} \pp{p'}{x_i} + \mu_{ij} v'_j = 0 \\
 	\label{eq:A.2}
 	&& \pp{p'}{t} + \frac{\gamma p^0}{\phi} \pp{v_i'}{x_i}= 0.
\end{eqnarray} 
We consider the resulting porous wave equation by introducing an acoustic potential that satisfies
\begin{equation} \label{eq:A.4}
	v'_i = \pp{\varphi}{x_i}.	
\end{equation}
Based on this expressions and using \eqn{eq:A.2} the time derivatives of pressure and velocity can be expressed via
\begin{eqnarray} \label{eq:A.3}
 	&& \pp{v'_i}{t}  = \ppsq{\varphi}{t}{x_i}\\
 	&& \pp{p'}{t}  = -\frac{\gamma p^0}{\phi} \ppsd{\varphi}{x_i}.
\end{eqnarray} 
By taking the time derivative of \eqn{eq:A.1} and substituting in this resulting expression all the occurring time derivatives of fluctuating pressure and velocity with the previous expressions, a scalar (wave) equation for the acoustic potential is obtained reading
\begin{equation} \label{eq:A.5}
	\pp{}{x_i} \left (  \ppsd{\varphi}{t} - c_0^2 \ppsd{\varphi}{x_i} \right ) + \mu_{ij} \ppsq{\varphi}{t}{x_j} = 0. 
\end{equation}
Note, the first term in brackets alone represents the wave equation for a medium at rest. The speed of sound\footnote{Due to almost constant perturbation variables within the characteristic filter volume---even across an interface between fluid and the porous medium---the Favre fluctuation of the speed of sound vanishes, i.e. $\left[ c_0^2 \right]^{\prime\prime} \to 0$. It finally is $p^\prime = c_0^2 \rho^\prime$.} in the free fluid\footnote{The sound propagation in the porous material is dispersive, i.e. the phase and the group velocity differ. Please see below} resulting therein is obtained from the mean-flow variables as usual as $c_0 = \sqrt{\gamma p^0/\rho^0}$. It is $\gamma=1.4$ the isentropic exponent of ambient air. The additional term represents the effect of the porosity. To solve this equation system we use coordinates aligned with the main axis of the porosity matrix $\mu_{ij}$. \par
For this we apply a Galelian transformation to the previous equation using the transformation rules $\partial/\partial t \to \partial/\partial t'$ and $\partial/\partial x_i \to \partial/\partial x_i'$. In the new coordinate system the porosity matrix is diagonal with the diagonal elements prescribed by the corresponding eigenvalues $\lambda_i$ of $\mu_{ij}$, i.e. $\mu_{ij} \to \mu'_{ij} = \mbox{diag}(\lambda_i)$ and $\mu'_{ij}\,\partial/\partial x_j'=\lambda_i\,\partial/\partial x_i'$. Hence, in the rotated coordinate system the equation becomes
\begin{equation} \label{eq:A.6}
	\pp{}{x_i^\prime} \left(  \ppsd{\varphi}{{t^\prime}} - c_0^2 \ppsd{\varphi}{{x_i^\prime}} \right) + \lambda_i \ppsq{\varphi}{t^\prime}{x_i^\prime} = 0. 
\end{equation} 
Eventually we make the ansatz to describe the acoustic potential as a superposition of $n$ contributions (in $n$-D) corresponding to each coordinate direction in the main axis system. For example, in 2-D, the acoustic potential is specified via
\begin{equation}
	\varphi( x_1^\prime,x_2^\prime,t^\prime) = \varphi_1(x_1^\prime,t^\prime) + \varphi_2(x_2^\prime,t^\prime).
\end{equation}
Introducing this ansatz into \eqn{eq:A.6} we obtain (in 2-D) two independent acoustic equations with individual damping term for each contribution to the acoustic potential, viz.
\begin{equation}
	\ppsd{\varphi_i}{{t^\prime}} - c_0^2 \ppsd{\varphi_i}{{x_j^\prime}} + \lambda_{(i)} \pp{\varphi_{i}}{t^\prime} = 0. 
\end{equation}
Note, indices with round brackets indicate suppression of the summation-over-equal-indices rule. Further note, whereas the acoustic particle velocity can be deduced directly by means of \eqn{eq:A.4} from the acoustic potential, the pressure fluctuations is related to the acoustic potential via \eqn{eq:A.1}. In the main axis system, the relationship(in 2-D) becomes
\begin{equation} \label{eqn:app_acoustic_pressure}
	 p' =  -\frac{\rho^0}{\phi}  \left( \pp{\varphi_1}{t} + \lambda_1 \varphi_1 + \pp{\varphi_2}{t} + \lambda_2 \varphi_2 \right) \punkt 
\end{equation}
If we consider harmonic signals with $\hat \varphi_j(x_j,t)  = \hat A_j \exp\left (i \left ( \omega t - k_{(j)} x_j \right )\right )$, the corresponding pressure fluctuations reads
\begin{equation}
	\hat p = -\frac{\rho^0}{\phi}  \left[ \left( i \omega + \lambda_1\right)\hat\varphi_1 + \left( i \omega + \lambda_2\right)\hat\varphi_2  \right] \punkt 
\end{equation}
Furthermore, by introducing the harmonic ansatz into the wave-equations we obtain the dispersion relations 
\begin{equation}
	-\omega^2 + k_j^2 c_0^2 + i\omega \lambda_j = 0 .
\end{equation} 
By means of the dispersion relations, the wave number $k$ can be expressed as a function of angular frequency $\omega$,
\begin{equation} 
  k_j = \frac{\omega}{c_0} \sqrt{1 - i \, \frac{\lambda_j}{\omega}} ,
\end{equation}
yielding the damped wave results
\begin{equation} \label{eqn:app_acoustic_potential_expanded}
	\hat \varphi_j = \hat A_j 
	\exp\left(i\omega \left( t - \frac{1}{c_0}\sqrt{\frac{1}{2}\left[ \sqrt{\frac{\lambda_j^2}{\omega^2} + 1} + 1 \right] } \; x_j\right) \right) %
	\exp\left( - \frac{\omega}{c_0}\sqrt{\frac{1}{2}\left[ \sqrt{\frac{\lambda_j^2}{\omega^2} + 1} - 1 \right] } \; x_j \right)
\end{equation}
%

%
\end{document}